\documentclass[useAMS,usenatbib]{mn2e}

 \usepackage{ifpdf}
\usepackage{amsmath}

\usepackage{textcomp}
\usepackage{adjustbox}
 \usepackage{blindtext}

\usepackage{rotating,graphicx}
 \usepackage{hyperref}
\usepackage{epstopdf}
\usepackage{epsfig}
\usepackage{float}
\usepackage{subfig}
\usepackage{textcomp}
\usepackage{lmodern}% http://ctan.org/pkg/lmodern
\usepackage{slantsc}% http://ctan.org/pkg/slantsc
\usepackage{amssymb}
\newcommand{\aap}{A\&A}          % Astronomy & Astrophysics
\newcommand{\aj}{AJ}             % Astronomical Journal
\newcommand{\apj}{ApJ}           % Astrophysical Journal
\newcommand{\apjs}{ApJS}         % The Astrophysical Journal Supplement Series
\newcommand{\apjl}{ApJL}         % The Astrophysical Journal Letters
\newcommand{\mnras}{MNRAS}       % Monthly Notices of the RAS
\newcommand{\pasp}{PASP}         % Publications of the Astronomical Society of the Pacific
\newcommand{\procspie}{Proc. SPIE} 

\title[Internal kinematics of the H{\textsl{\textsc{ii}}} regions in interacting and isolated galaxies]
{Comparative internal kinematics of the H{\sc{ii}} regions in interacting and isolated galaxies: 
implications for massive star formation modes.}
\author[Zaragoza-Cardiel et al.]{Javier Zaragoza-Cardiel$^{1,2}$\thanks{E-mail:jzc@iac.es},
 John E. Beckman$^{1,2,3}$, Joan Font$^{1,2}$,\newauthor 
 Bego\~na Garc\'ia-Lorenzo$^{1,2}$,
 Artemi Camps-Fari\~na$^{1,2}$, Kambiz Fathi$^{4}$, Philip A. James$^{5}$, \newauthor Santiago Erroz-Ferrer$^{1,2}$,
Jorge Barrera-Ballesteros$^{1,2}$, 
 Mauricio Cisternas$^{1,2}$\\
$^{1}$Instituto de Astrof\'isica de Canarias, C/ V\'ia L\'actea s/n, 38205 La Laguna, Tenerife, Spain\\
$^{2}$Department of Astrophysics, University of La Laguna, E-38200 La Laguna, Tenerife, Spain\\
$^{3}$CSIC, 28006 Madrid, Spain\\
$^{4}$Stockholm Observatory, Department of Astronomy, Stockholm University, AlbaNova, 106 91 Stockholm, Sweden\\
$^{5}$Astrophysics Research Institute, Liverpool John Moores University, 146
Brownlow Hill, Liverpool L3 5RF, UK\\
}

\begin{document}

\date{}

\pagerange{\pageref{firstpage}--\pageref{lastpage}} \pubyear{}

\maketitle

\label{firstpage}

\begin{abstract}
We have observed 10 interacting galaxy pairs using the 
Fabry-Perot interferometer GH$\alpha$FaS (Galaxy H$\alpha$ Fabry-Perot system) on
the $4.2\rm{m}$ William Herschel Telescope (WHT) at
the Observatorio del Roque de los Muchachos, La Palma. 
We present here the H$\alpha$ surface brightness, velocity and 
velocity dispersion maps for the 10 systems we have not previously observed using this technique, 
as well as the physical properties (sizes, H$\alpha$ 
luminosities and velocity dispersion) of 1259 H{\sc ii} regions from the full sample.
We also derive the physical properties of 1054 H{\sc ii} regions in a sample of 
28 isolated galaxies observed 
with the same instrument in order to compare the two populations of H{\sc ii} regions. 
We find a population of the brightest H{\sc ii} regions for which the 
scaling relations, for example the relation between the H$\alpha$ luminosity and the radius, 
are clearly distinct from the relations for the regions of lower luminosity. 
The regions in this bright population are more frequent in the interacting galaxies. 
We find that the turbulence, and also the star formation rate, are enhanced in the 
H{\sc ii} regions in the interacting galaxies. 
 We have also extracted the H$\alpha$ equivalent widths for the H{\sc ii} regions of both samples, and we have found that
the distribution of H{\sc ii} region ages coincides for the two samples of galaxies.
We suggest that the SFR enhancement is 
brought about by gas flows induced by the interactions, which give rise to 
gravitationally bound gas clouds which grow further by accretion from the 
flowing gas, producing conditions favourable to star formation.
\end{abstract}


\section{Introduction}

Galaxy mergers play an important role in galaxy evolution \citep{1978MNRAS.183..341W}, but it is
still not clear whether, and if so how, they trigger star formation 
\citep{Somerville08,2011EAS....51..107B,Tadhunter11},  
 whether they stimulate nuclear activity 
\citep{Canalizo07,Bennert08,Georgakakis09,2011ApJ...726...57C,Ramos11,Ramos12,Bessiere12}, 
and what are the mechanisms by which they produce 
new structures such as tails, bars and warps. They are important sites for the  
feedback processes, which are considered to be of considerable importance in producing realistic galaxies 
in the context of $\Lambda$CDM cosmological models. 
\citep{2010MNRAS.409.1088B,2013arXiv1311.2073H}

Because mergers imply distortions quantifiable most accurately by kinematical studies of the galaxies as a whole, 
3D studies of galaxy mergers are extremely important. Recently, 
\cite{2014arXiv1409.6791W} published a 3D study of galaxies 
with redshift $0.7<z<2.7$ claiming that the majority of 
galaxies are star forming and turbulent discs dominated by rotation and therefore not strongly affected by mergers. 
However, nearby galaxy studies can give us better 
clues in practical terms about how to study star forming systems at higher redshifts, since we 
can achieve far better spatial resolution.

Star formation enhancement in galaxy collisions is theoretically well predicted  
for the central parts of the galaxies since 
collisions of galaxies tend to produce linear structures, inducing strong gas
inflows towards the nucleus \citep{1996ApJ...464..641M}. However, the off-nuclear peaks of star formation, such 
as are clearly present in the Antennae galaxies, are much less well understood, and they are likely to be at least 
as important \citep{2010MNRAS.409.1088B,2014AJ....147...60S}.

H$\alpha$ emission is one of the most useful indicators of massive star formation in galaxies. 
 Observing its full spectral profile is a very powerful tool for disentangling kinematics, but simply mapping the 
H$\alpha$ surface brightness of a galaxy is already a comprehensive way to gain insight into global star formation.  
\cite{1989ApJ...337..761K,2006A&A...459L..13B} studied the properties of the H{\sc ii} region populations, and in particular their 
luminosities, finding a break in the luminosity function at  $\log L_{\rm{H\alpha}}=38.6\thinspace\rm{dex}$, which some 
authors \citep{2000AJ....119.2728B} proposed could be used as a distance calibrator for 
galaxies. However, a physical basis for the break is desirable in order 
to strengthen its use as a distance calibrator, and while scenarios have been proposed 
\citep{2000AJ....119.2728B,zaragoza13}, a clear explanation has not so far been agreed on. 

\cite{1981MNRAS.195..839T} studied the scaling relations of extragalactic H{\sc ii} regions 
measuring the sizes, luminosities, and velocity dispersions. They found that 
the luminosity, the size, and the velocity  dispersion  were correlated ($L\propto R^2$ and $L\propto \sigma_v^4$).
The $L-\sigma_v$ relation has also been proposed as a distance calibrator 
\citep{1981MNRAS.195..839T,chavez}. For this it is desirable 
 to agree on a physical basis for the driver of the turbulence in H{\sc ii} regions, 
but there are a number of possible processes, which may combine in different 
proportions to produce the observed line widths. The studies by \citep{1981MNRAS.195..839T,zaragoza13,zaragoza14,chavez} 
 point towards self-gravity, at least in the more luminous regions,
while \cite{1979A&A....73..132D,1984A&A...141...49H,1984A&A...130...29R,1994ApJ...425..720C} favour supernova explosions and stellar winds.
\cite{gutierrez11} used HST data from M51 to extract sizes and luminosities 
with unprecedented resolution 
and found similar scaling relations to those cited in the $L_{\rm H\alpha}-R$ relation. 

 However, our recent results \citep{zaragoza13} found different scaling relations for the 
brightest set of H{\sc ii} regions  in the interacting pair of galaxies Arp 270. We have found 
the same discrepancy 
in the H{\sc ii} regions of the Antennae galaxies \citep{zaragoza14} who explain this as 
arising from two distinct populations of molecular clouds as they find an equivalent dichotomy 
in the molecular cloud population of the Antennae. This change in the scaling relations changes the behaviour 
of the star formation rate (SFR, derived from H$\alpha$ luminosity) dependency 
on the parameters (mass, radius) of the star forming regions.

Here we present new observations of 10 interacting galaxies with the Fabry-Perot interferometer 
GH$\alpha$FaS (Galaxy H$\alpha$ Fabry-Perot System). We have measured the basic physical parameters
($L_{\rm H\alpha}$, $R$, and $\sigma_v$) 
of each H{\sc ii} region
and add equivalent data from the previous studies of interacting galaxies, 
Arp 270 and the Antennae galaxies, \cite{zaragoza13,zaragoza14}, 
respectively. We have extracted the same parameters for the H{\sc ii} regions of 
28 isolated galaxies from \cite{2015arXiv150406282E} in order to compare results from the two samples. 
 
For practical purposes we have called the galaxies
which have no close neighbours and which therefore suffer negligible 
interaction with other galaxies \textgravedbl isolated\textacutedbl  even though the term is used in this
article in a somewhat looser way than in studies dedicated to the careful 
definition of galaxies as isolated.
 The sample of isolated galaxies was selected to spread the whole morphological types, and 
different absolute magnitudes ranging from $-20.6$ to $-17.7$.

In section \S2 we present the galaxy sample, the observations, and the data reduction. 
In section \S3 we present the moment maps of the interacting galaxies. 
In section \S4 we explain how we derive the parameters of the H{\sc ii} regions from 
those extracted directly from the observations. 
In section \S5 we present the physical properties of the H{\sc ii} regions in the interacting galaxies and 
in \S6 those from the sample of isolated galaxies. In section 
\S7 we discuss the differences between the two samples, making comparison with previous studies, and 
in section \S8 we present a hypothesis to explain the  zones of  strong star formation well away from the galactic nuclei. 
Finally, in section \S9 we present our conclusions.

\section{Observations and galaxy sample}

\begin{figure*}

\centering

\begin{tabular}{cc}

\includegraphics[width=0.42\linewidth]{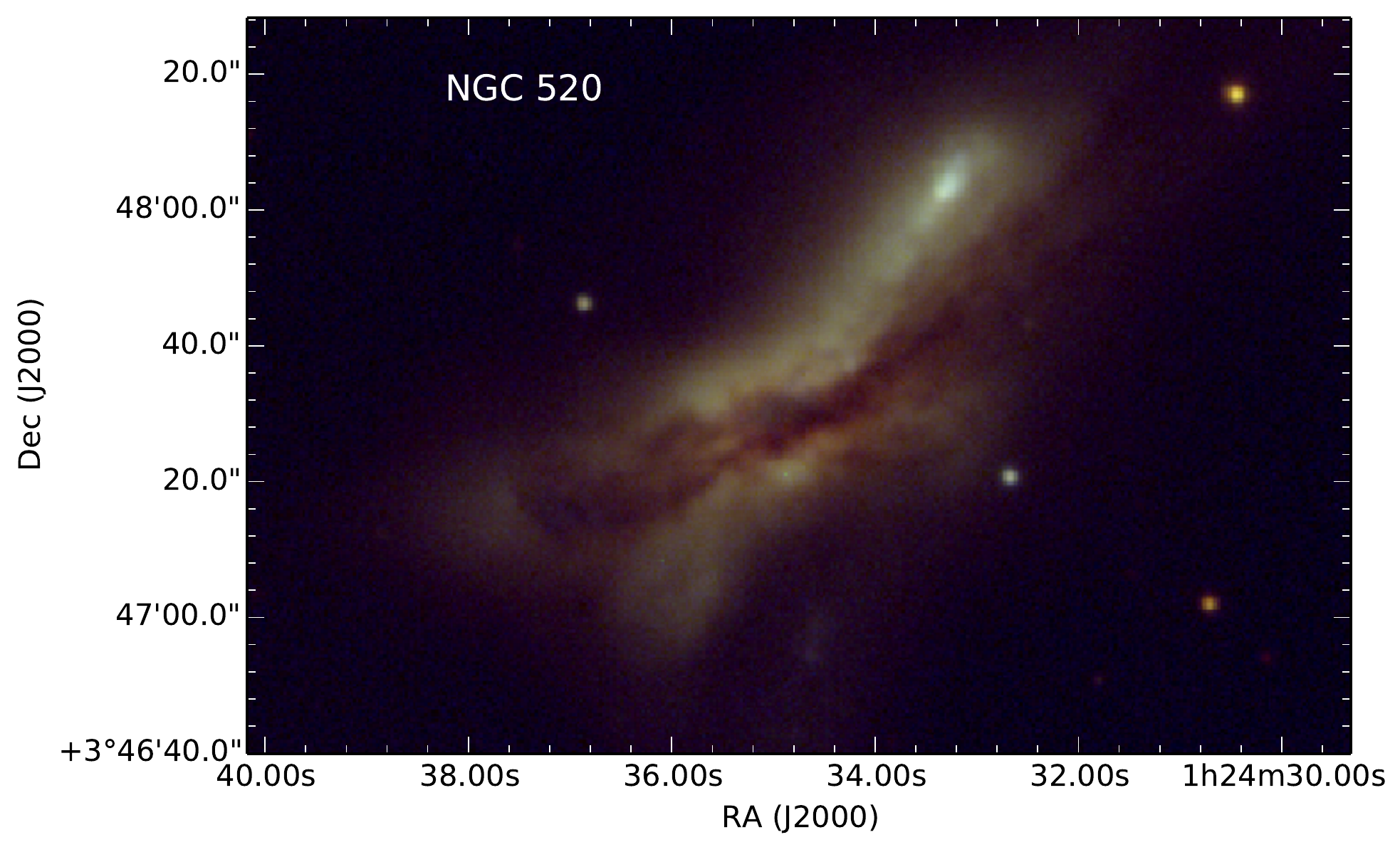} &
\includegraphics[width=0.42\linewidth]{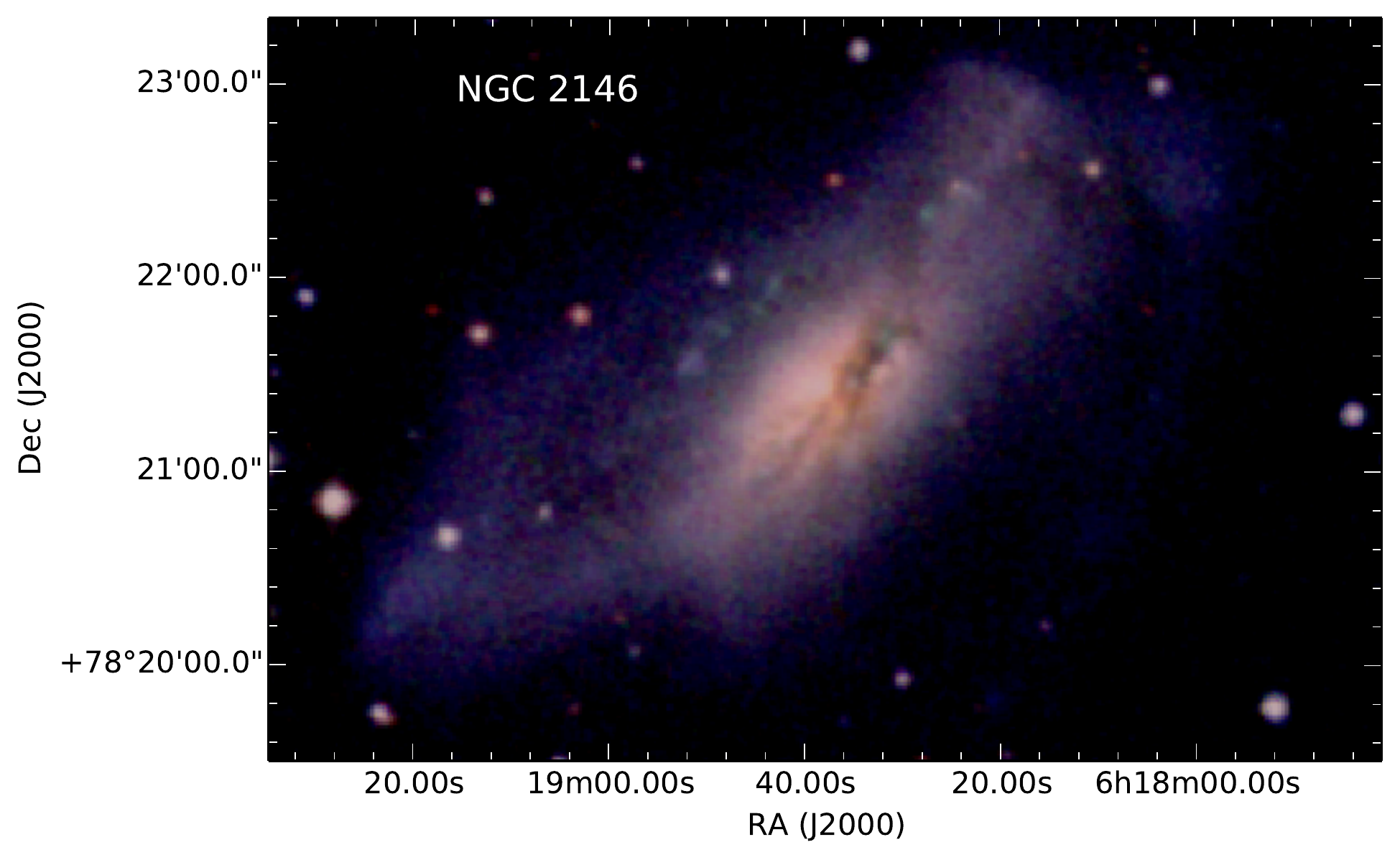}\\
\includegraphics[width=0.42\linewidth]{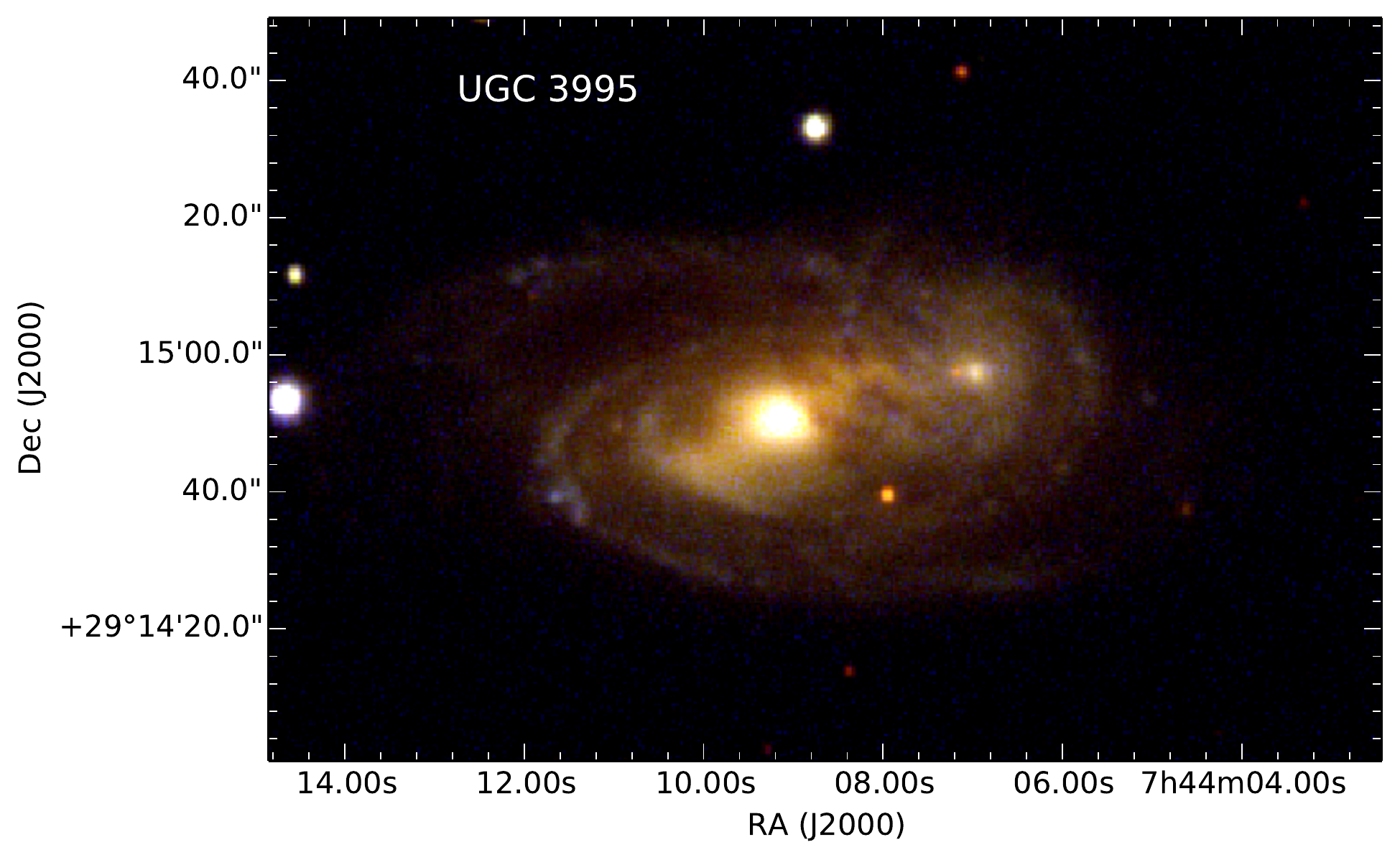}&
\includegraphics[width=0.42\linewidth]{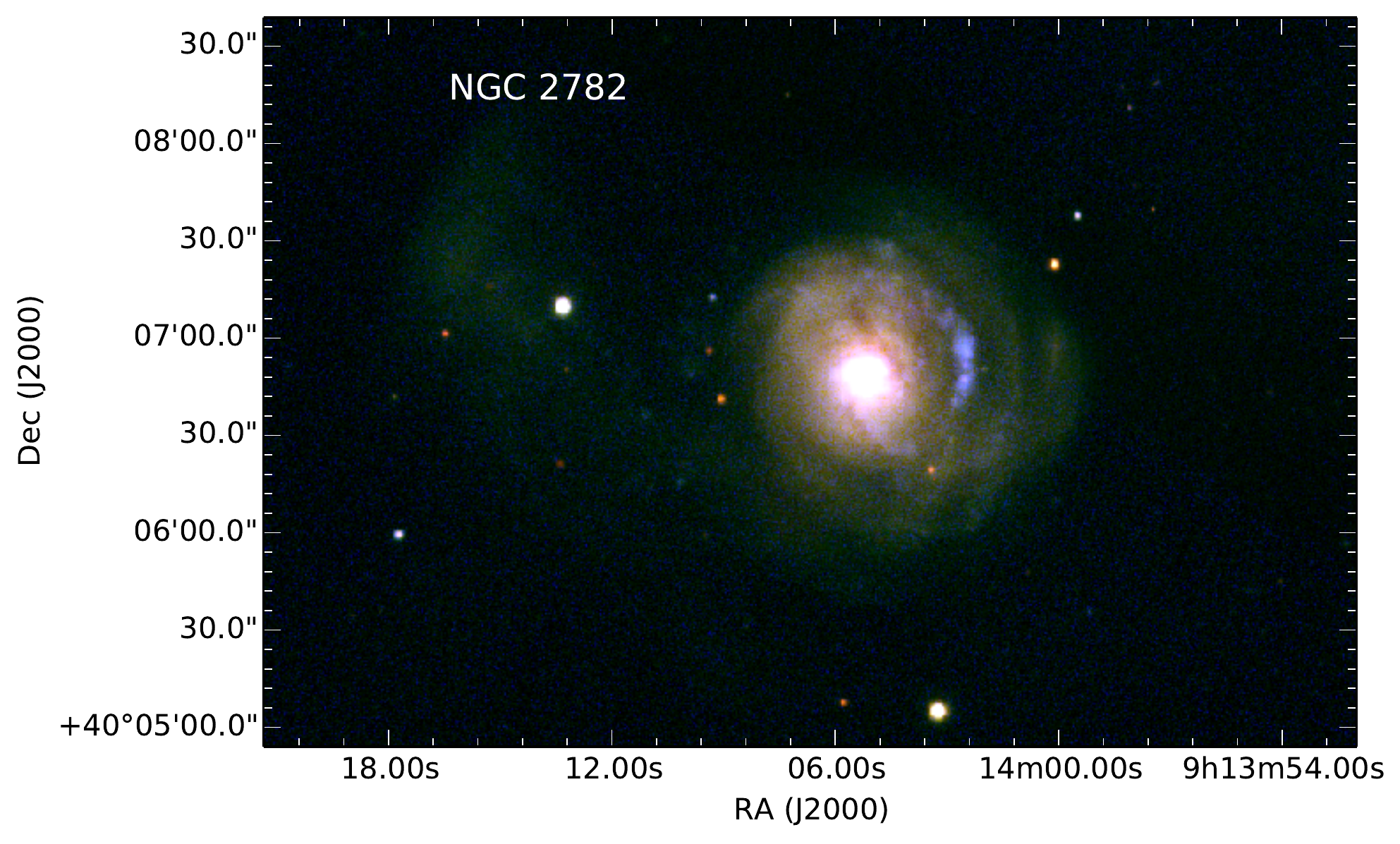}\\
\includegraphics[width=0.42\linewidth]{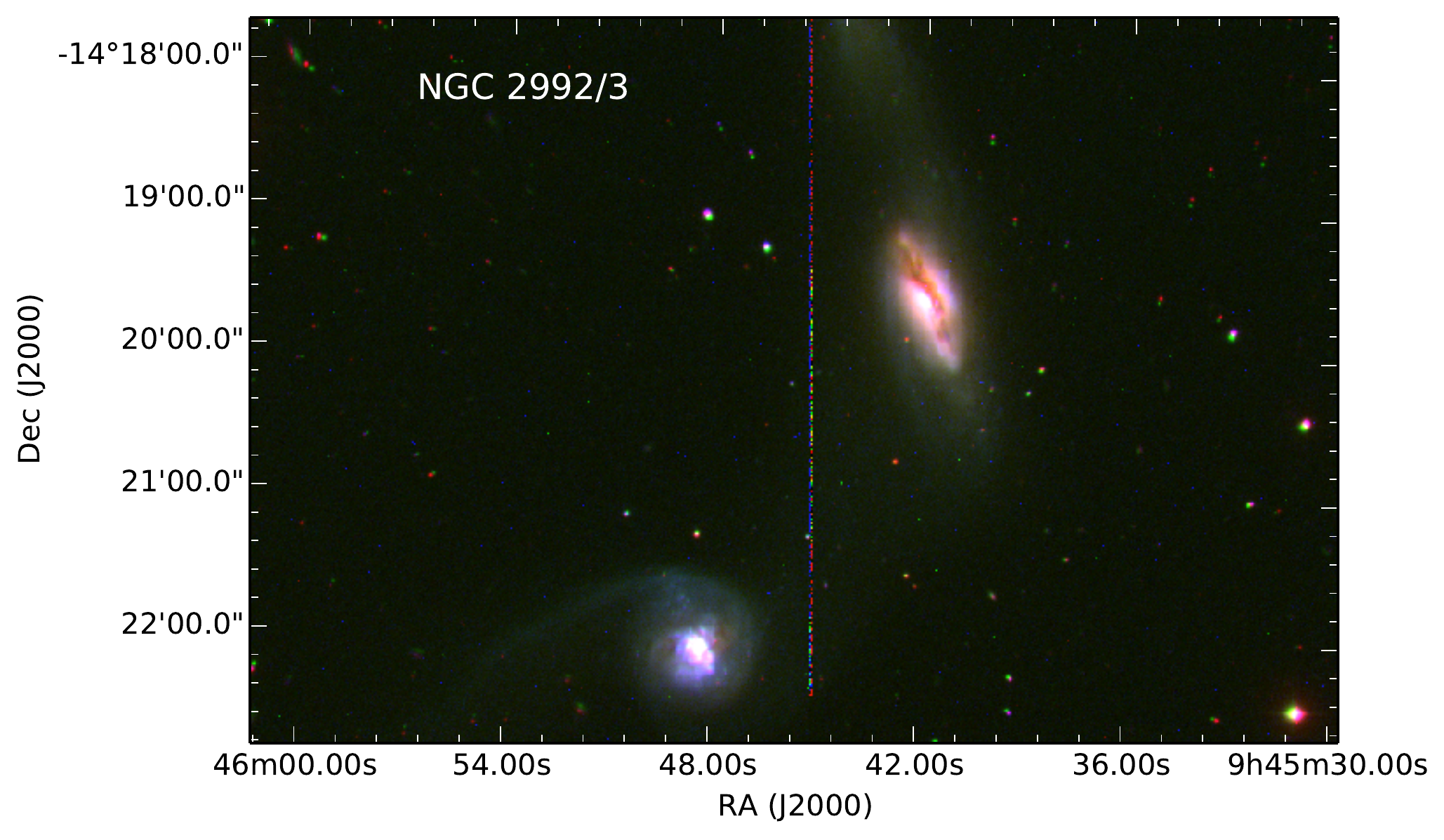} &
\includegraphics[width=0.42\linewidth]{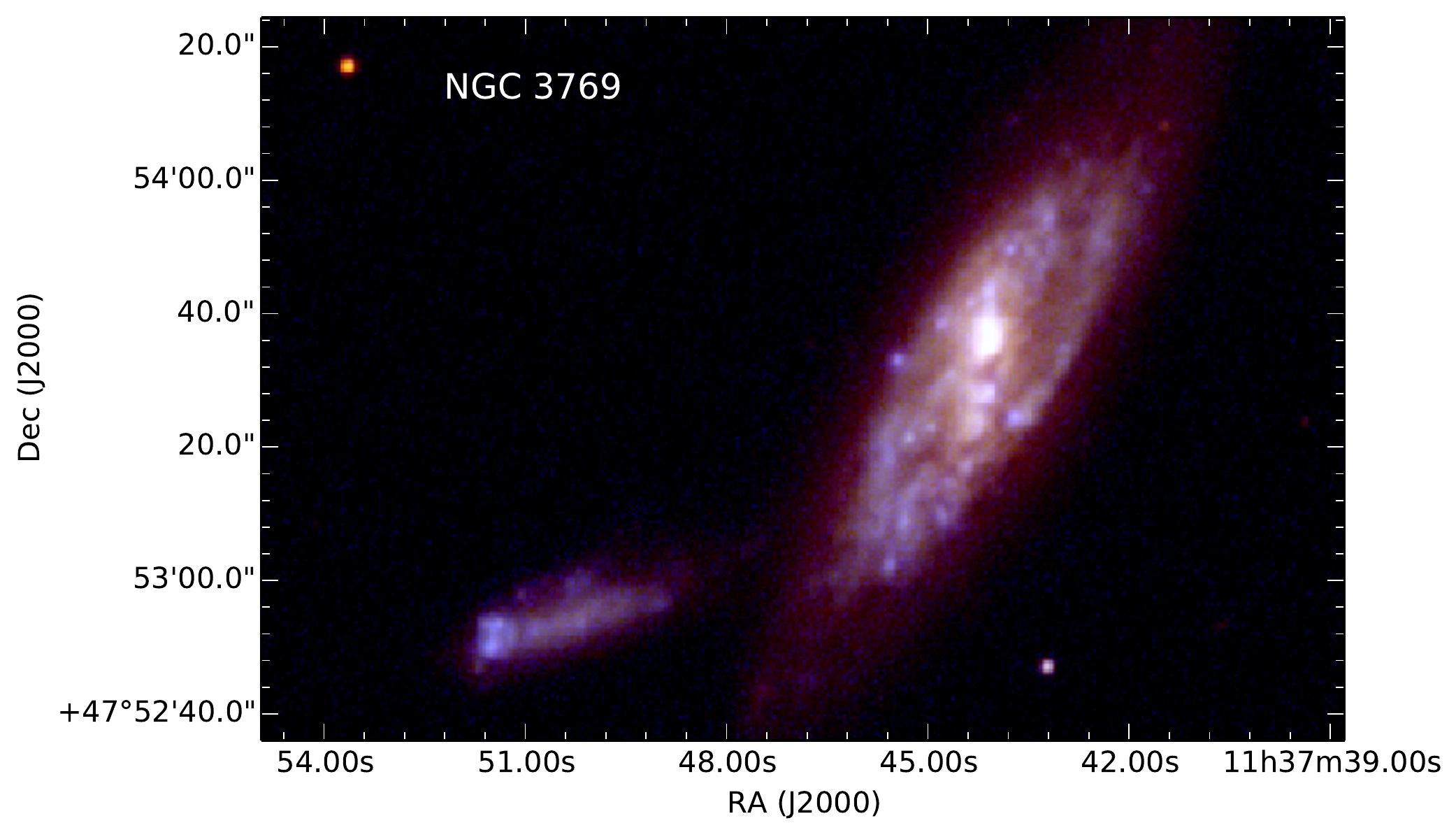}\\
\includegraphics[width=0.42\linewidth]{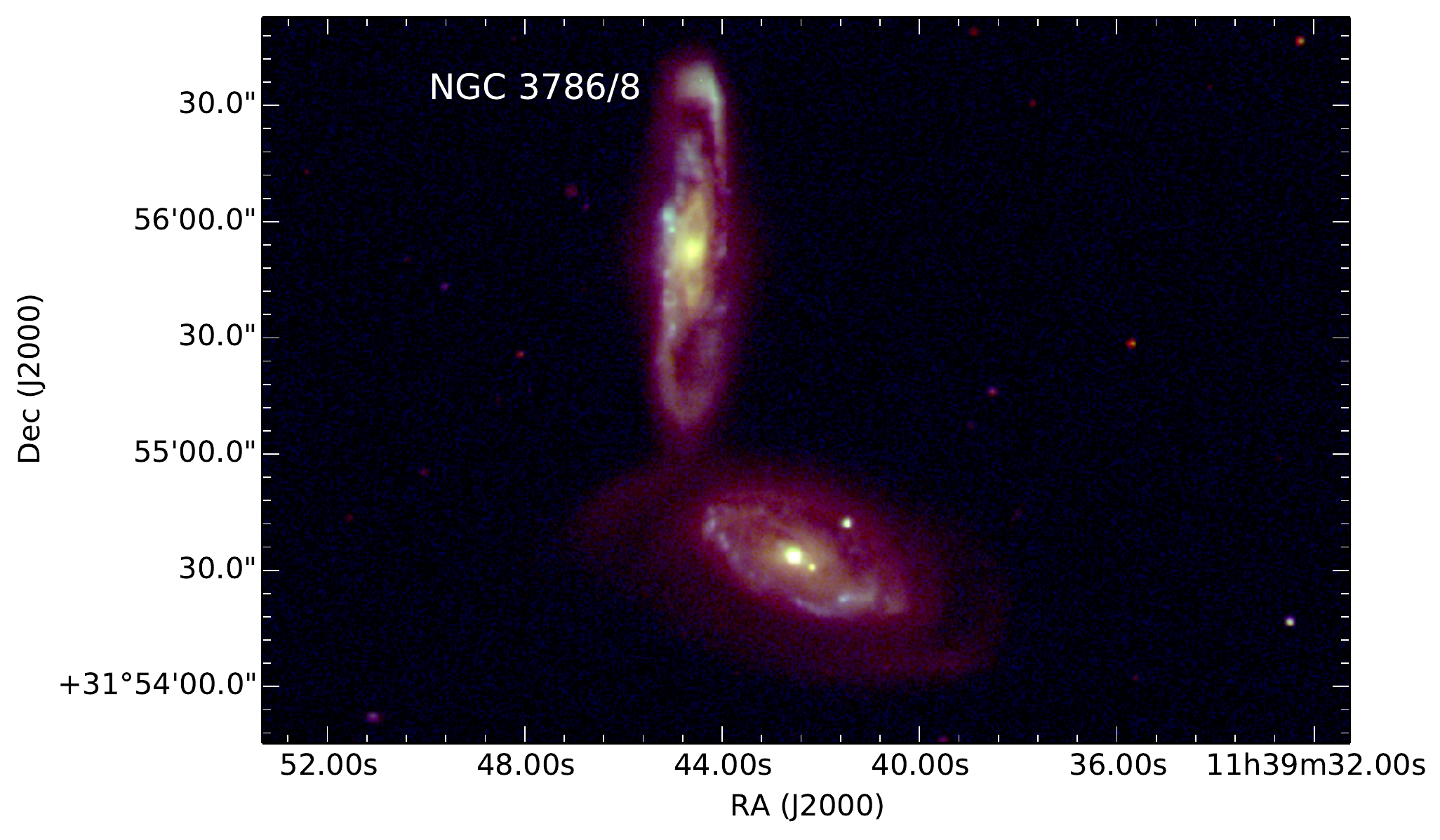}&
\includegraphics[width=0.42\linewidth]{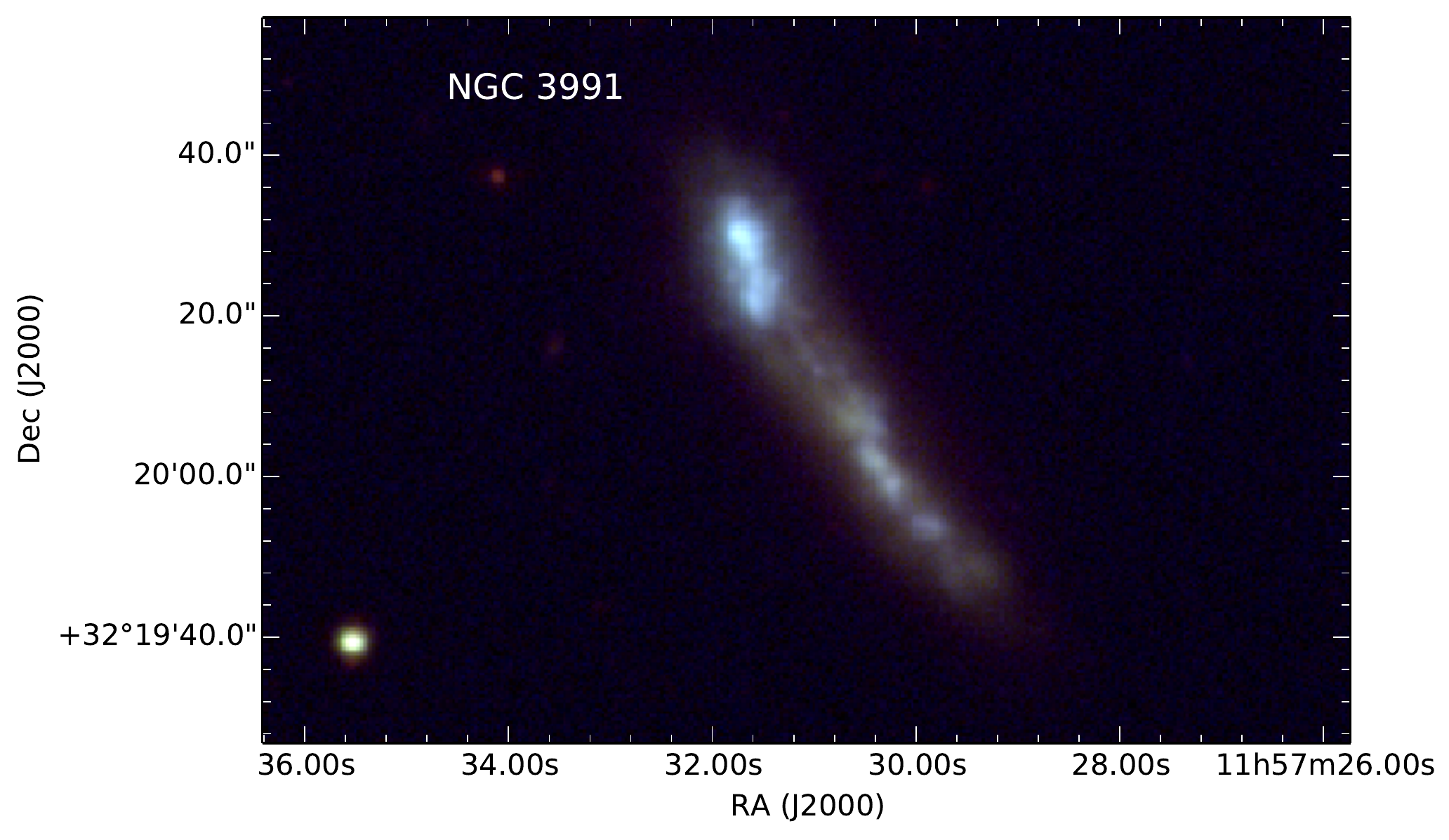}\\
\end{tabular}

\caption{Color composite images for the part of the sample 
of interacting galaxies we present here new observations. All of them are combined using SDSS data, except NGC 2146 
(using DSS), and NGC 2992/3 (using data from du Pont telescope at 
Las Campanas observatory).}
\label{fig_obs}
\end{figure*}

We observed the interacting galaxies with GH$\alpha$FaS \citep{2008PASP..120..665H}, a Fabry-Perot 
interferometer mounted on the WHT (William Herschel Telescope) 
at the Roque de los Muchachos Observatory, La Palma. 
The optical color composite images 
of the sample are in Fig. \ref{fig_obs}. 
The observations were taken during several observing runs  
between December 2010 and February 2013 (see Table \ref{table_obs}).

\begin{table*}
 \begin{tabular}{ccccccc}
   \hline
 Name &RA (J2000)&Dec (J2000) & Date &Exp time & Seeing & Dist \\
   &     (hh:mm:ss)         &     ($^{\circ}$ $\mathrm{\prime}$ $\mathrm{\prime\prime}$) & dd/mm/yyyy& (min) &(arcsec) & (Mpc)\\
  \hline
NGC 520       &     01:24:35.1& +03:47:33 &18/11/2012 & 181  & 0.9    &  34.4    \\
NGC 2146      &     06:18:37.7& +78:21:25 &23/12/2010 & 181  & 0.9    &  21.9    \\
UGC 3995      &     07:44:09.3& +29:14:48 &19/11/2012 & 181  & 1.0    &  60.6    \\
NGC 2782      &     09:14:05.1& +40:06:49 &17/11/2012 & 181  & 1.0    &  23.4     \\
NGC 2992      &     09:45:42.0& -14:19:35 &19/11/2012 & 156  & 1.0    &  31.6     \\
NGC 2993      &     09:45:48.3& -14:22:06 &18/11/2012 & 148  & 0.9    &  30.5   \\
NGC 3769      &     11:37:44.1& +47:53:35 &29/05/2011 & 156  & 0.8    &  16.1    \\
NGC 3786      &     11:39:42.5& +31:54:33 &28/02/2012 & 205  & 1.0    &  50.9      \\
NGC 3788      &     11:39:44.6& +31:55:52 &28/02/2012 & 205  & 1.0    &  50.1    \\
NGC 3991      &     11:57:31.1& +32:20:16 &04/02/2013 & 181  & 1.2    &  43.7   \\
\hline
\end{tabular}
\caption{Interacting galaxies observations log.}
\label{table_obs}
\end{table*}

These galaxies were selected from the catalog of
interacting galaxies by \citet{1966ApJS...14....1A} and \citet{2003yCat.7236....0V}
(except NGC 2146 and UGC 3995, which are classified as  Galaxy pairs in 
\citet{1976RC2...C......0D} and \citet{1985BICDS..29...87K}, respectively) 
according to the following criteria: 
\begin{enumerate}

 \item  Nearby systems (V$_{sys} < 10000$ km/s),
in order to have adequate spatial resolution. From $\sim30\thinspace\mathrm{pc}$ to $\sim100\thinspace\mathrm{pc}$ 
in our sample.
 \item  Angular size which fits the GH$\alpha$FaS FOV and allows satisfactory
sky subtraction.
 \item Declination -20$^{0} < \delta < 80^{0}$, to ensure observations with air mass smaller than 1.6. 

\end{enumerate}
GH$\alpha$FaS has a circular FOV of $3.4$ arcmin diameter. The etalon has a FSR (Free Spectral Range) 
of $8\mathrm{\AA}$ in H$\alpha$, which corresponds to a $390\mathrm{km/s}$ with a spectral 
resolution of  $~8\mathrm{km/s}$. The pixel size is $0.2$ arcsec, which gave us the values of 
 seeing limited angular resolution given in Table \ref{table_obs}. Further details 
about each galaxy observation are listed in Table \ref{table_obs}. 
The morphological type, apparent and absolute magnitude of the whole sample of interacting 
galaxies are in Table \ref{table_gal}. The comparison of these properties with those of the sample 
of isolated galaxies from \citet{2015arXiv150406282E}, yields that both of them spread over the 
whole morphological types, and that the interacting galaxies sample is brighter in average, although 
less than an order of magnitude. 
Thus, the galaxy masses (derived from absolute magnitude) in 
the interacting galaxies are higher in average compared to the masses in the isolated galaxies.

\begin{table}
 \begin{tabular}{cccc}
   \hline
 Name &  $m_B$ & $M_B$ & morphology  \\
  \hline
NGC 520 & 12.2 & -20.48 & Sa\\ 
NGC 2146 & 10.53 & -21.17 & SBab\\ 
UGC 3995 & 13.52 & -20.39 & Sbc\\ 
NGC 2782 & 12.32 & -19.53 & SABa\\ 
NGC 2992 & 13.06 & -20.22  & Sa \\  
NGC 2993 & 13.11 & -19.81 & Sa \\
NGC 3395 (Arp 270) & 12.4 & -19.32 & Sc\\ 
NGC 3396 (Arp 270) & 12.43 & -19.29 & SBm\\ 
NGC 3769 & 12.54 & -18.49 & Sb\\ 
NGC 3786 & 13.47 & -20.06 & SABa\\ 
NGC 3788 & 13.44 & -20.06 & SABa\\ 
NGC 3991 & 13.51 & -19.69 & Im\\ 
NGC 4038 (Antennae) & 10.85 & -20.86 & SBm\\ 
NGC 4039 (Antennae) & 11.04 & -20.67 & SBm\\ 
\hline
\end{tabular}

\begin{minipage}{\linewidth}
\caption{Whole interacting galaxies sample. The apparent magnitude and the morphology are taken 
from Hyperleda (\url{http://leda.univ-lyon1.fr} \citet{2003A&A...412...45P}).}
%\protect\cite{2003A&A...412...45P}.
\label{table_gal}
\end{minipage}

\end{table}

\subsection{Data reduction}
GH$\alpha$FaS is mounted at the Nasmyth focus of the WHT without an optical derotator.  
  Correction for field rotation is therefore necessary, and was performed using the technique described 
  in detail in \citet{2010MNRAS.407.2519B}. 
The observations are divided into individual cycles of $8\thinspace\mathrm{min}$, and each cycle is divided 
into 48 channels sepparated by $8\thinspace\rm{km/s}$ and  spanning the whole spectral range, which allows us to calibrate in velocity before 
applying digital derotation.  We applied to the datacube for a given galaxy the procedures detailed 
in \citet{2006MNRAS.368.1016D}, including 
phase-correction, spectral smoothing, sky subtraction and  gaussian spatial smoothing.

The flux calibration of the GH$\alpha$FaS datacubes is performed using 
a continuum-subtracted and flux-calibrated ACAM (Auxiliary-port CAMera, \cite{2008SPIE.7014E..6XB}) 
H$\alpha$ image.  
ACAM is an instrument
mounted permanently at the WHT used for broad-band and narrow-band imaging.
We have followed the procedure explained in \cite{2012MNRAS.427.2938E}, which 
consists in measuring the flux from a set of selected H II regions in both the GH$\alpha$FaS 
cube and the ACAM image, and calibrating by direct comparison.

\begin{figure*}

\centering

\begin{tabular}{cc}

\includegraphics[width=0.42\linewidth]{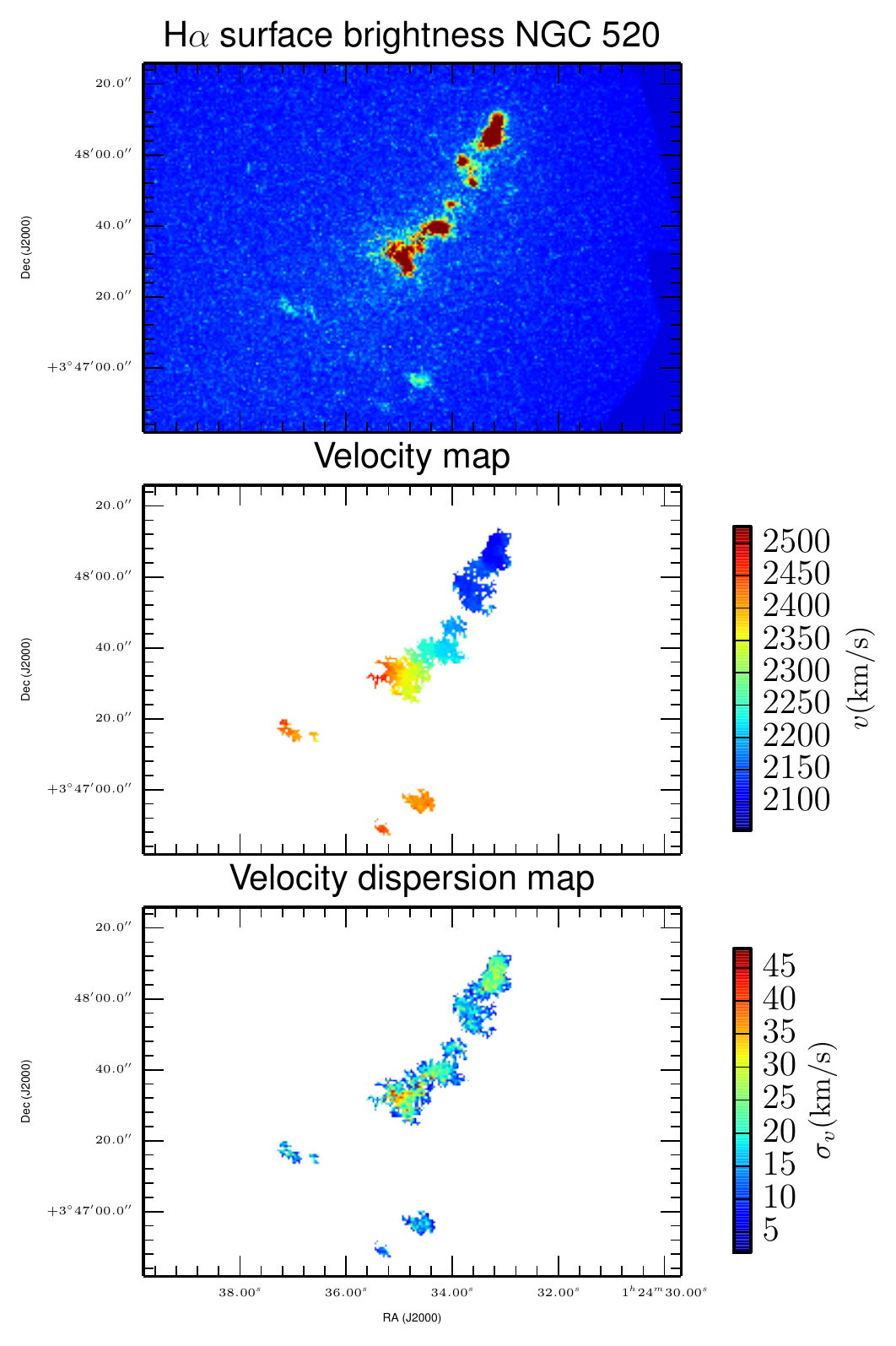} &
\includegraphics[width=0.42\linewidth]{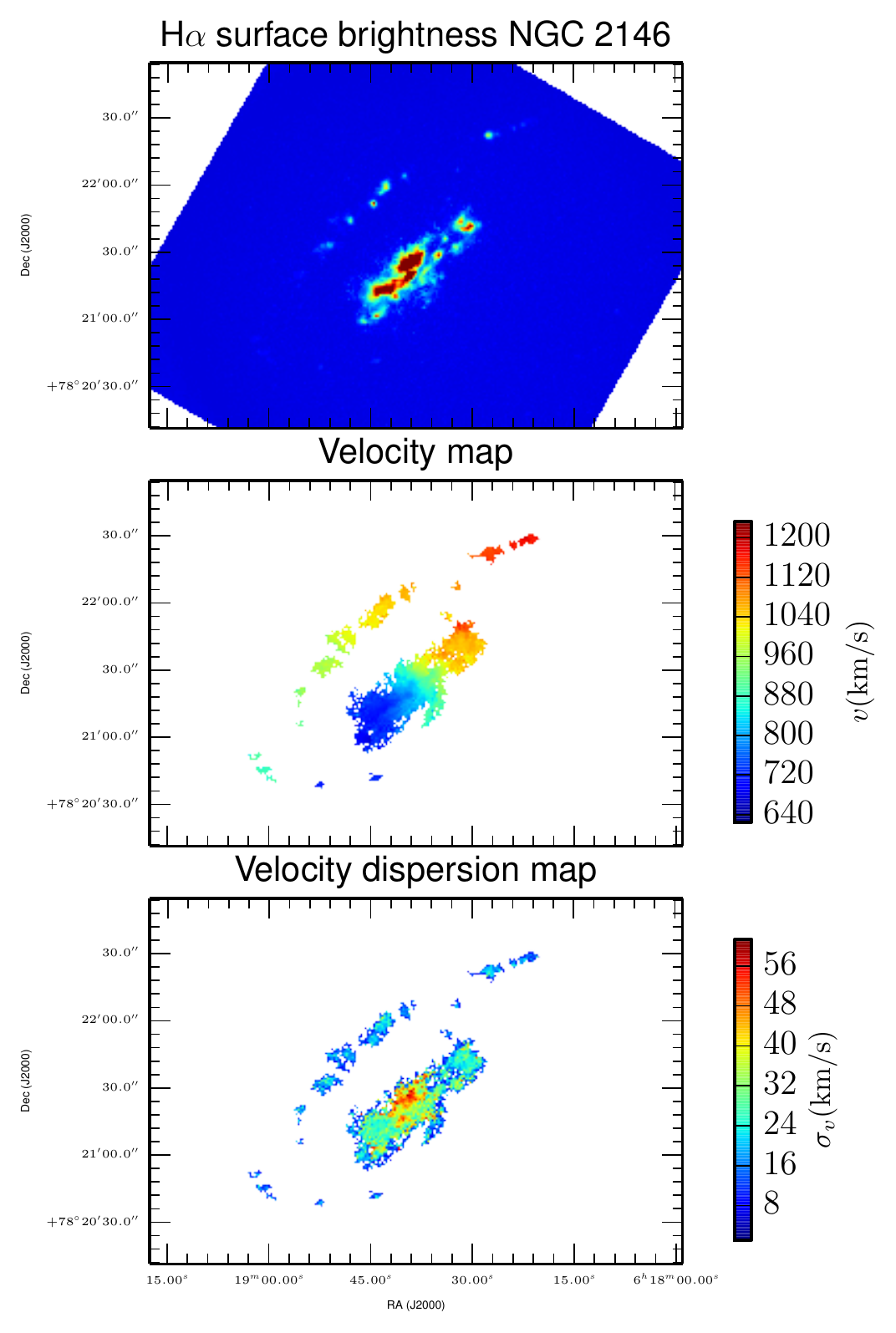}\\
\includegraphics[width=0.42\linewidth]{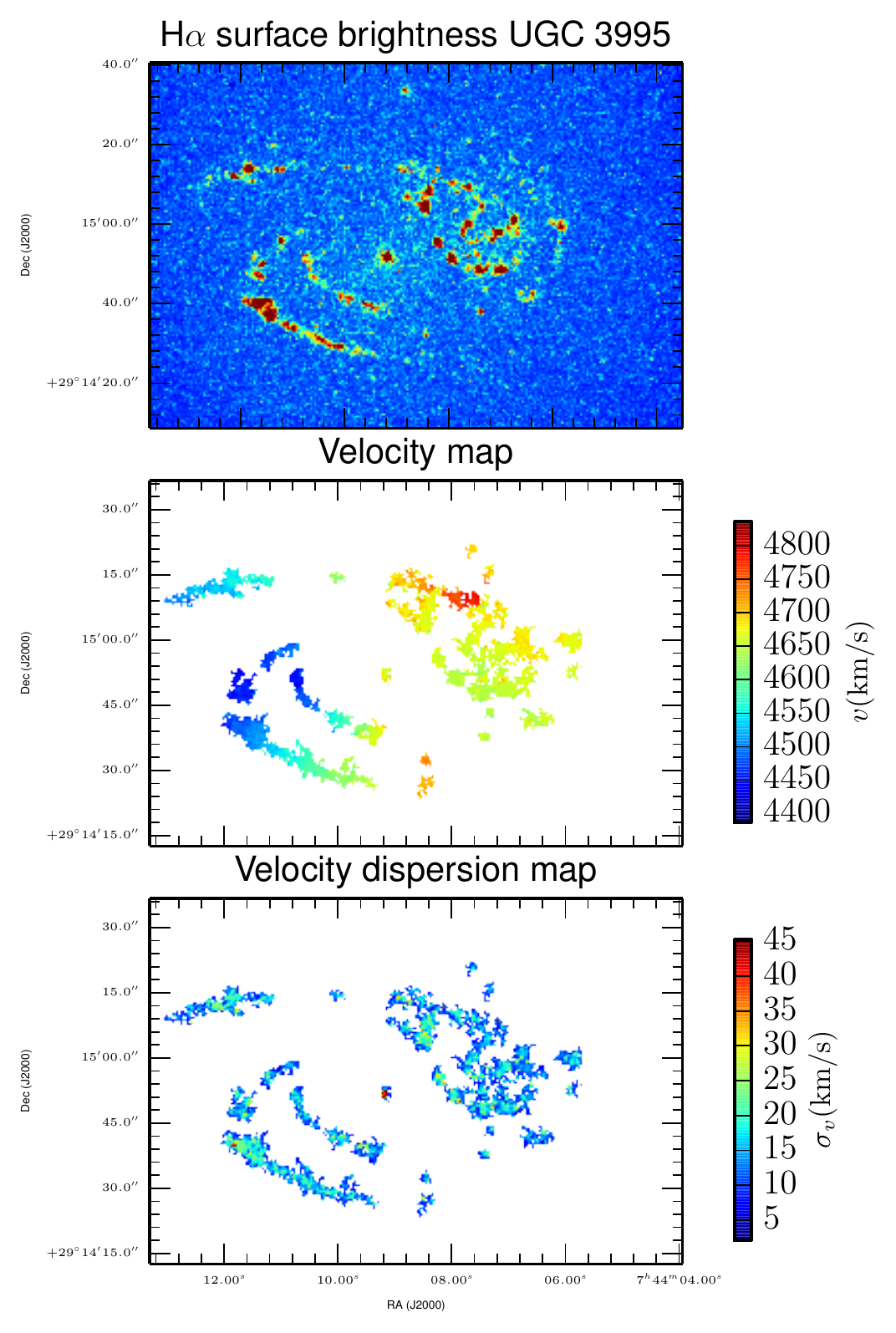}&
\includegraphics[width=0.42\linewidth]{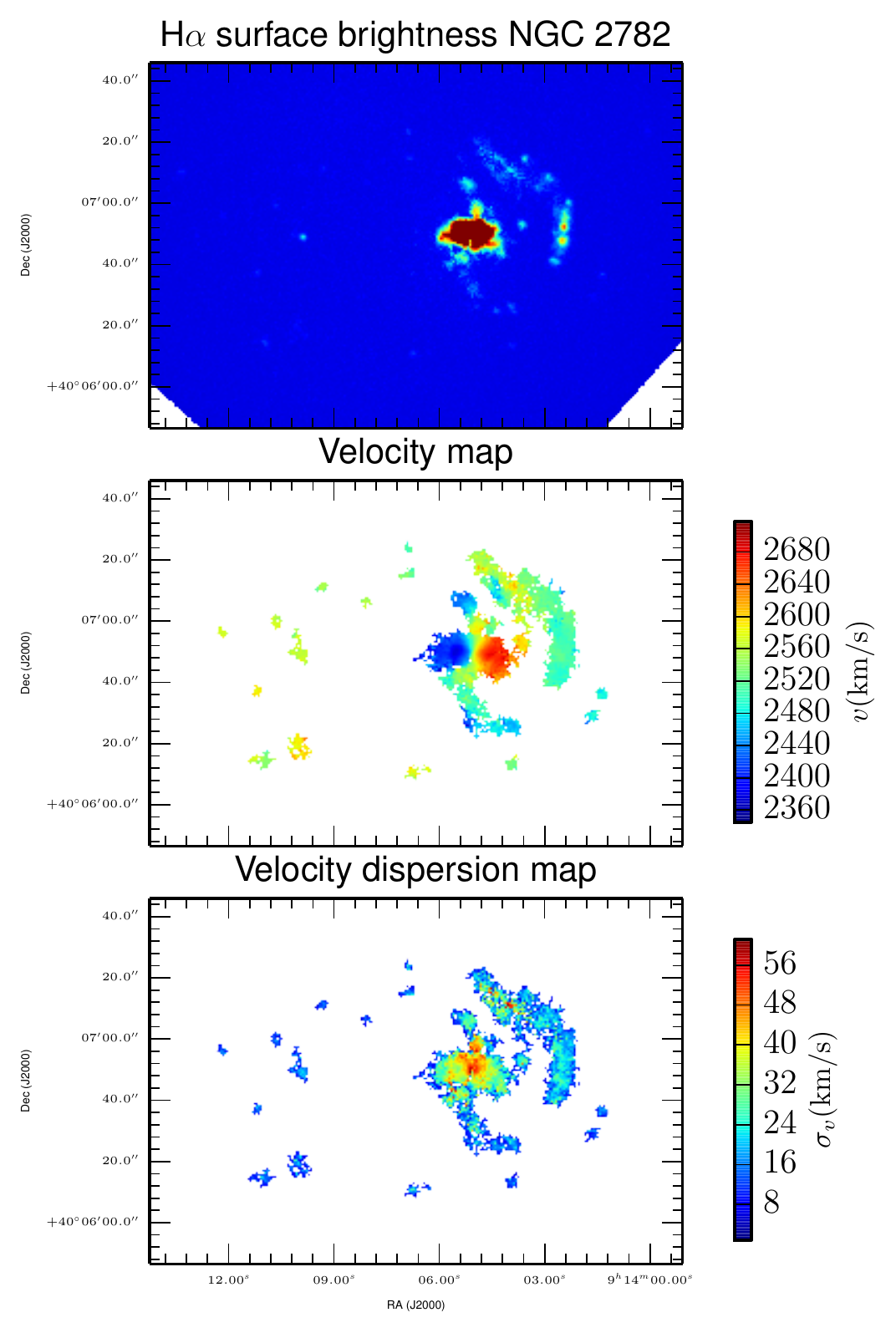}\\

\end{tabular}

\caption{Moment maps of the interacting galaxies sample derived from the GH$\alpha$FaS datacube. 
The colour versions of the maps are available in the electronic version
of the article and the moment maps are available through CDS.}
\label{fig_moments1}
\end{figure*}

\begin{figure*}

\centering

\begin{tabular}{cc}

 \includegraphics[width=0.42\linewidth]{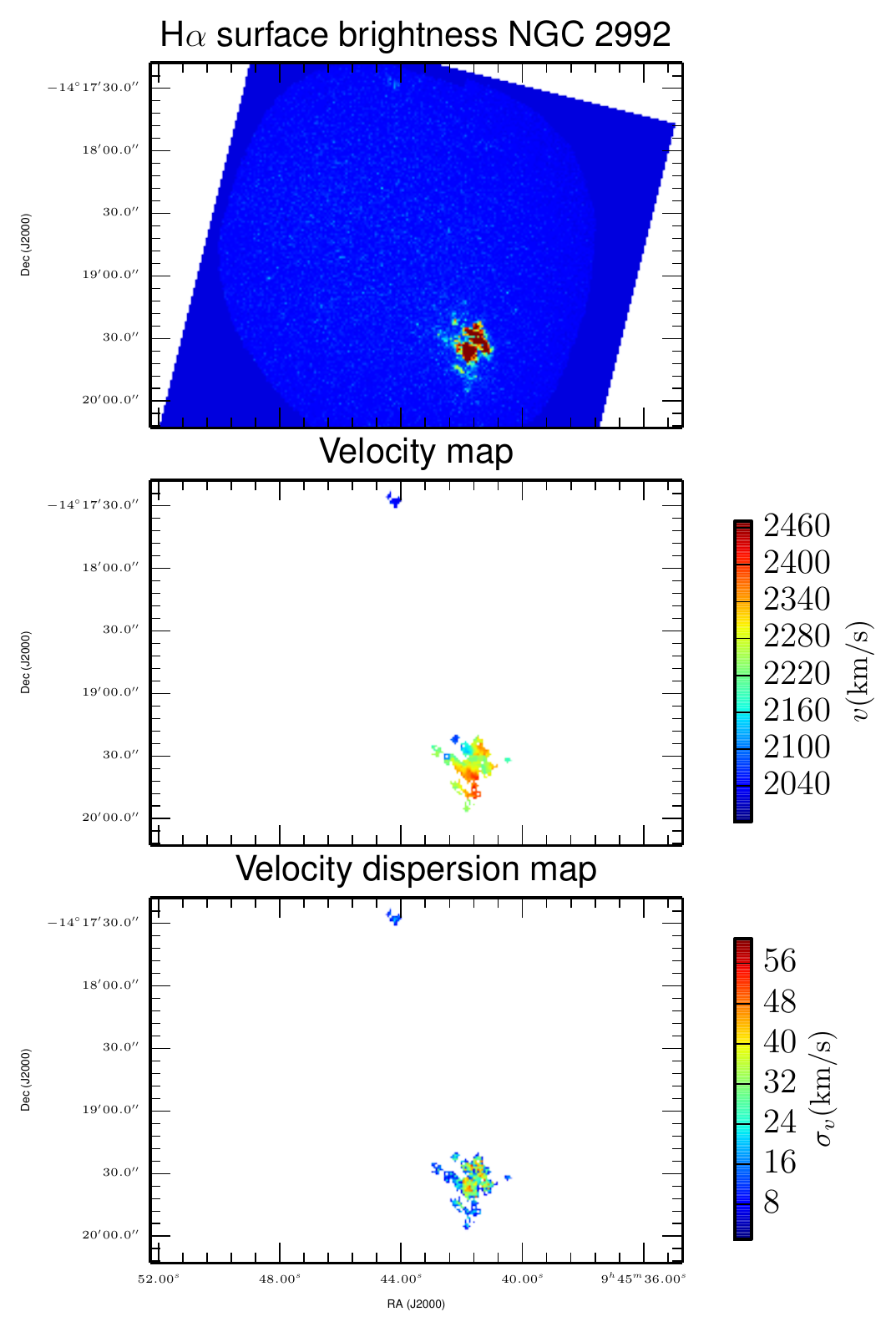}&
\includegraphics[width=0.42\linewidth]{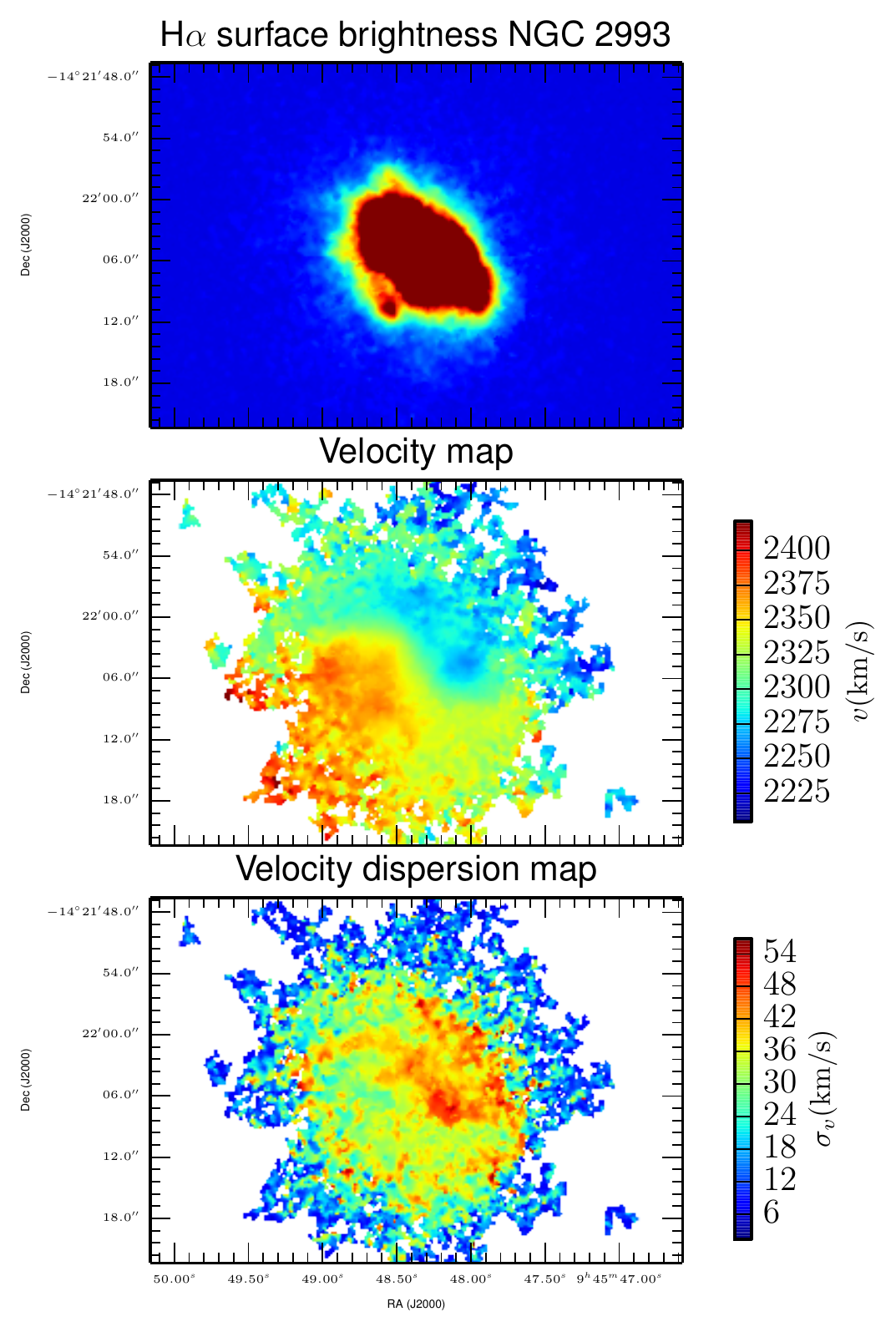}\\
 \includegraphics[width=0.42\linewidth]{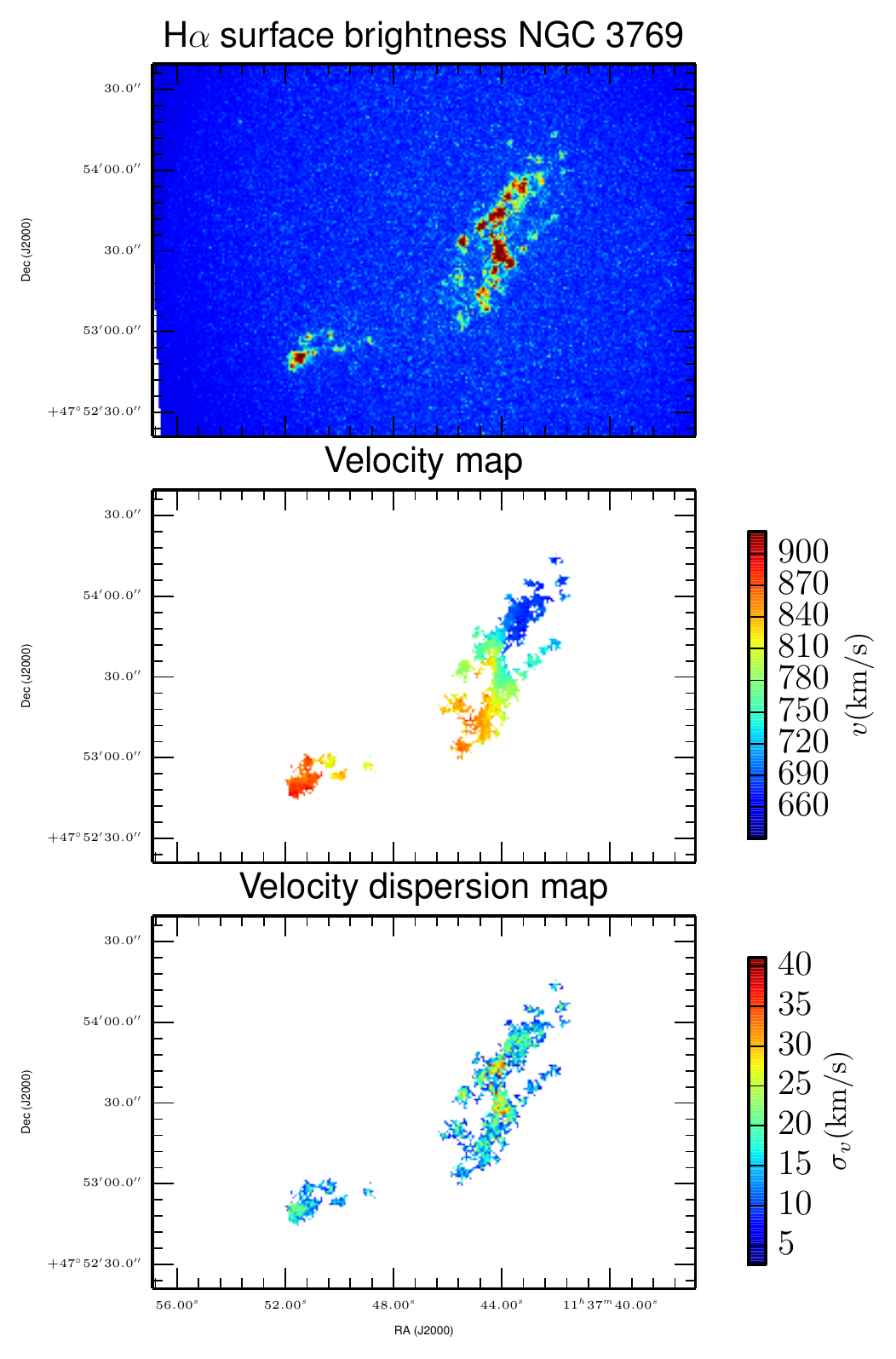}&
\includegraphics[width=0.42\linewidth]{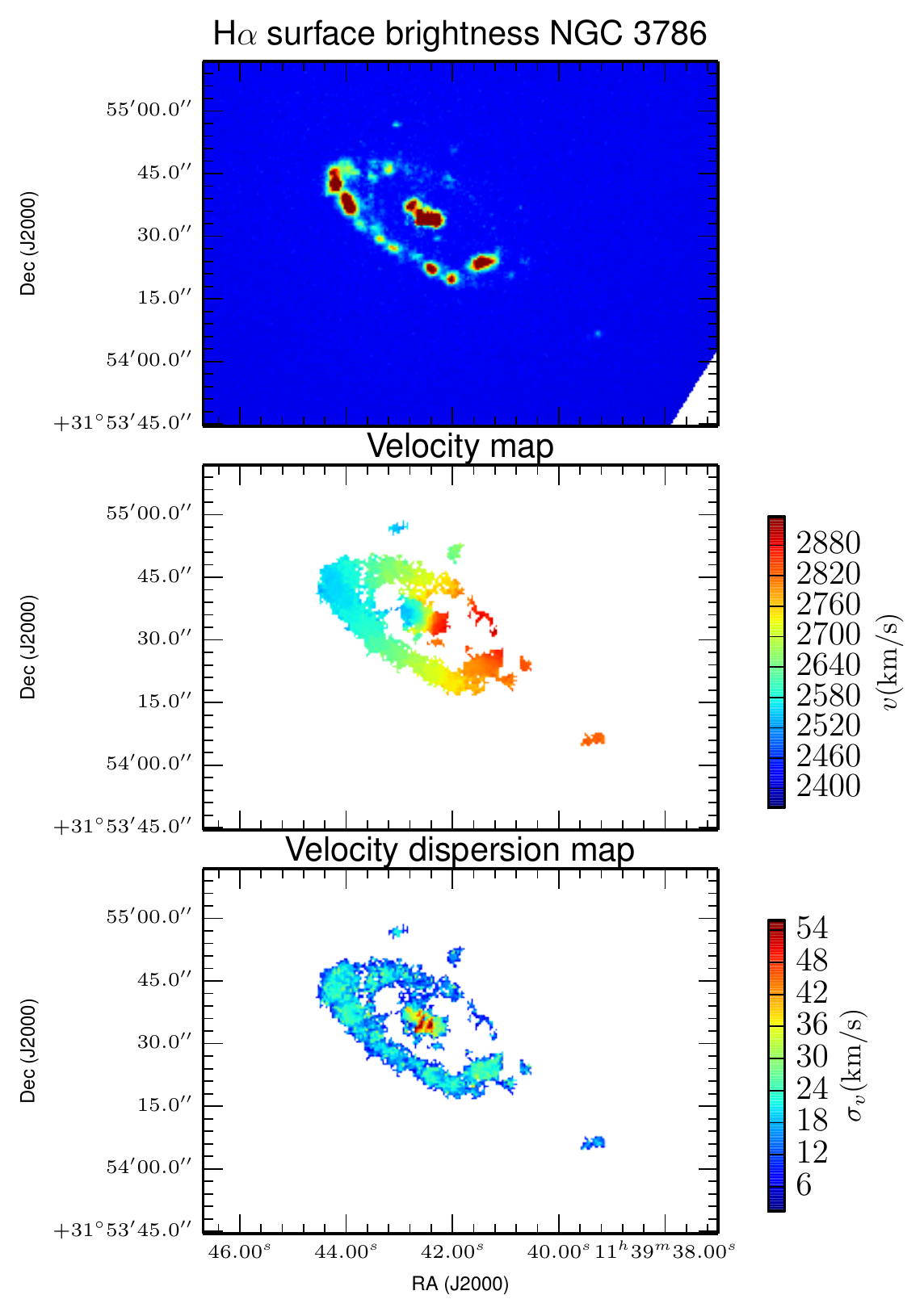}\\

\end{tabular}

\contcaption{}
\label{fig_moments2}
\end{figure*}

\begin{figure*}

\centering

\begin{tabular}{cc}

\includegraphics[width=0.42\linewidth]{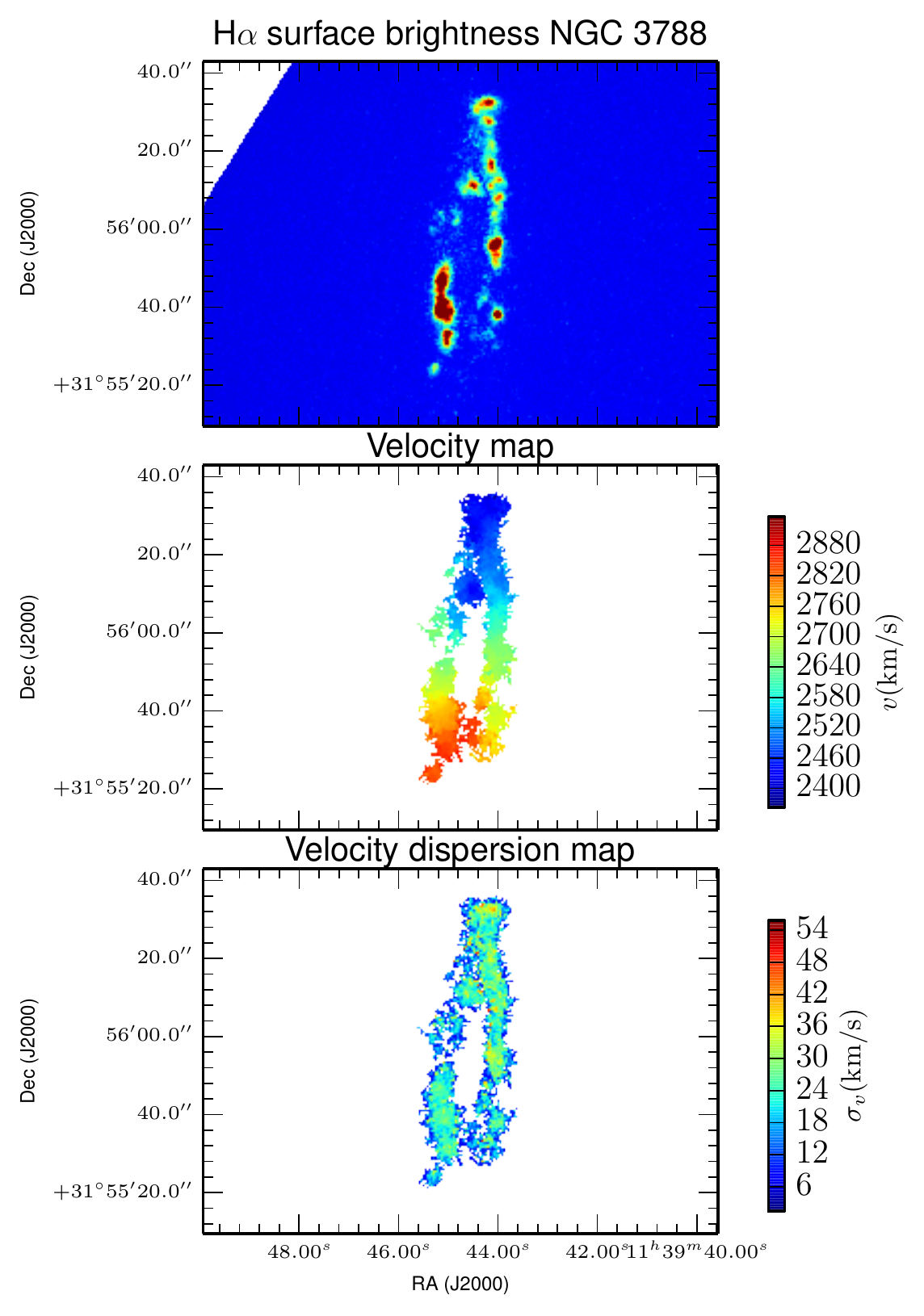}&
\includegraphics[width=0.42\linewidth]{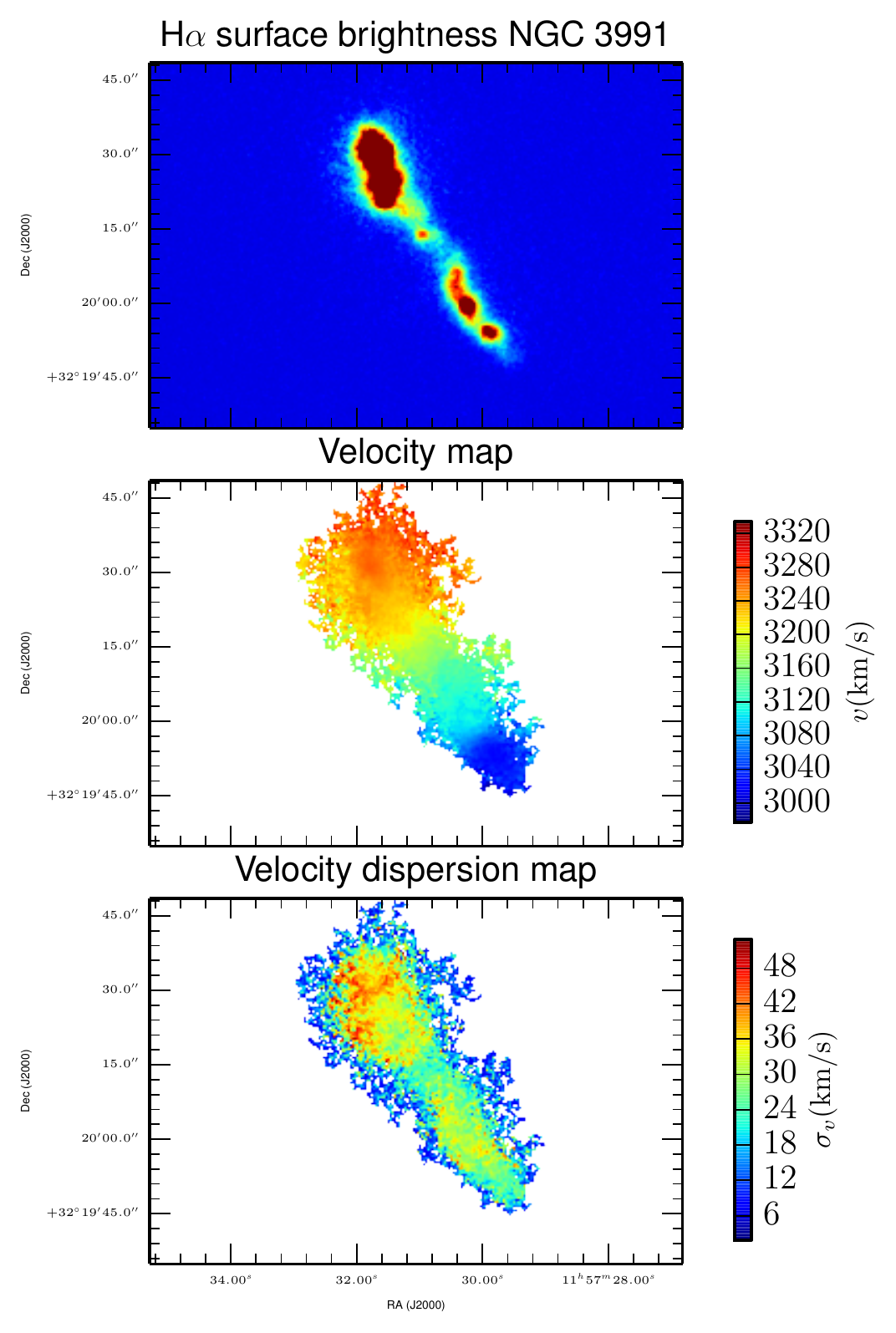}\\

\end{tabular}

\contcaption{}
\label{fig_moments3}
\end{figure*}

\section{Moment maps of interacting galaxies}

For each galaxy datacube, we have extracted the H$\alpha$ surface brightness, velocity, and 
velocity dispersion maps using the tools described in 
\citet{2006MNRAS.368.1016D}. We show the maps in Fig. \ref{fig_moments1}. 
The complete sample of the moment maps is available through CDS. We have also included 
in the information sent to CDS the continuum subtracted H$\alpha$ emission datacube.

\section{Physical properties of H{\sc ii} regions}

We used \textsc{astrodendro}\footnote{\href{http://www.dendrograms.org}{http://www.dendrograms.org}}, 
a Python package to compute \textgravedbl dendrograms\textacutedbl of Astronomical data
\citep{2008ApJ...679.1338R} to extract the relevant parameters of the H{\sc ii} regions from the datacubes. 
The parameters extracted are the H$\alpha$ luminosity, $L_{\rm H\alpha}$, the radius of the 
region, $R$, and the velocity dispersion, $\sigma_v$ as explained in 
\citet{2006PASP..118..590R,2008ApJ...679.1338R}, from where we have extracted a brief description in order  
to make it easier to understand the present article.

We can describe a GH$\alpha$FaS datacube as a collection 
of values, $T_i$, at a given point in the 3 dimensions which define 
the coordinates of the datacube, two spatial dimension, ($x_i,y_i$), and the  third the velocity 
along the line of sight, $v_i$. The method 
considers that a region is separated from the 
rest of the cube by an isosurface of 
value $T_{\mathrm{edge}}$ where $T>T_{\mathrm{edge}}$ inside the region and  
 $T<T_{\mathrm{edge}}$ just outside the region. The method then estimates 
the major and minor axes as the mean values of the second spatial moments 
in the directions of 
these axes, respectively;

\begin{equation}
\sigma_{\mathrm{maj,min}}=\sqrt{\frac{\sum_i{{T_i(x_i-\overline{x})^2}}}{\sum_i{T_i}}} 
\end{equation}

where the sum is over the pixels inside the region, and $x$ are the 
points lying on the major or the minor axis of the region. 

The equivalent radius is then, $R=\eta\sqrt{\sigma_{\mathrm{maj}}\thinspace\sigma_{\mathrm{min}}}$ 
where it is assumed that the region is spherical, so $\eta=1.91$.

\begin{figure*}
\centering
\begin{tabular}{cc}
\includegraphics[width=0.42\linewidth]{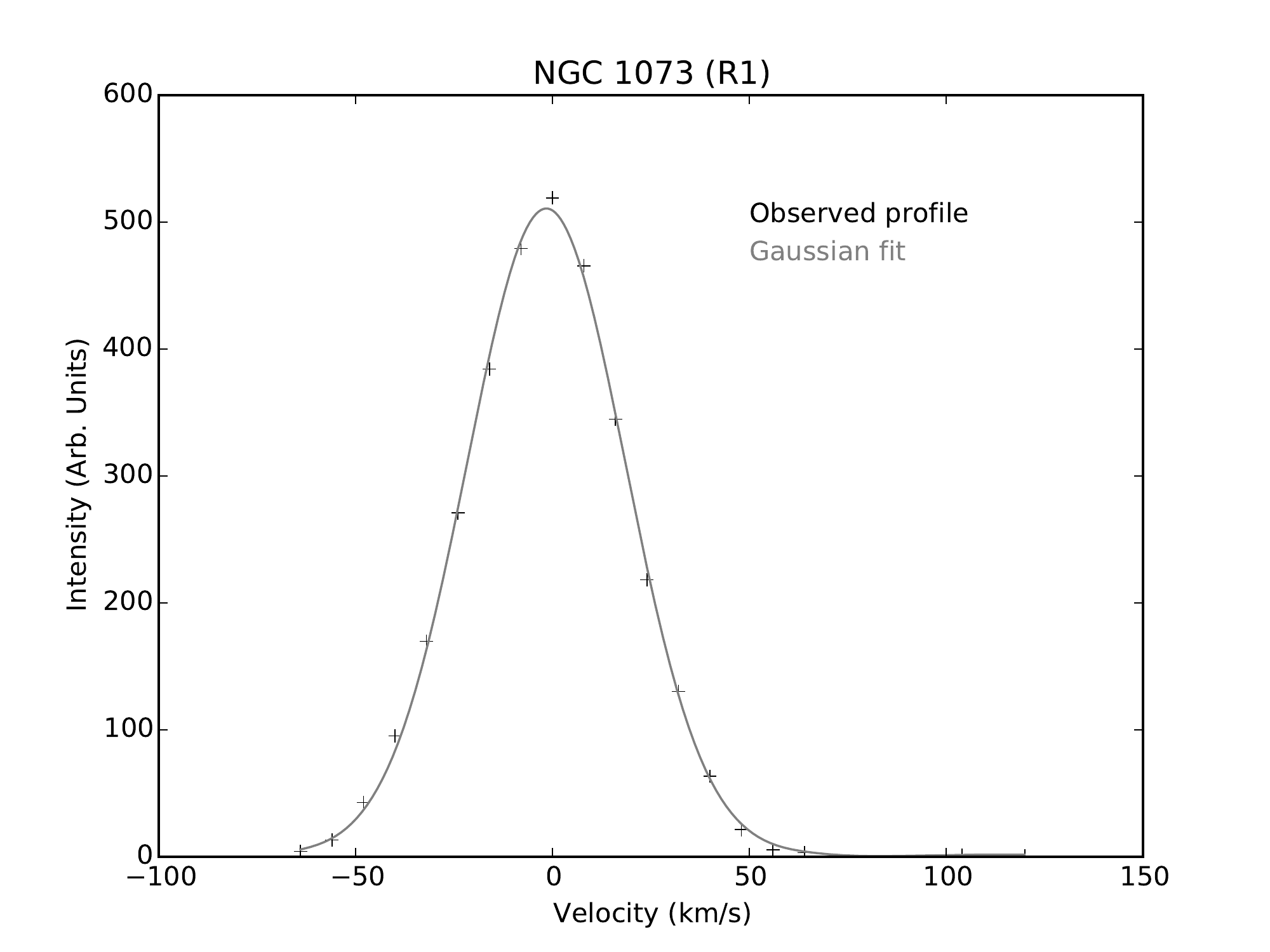}&
\includegraphics[width=0.42\linewidth]{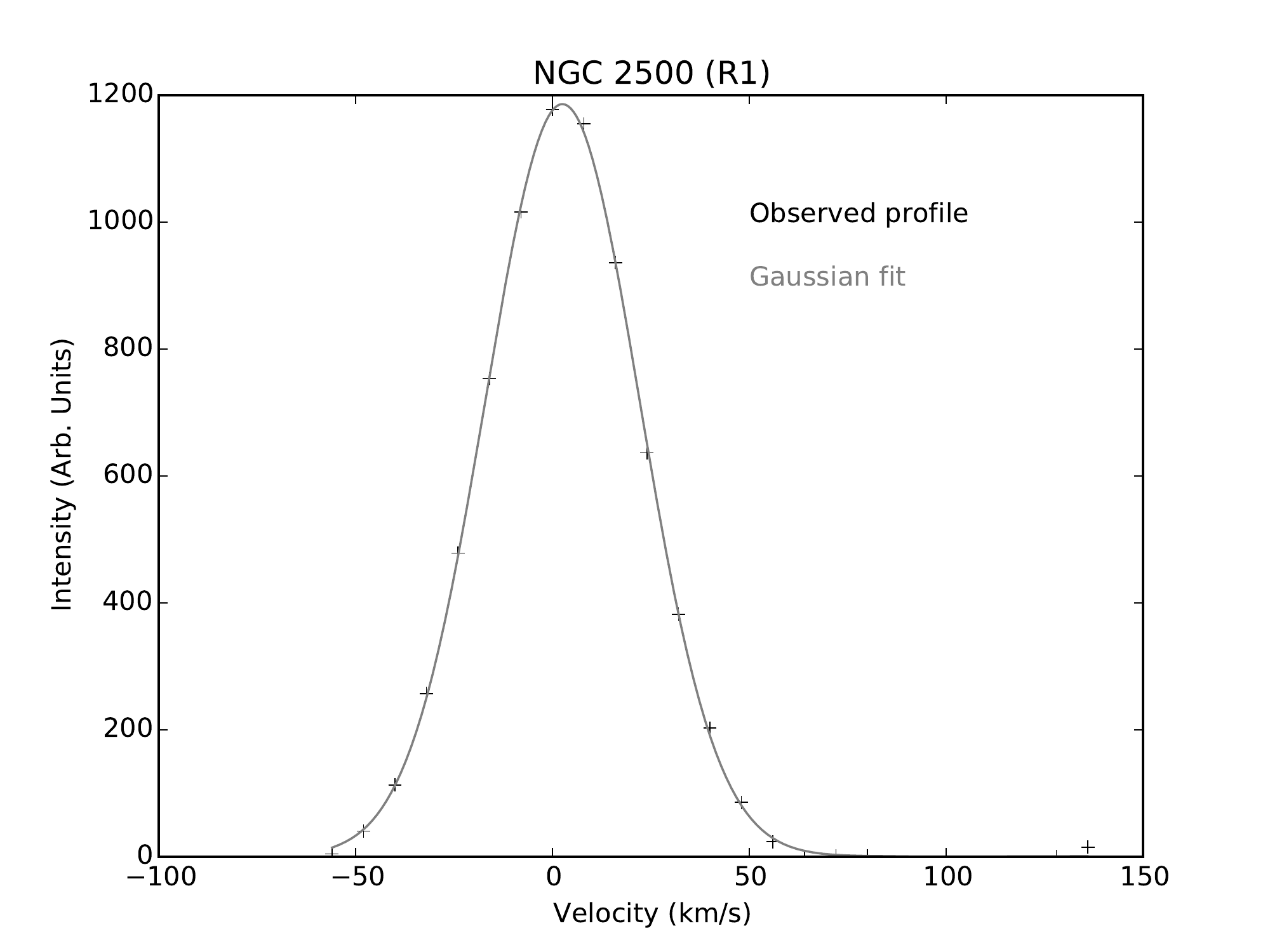}\\
\includegraphics[width=0.42\linewidth]{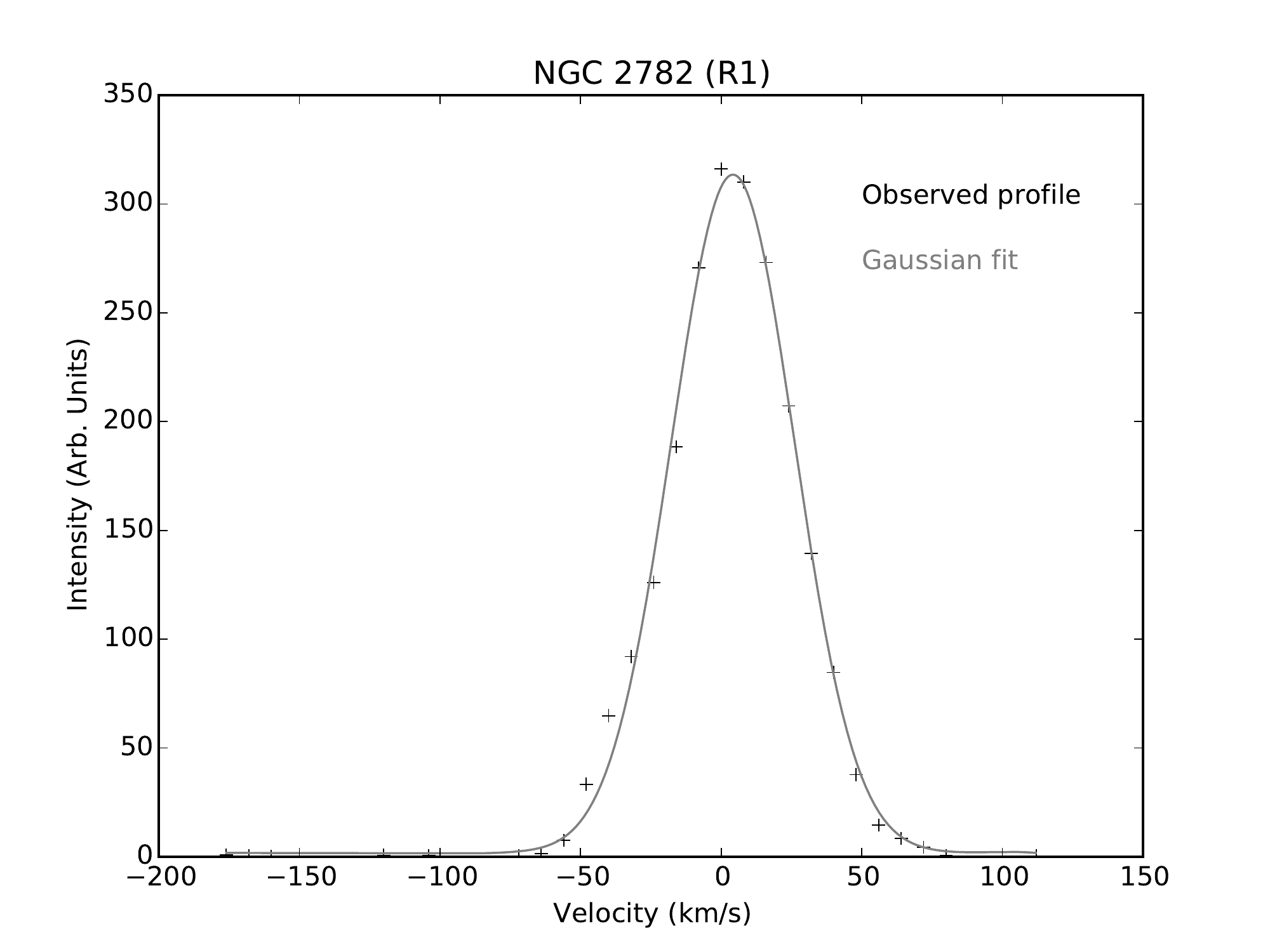}&
\includegraphics[width=0.42\linewidth]{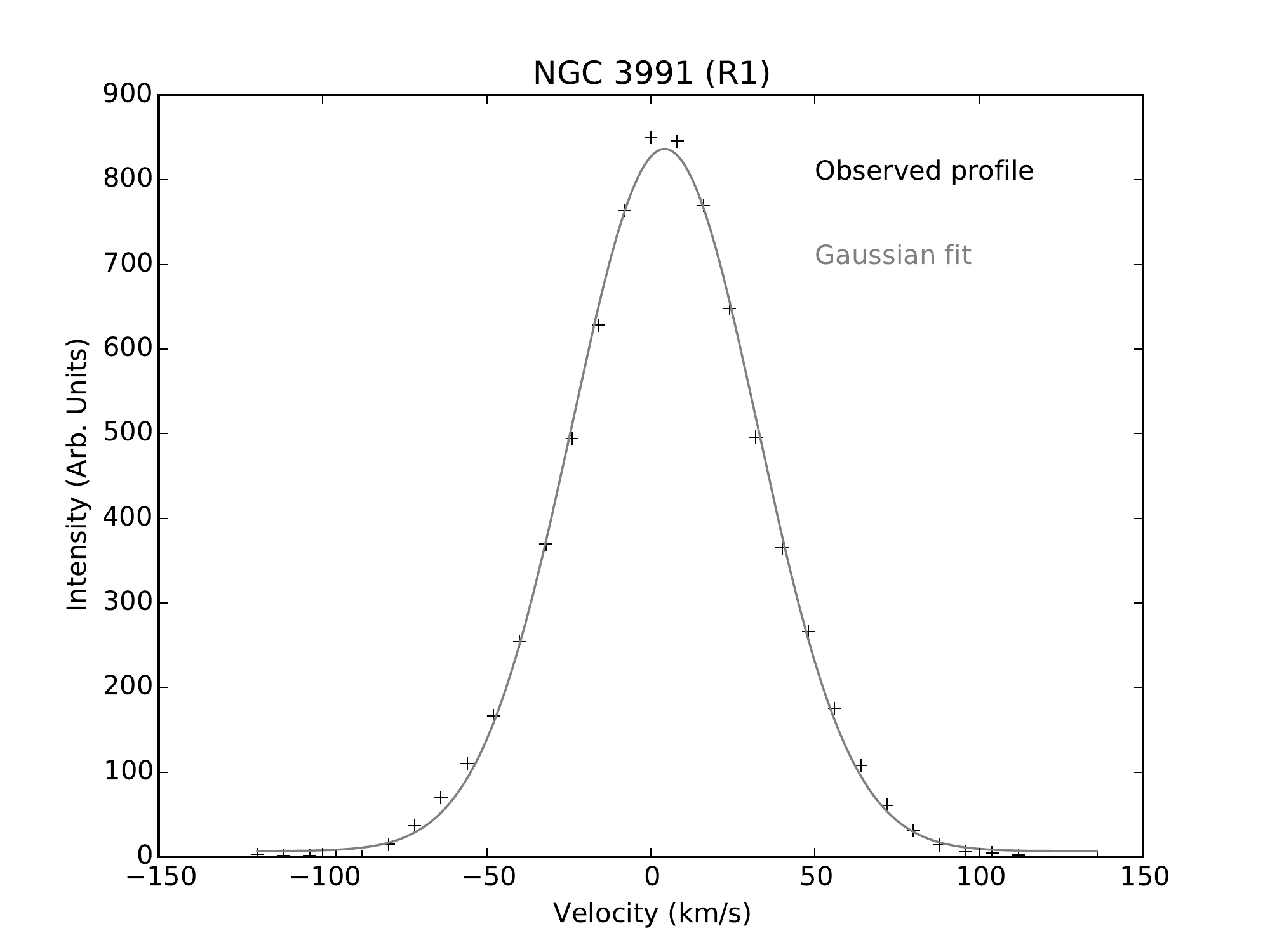}\\

\end{tabular}
\caption{H$\alpha$ profile for a sample of the brightest H{\sc{ii}} regions (Region 1 in tables \ref{table_hii_int}1 and \ref{table_hii_iso}2) of a subsample of galaxies.}
\label{fig_profiles_hii}
\end{figure*}

The second moment is used to estimate the velocity dispersion along the velocity 
axis weighted by the datacube values
\begin{equation}
 \sigma_v=\sqrt{\frac{\sum_i ^{\mathrm{region}} T_i(v_i-\overline{v})^2}{\sum_i ^{\mathrm{region}} T_i}},
 \label{eq_sigma}
\end{equation}
where 
\begin{equation}
 \overline{v}=\frac{\sum_i ^{\mathrm{region}}T_i v_i}{\sum_i ^{\mathrm{region}}T_i},
\end{equation}

and the sum is over all the pixels inside the region defined by $T_{\mathrm{edge}}$. 
 Equation \ref{eq_sigma} assumes that the observed H$\alpha$ profiles are gaussians. We plot a subsample of H$\alpha$ profiles 
for a selection of the regions from our sample of galaxies observed with GH$\alpha$FaS where we can check how Gaussian are the emission 
profiles in Fig. \ref{fig_profiles_hii}. We might expect to find multiple components in the emission profiles as well as broad wings typical of H{\sc{ii}} regions, however, the identification method differentiates 
between them if they are separated in the datacube by a few times the rms, and selects the Gaussian core.  A more detailed study of multiple components and expansive superbubbles using 
GH$\alpha$FaS observations is described in \cite{2015MNRAS.447.3840C}.

The zeroth moment is used to estimate the flux
$F=\sum_i{T_i \thinspace\delta v\thinspace\delta x\thinspace \delta y}$, 
where $\delta x$, $\delta y$, and $\delta v$ are the pixel sizes. 

A key reason to use this method is that it 
is unbiased with respect to $T_{\mathrm{edge}}$.  
We adopt $T_{\mathrm{edge}}=4\times l_{\mathrm{rms}}$, set the intervals used in the process 
of searching for different H{\sc ii} regions to the 
same value as  $T_{\mathrm{edge}}$,  and consider 
that the minimum area in pixels is the resolution of each observation. 
Three paradigms can be used with these method described in 
\cite{2008ApJ...679.1338R}. We use here the ``bijection'' paradigm.

Although correcting H$\alpha$ for dust attenuation 
would strengthen our conclusions, making this correction 
would not affect the scaling relation behaviour, as we showed 
for the Antennae galaxies \citep{zaragoza14} where 
there is quite heavy and variable dust attenuation \citep{2005ApJ...635..280B}. 
In a very detailed study using HST H$\alpha$ and continuum images of M51, 
\citet{2010ihea.book...77G} found 
that the dust attenuation is not a systematic function of H{\sc ii} region 
luminosity, or size. This is explained using an inhomogeneous model in 
which the bulk of the ionization occurs in denser clumps, in such a 
way that the dust attenuation depends on the mean clump size rather 
than on the overall size of the region.  This implies that a statistical 
study of the type presented here should not be significantly affected.

Given the H$\alpha$ luminosity, $L_{\rm H\alpha}$, and the size of the region, $R$, we can derive 
the mean electron density, $n_e$, following \citet{2005A&A...431..235R}. 
Making a spherical first order approximation for the  H{\sc ii} regions  \citep{1978ppim.book.....S}:

\begin{equation}
\frac{L_{\rm H\alpha}}{\pi {R_{\rm{cm}}}^2}=h\nu_{\rm H\alpha}\alpha_{\rm H\alpha}^{\rm{eff}}(H_0,T)2.46\cdot 10^{17}\cdot n_e^2{R_{\rm{cm}}}
\label{eq:density}
\end{equation}

where $h\nu_{\rm H\alpha}$ is the energy of an H$\alpha$ photon, $\alpha_{\rm H\alpha}^{\rm{eff}}(H_0,T)$ is the effective recombination 
coefficient of the H$\alpha$ emission, and $R_{\rm{cm}}$ is the radius in cm. 
  Equation \ref{eq:density} assumes no variation of the filling factor and of the ionizing photon scape probability with the properties of the regions. 
However, a comprehensive study of the filling factor for the H{\sc{ii}} regions in 
NGC 6946 by \cite{2013ApJ...765L..24C} implies that there should be little variation in the effect of the filling factor on the mean electron density between larger and smaller H{\sc ii} regions, while a detailed study by 
\cite{2002A&A...386..801Z} showed that the ionizing photon escape fraction does not show strong variations with luminosity (and size), 
as would be expected from regions with essentially clumpy structure \citep{2004A&A...424..877G}

Equation \ref{eq:density} also assumes 
that the source of ionization is only ionizing photons emitted by the massive stars, and does not 
include other sources of ionization such as shocks. Previous  results reported in \cite{2004AJ....127.1405C} for a 
selection of galaxies suggest that 
the proportion of ionization by shocks is rather small compared to the effect of photoionization. The diagnostics used 
to disentangle the two possible contributions to the ionization are in any case  degenerate, but previous work gives an 
upper limiting value of 33\%  for the contribution due to the ionization by shocks, and a canonical value of ~15\% 
\citep{2011ApJ...731...45H}. 

From Equation \ref{eq:density}, we can derive the ionized gas mass, $M_{\rm{HII}}$

\begin{equation}
M_{\rm{HII}}(\mathrm{M_{\odot}})  =  \frac{4}{3}\pi\thinspace R^3 \thinspace n_e \thinspace m_p %\\
= 1.57\times 10^{-17} \sqrt{L_{\rm H\alpha}\times R^3},
\label{eq2}
\end{equation}

where $R$ is in pc, $L_{\rm H\alpha}$ is in erg/s, and $m_p=1.67\times 10^{-27}\mathrm{kg}$ is the proton mass.

\begin{figure}
 \centering

\epsfig{file=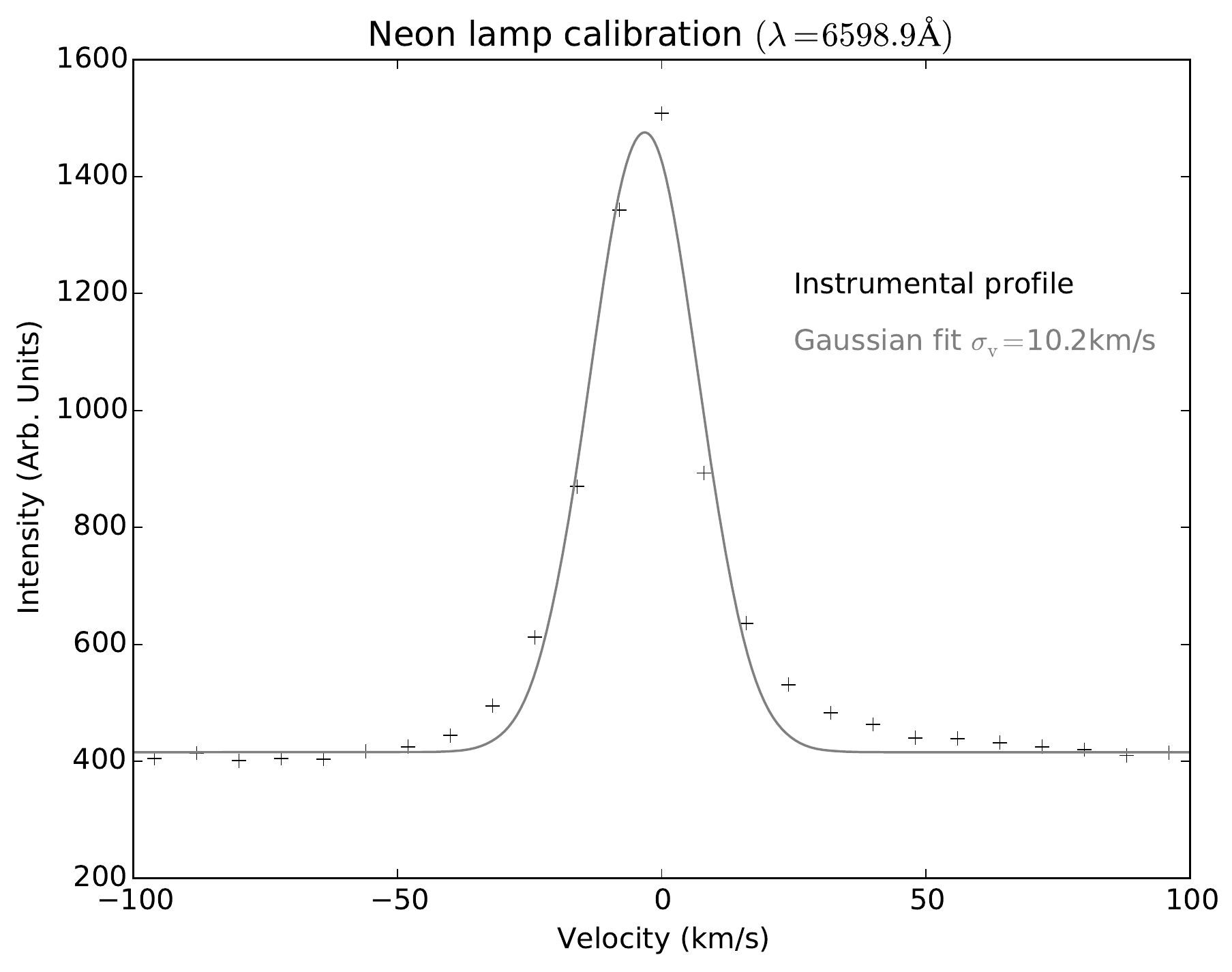,width=0.9\linewidth}
\caption{Emission profile of the Neon calibration lamp at 
6598.9$\rm{\AA}$.}
\label{fig_inst}
\end{figure}

As in \cite{1981MNRAS.195..839T,2010MNRAS.407.2519B,zaragoza13,zaragoza14}, 
the use of the velocity dispersion of the H{\sc ii} region gives us 
really new information about the physics in those regions when compared to previous purely morphological studies 
\citep{1989ApJ...337..761K,2006A&A...459L..13B}. We corrected the velocity dispersion for the instrumental, 
the natural ($\sigma_{v,n}=3\mathrm{km/s}$), and the thermal ($\sigma_{v,th}=9.1\mathrm{km/s}$ linewidths, assuming 
an isothermal H{\sc ii} region temperature of $T=10^4 \mathrm{K}$), and subtracting them in quadrature from the observed width. 
 The instrumental velocity dispersion is derived from the Neon calibration lamp using the emission line at 
6598.9$\rm{\AA}$. For each GH$\alpha$FaS observation of one galaxy, a calibration datacube is taken, so we 
derive the instrumental velocity dispersion fitting a Gaussian to the brightest pixel of the calibration datacube (see Fig. \ref{fig_inst})
We only use as a valid velocity dispersions those with line widths
greater than $8\thinspace\mathrm{km/s}$, i.e., velocity dispersions greater than $4\thinspace\mathrm{km/s}$, which is the effective
velocity resolution of the
instrument ($8\thinspace\mathrm{km/s}$ for GH$\alpha$FaS).

We estimated the virial parameter, $\alpha_{\rm{vir}}=5\frac{\sigma_v^2 R}{G M_{\rm{gas}}}$ \citep{1992ApJ...395..140B}, 
using the corrected velocity dispersion,  the derived radius of the H{\sc ii} regions, and using the mass of the ionized gas as the gas mass, $M_{\rm{gas}}=M_{\rm{HII}}$ as a first 
approximation.  The stellar mass inside H{\sc{ii}} regions is small (less than 4\%) compared with the total H{\sc{ii}} region mass \citep{2005A&A...431..235R}, so we use the size of the H{\sc{ii}} region and not that of the 
stellar ionizing cluster to estimate the virial parameter. It is true that H{\sc ii} regions, particularly large, luminous regions, 
have considerable neutral  and molecular gas as well as ionized gas. In the larger more luminous regions, the ionized gas is an order of magnitude smaller than the neutral plus molecular gas \citep{1996AJ....112..146Y,2004A&A...424..877G}. 
The results presented in \citet{zaragoza14} suggest that a constant ionization fraction 
is a good approximation, at least, for the correlations we are going to study here. Thus, 
the real virial parameters for all the regions will be a nearly constant fraction of those estimated here,  between $0.05$ \citep{zaragoza14} and $0.1$ \citep{1996AJ....112..146Y,2004A&A...424..877G}, 
depending on the ionization fraction value. Since 
the virial parameter is  in any case uncertain by a factor which depends in detail on the (non uniform) density 
distribution and deviations from sphericity \citep{1992ApJ...395..140B}, the extra scatter in the uncertainty entailed by using 
ionized gas and a constant factor to make up the total gas mass will not seriously affect our results.

\subsection{Uncertainties}

We have estimated the uncertainties in the measured parameters following 
the bootstrapping method explained in \cite{2006PASP..118..590R}. 
The method estimates the uncertainties by randomly sampling the values of the datacube for 
each region allowing repeated values, and then subtracts the parameter (Luminosity, size or 
velocity dispersion) several times to estimate the standard deviation of these derived 
parameters. The uncertainty for each parameter is its standard deviation scaled up 
by the square root of the number of pixels in one resolution element. 
This uncertainty does not include the intrinsic error 
of the flux in the data cubes.

\section{H{\sc{ii}} regions in the interacting galaxies}
\begin{figure*}
 \centering
\epsfig{file=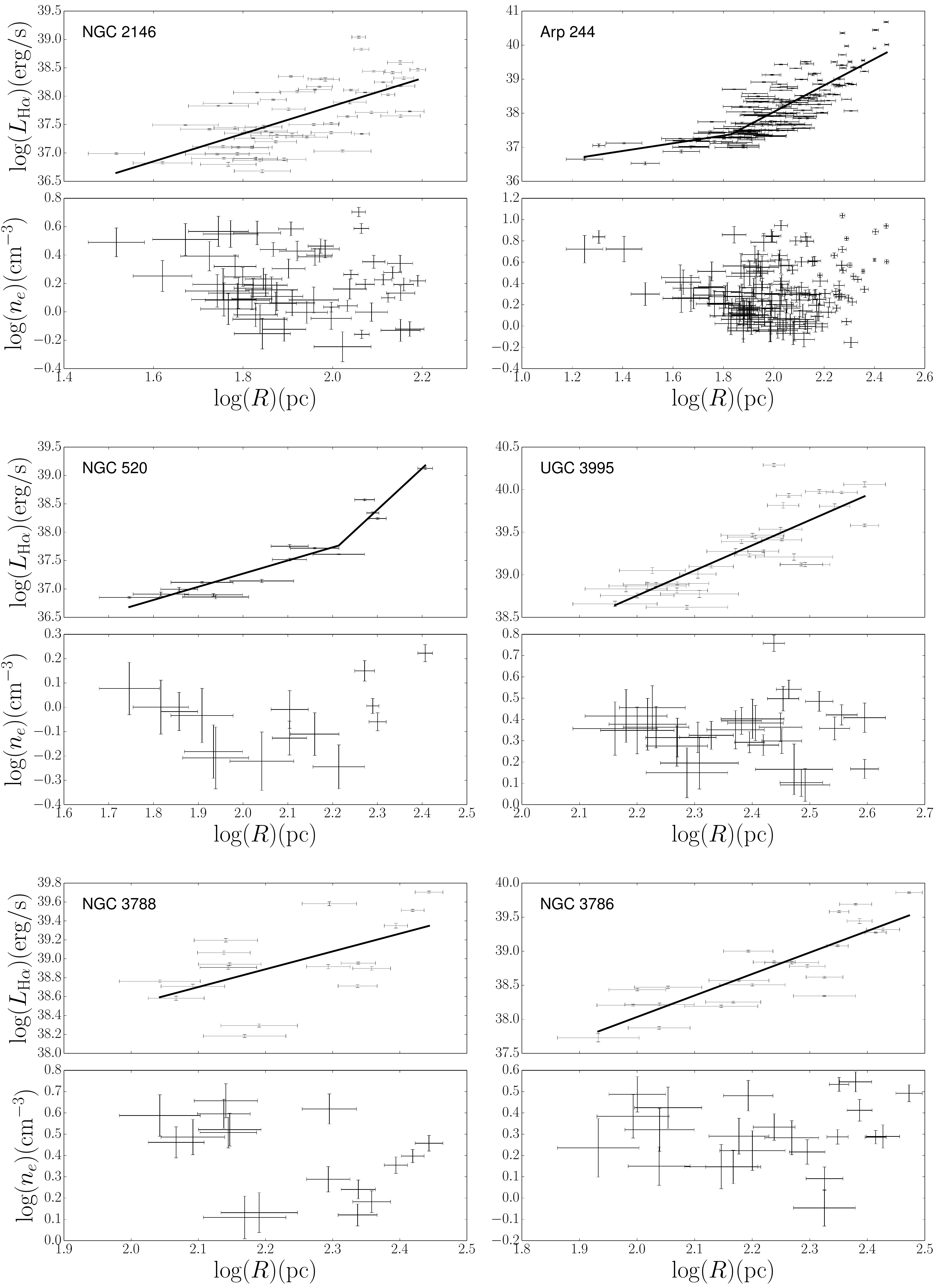,width=0.9\linewidth}
% \begin{tabular}{cc}
% 
% \epsfig{file=l-r_int.pdf,width=0.45\linewidth} &
% \epsfig{file=dens_int.pdf,width=0.45\linewidth}\\
% \end{tabular}
% 
% %\includegraphics[width=0.9\linewidth]{lum_func.eps}
% \caption{Left: $L_{\rm H\alpha}$ versus $R$ of H{\sc ii} regions in interacting galaxies. The result of the double linear fit is drawn as a 
% solid line. Right: $n_e$ versus $R$ of H{\sc ii} regions in interacting galaxies. }

\caption{H$\alpha$ luminosity, $L_{\rm H\alpha}$, versus H{\sc ii} size (radius, $R$) for interacting galaxies. The result of the double (or single) linear fit is drawn as a 
 solid line. The $n_e$ versus $R$ of H{\sc ii} regions in interacting galaxies is plotted below the $L_{\rm H\alpha}$-$R$ plot. }
\label{scalrel_int}
\end{figure*}

\begin{figure*}

 \centering
\epsfig{file=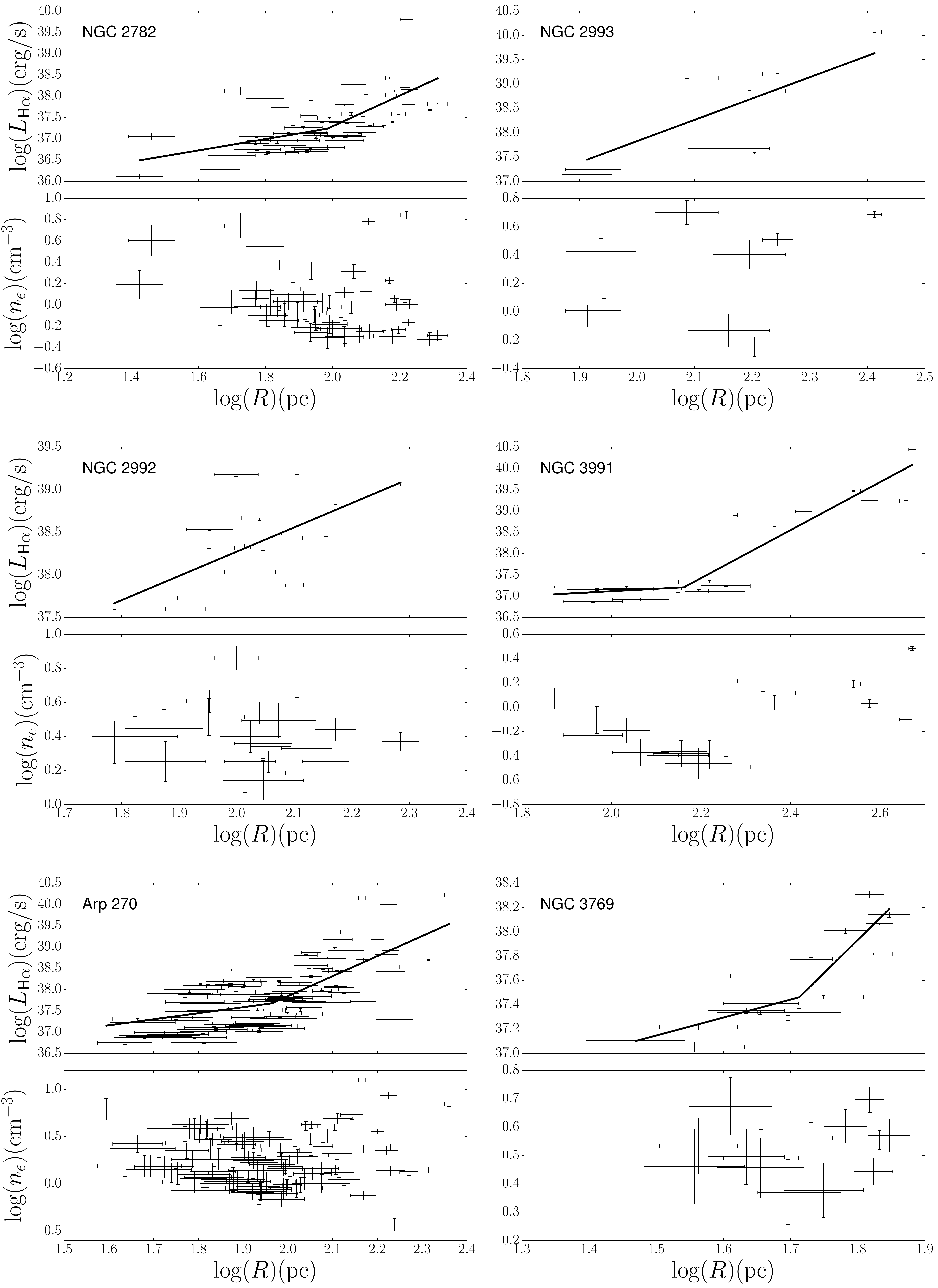,width=0.94\linewidth}
\contcaption{}
\end{figure*}

We present the parameters of the H{\sc ii} regions belonging to the interacting galaxies in Table \ref{table_hii_int}1. We have included 
the data for the interacting pair Arp 270 presented in \cite{zaragoza13}, where we have reanalyzed the H{\sc ii} region parameters 
using \textsc{astrodendro} package in substitution of \textsc{clumpfind} in order 
to produce a homogeneous study for the full set of galaxies. 
We present the reanalyzed H{\sc ii} regions of Arp 270 in Table \ref{table_hii_int}1. However, we have also used in the subsequent 
analysis the parameters of the H{\sc ii} regions of the Antennae galaxies presented in \cite{zaragoza14} in order to obtain a larger 
statistical sample of regions 
for the interacting galaxies. For these objects, we have 1259 H{\sc ii} regions in total. 
The parameter most sensitive to the noise level 
is the radius of the region, so in the following analysis we have removed the regions where the relative error in the radius is larger 
than $15\%$.  After applying this exclusion level we were left with 537 H{\sc ii} regions where the error in radius is sufficiently small, 
and for this sample the relative errors in the 
derived parameters: 
ionized gas mass, and electron density, are less than $20\%$.

\subsection{Scaling relations of H{\sc ii} regions}

\begin{table}
 \begin{tabular}{cccccc}
\hline
Name&$N_1$&$L_1$&$N_2$& $L_2$&$\log R_1$\\
\hline
NGC 2146 & 2.4 & 32.9 &  &  & \\  
Arp 244 & 1.1 & 35.3 & 3.8 & 30.2 & 1.8\\  
NGC 520 & 2.3 & 32.6 & 7.3 & 21.5 & 2.2\\  
UGC 3995 & 2.9 & 32.2 &  &  & \\  
NGC 3788 & 1.8 & 34.7 &  &  & \\  
NGC 3786 & 3.1 & 31.7 &  &  & \\  
NGC 2782 & 1.3 & 34.5 & 3.6 & 30.0 & 1.9\\  
NGC 2993 & 4.3 & 29.0 &  &  & \\  
NGC 2992 & 2.8 & 32.5 &  &  & \\  
NGC 3991 & 0.5 & 35.9 & 5.6 & 25.0 & 2.1\\  
Arp 270 & 1.4 & 34.8 & 4.7 & 28.4 & 1.9\\  
NGC 3769 & 1.4 & 34.9 & 5.4 & 28.1 & 1.7\\ 
\hline

 \end{tabular}
\caption{Results of the single and double (when applicable) linear fits for interacting galaxies
as defined in eq. \ref{polyfit_hii}.}
\label{tab_fit_int}
\end{table}

In Fig. \ref{scalrel_int} we plot the $L_{\rm H\alpha}$-$R$ relations for the H{\sc ii} regions for individual 
galaxies.  We fit an x-axis error weighted single linear fit, and also a double continuous x-axis error weighted linear fit (4 free parameters) using 
non-linear least squares to fit the double linear fit function to the data. We weight the error on the x-axis since the radii of the regions are 
the strongest source of uncertainty.  
We have chosen between a continuous double linear fit or a single one 
depending on the $\chi_{\mathrm{red}}^2$ value, taking into account the addition of 
two free parameters more in the double linear fit. In the cases where 
$\chi_{\mathrm{red}}^2$ is smaller for the double linear fit than 
in the single one, we choose the double continuous 
 x-axis error weighted linear fit:

\begin{equation}
\label{polyfit_hii}
\begin{split}
\log(L_{\rm H\alpha})  &= 
L_1+N_1\log(R) \thinspace \mathrm{;}%\\ 
 \thinspace\thinspace\thinspace\thinspace \mathrm{for}\thinspace \log(R)< R_1
\\
  \log(L_{\rm H\alpha}) & = 
L_2+N_2\log(R) \thinspace \mathrm{;}%\\ 
 \thinspace\thinspace\thinspace\thinspace \mathrm{for}\thinspace \log(R)> R_1.
 \end{split}
\end{equation}

In the cases where $\chi_{\mathrm{red}}^2$ is smaller for the single linear fit than 
in the double one, we choose the single 
linear fit $\log(L_{\rm H\alpha})  = L_1+N_1\log(R)$. The 
results are in Table \ref{tab_fit_int}. For 8 of the 12 galaxies in the sample, we found 
a regime where the exponent  $N_1$ and/or $N_2$ in the $L_{\rm H\alpha}-R$ relations is larger than three.
 We have found that for the galaxies with a double linear fit, the exponent $N_2$ is larger than three. 
Thus, the results for these larger (and brighter) H{\sc ii} regions
are different, when compared to the results of \citet{1981MNRAS.195..839T,gutierrez10} for M51, 
where the maximum luminosity of its 
H{\sc ii} regions is not as high as for our sample of interacting galaxies, 
and similar to recent results of \cite{zaragoza13,zaragoza14} for 
two of the systems included here, the Antennae and Arp 270. 
For two galaxies, NGC 2993, and NGC 3786 there are not enough fainter H{\sc ii} regions to perform 
a double linear fit, although they show an exponent in the $L_{\rm H\alpha}-R$ relation also larger than three, 
but we can not separate them into two regimes. 
 The Table \ref{tab_fit_int} and the Fig. \ref{scalrel_int} are 
sorted by absolute magnitude, from brighter to fainter. 
The brightest regime where the exponent is larger than three is independent of the 
absolute magnitude.

An exponent larger than $3$ in the $L_{\rm H\alpha}$-$R$ relation   
implies that the electron density (or the density of ionized gas) increases with 
the radius (and the luminosity) of the region.
We show the $n_e$-$R$ relation in Fig.\ref{scalrel_int} (below each $L_{\rm H\alpha}-R$ plot), where the electron 
density decreases with radius for the set of smaller (and fainter)
H{\sc ii} regions for the interacting galaxies. In those cases we find a two valued behaviour, density decreasing with radius in the small radius (i.e. low luminosity) part of the parameter space, and increasing with radius for large radii (i.e. large luminosities). This change is clearest for those systems with large numbers of luminous regions.

\subsection{$L_{\rm H\alpha}$-$\sigma_v$ envelope}

\begin{figure*}
 \centering
 \begin{tabular}{cc}
\epsfig{file=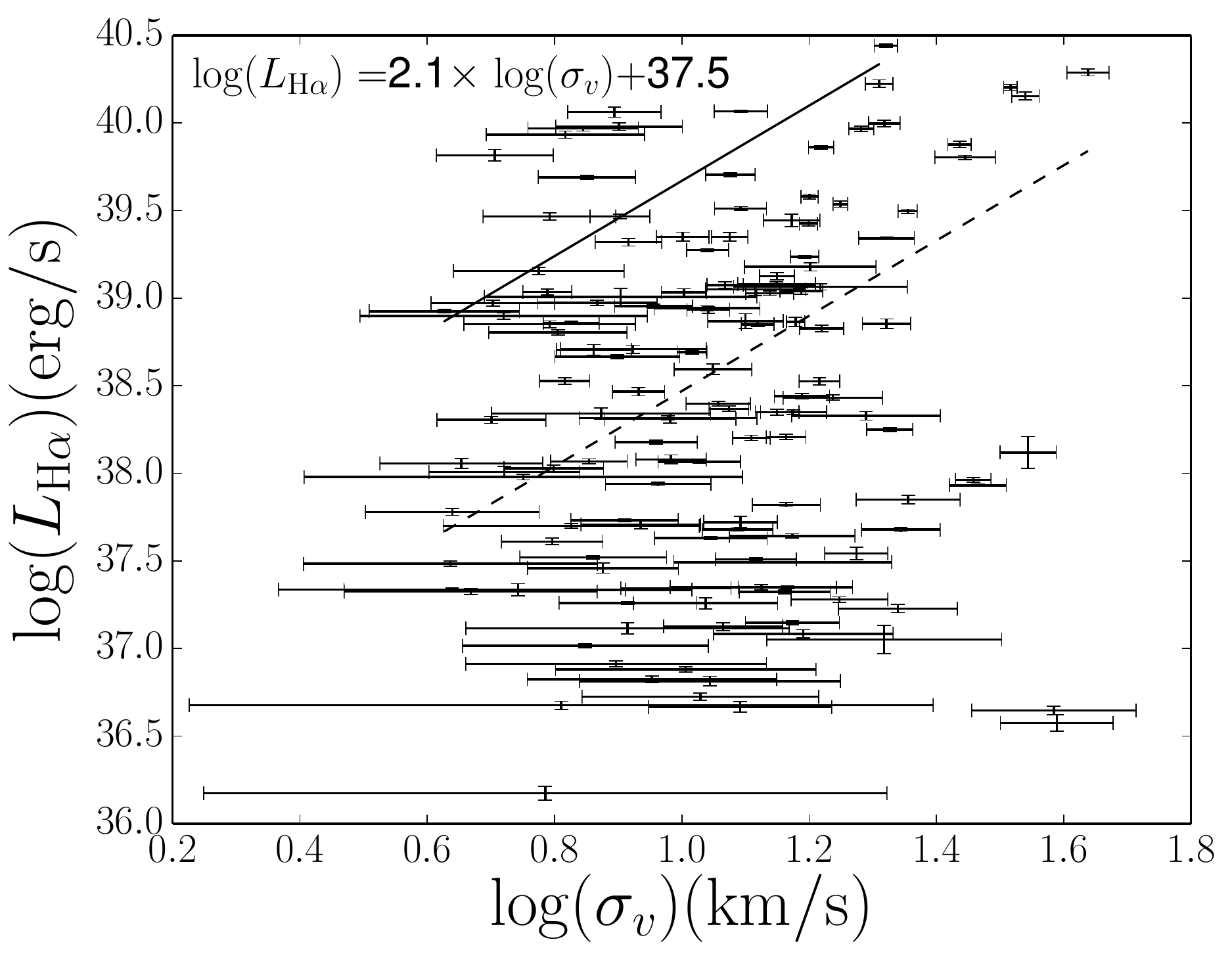,width=0.45\linewidth}&
\epsfig{file=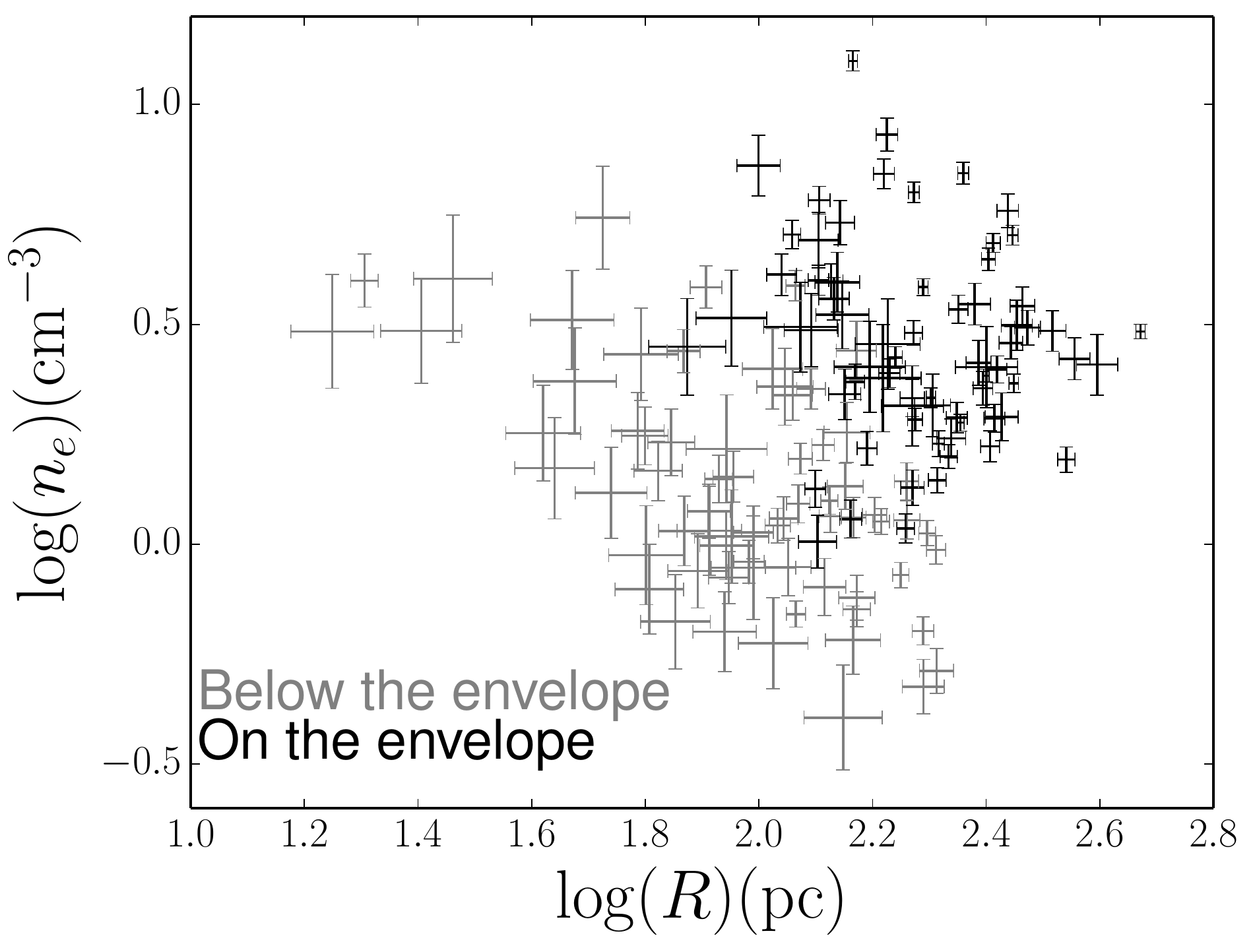,width=0.45\linewidth}\\
  \end{tabular}
\caption{Left: H$\alpha$ luminosity, $L_{\rm H\alpha}$, versus velocity dispersion, $\sigma_v$, 
for H{\sc ii} regions in interacting galaxies. 
We have plotted the fitted envelope as a solid line, and 
the displaced envelope condition explained in the paper, as a dashed line. 
Right: electron density, $n_e$, versus the radius,
$R$, for H{\sc ii} regions on the envelope (black) and those regions below the envelope (grey).}
\label{fig_env_int}
\end{figure*}

Since we have measurements of the velocity as a third dimension, we have used this to deepen our study of the 
 H{\sc ii} regions. We plot the $L_{\rm H\alpha}$-$\sigma_v$ envelope in Fig.\ref{fig_env_int} (left) for 
all the H{\sc ii} regions from the sample of interacting galaxies. 
\citet{2005A&A...431..235R} suggested 
that the H{\sc ii} regions on the envelope are virialized in the sense that they are gravitationally dominated since 
those regions are the ones showing the minimum velocity dispersion for a given luminosity, 
and the excess for the regions away from the envelope can be attributed to internal motions 
which are not in quasi-equilibrium, such as recent contributions from stellar winds and supernovae. 
We have checked if the distance of an H{\sc ii} region, or a set of H{\sc ii} regions, 
from the envelope could be used as a parameter to distinguish between two density regimes. We 
estimated the envelope as described in \citet{2005A&A...431..235R}, taking 
the H{\sc ii} region with the minimum velocity dispersion for each luminosity bin, and fitting 
a linear relation. The result is 

\begin{equation}
\log(L_{\rm{H\alpha}\thinspace \mathrm{env}})=2.1\times \thinspace \log(\sigma_{v})+37.5
\label{eqenv}
\end{equation}

where $L_{\rm{H\alpha}\thinspace \mathrm{env}}$ is in erg/s and $\sigma_v$ is in km/s. We have divided 
the populations of H{\sc ii} regions, those which are on the envelope and obey 
$L_{\rm H\alpha}>\frac{L_{\rm{H\alpha}\thinspace \mathrm{env}}}{40}$, and those which are not 
on the envelope and do not obey the previous inequality. This condition 
is plotted as a dashed line in Fig.\ref{fig_env_int} (left). We have plotted 
the electron density versus radius for all the H{\sc ii} regions from the sample of interacting galaxies 
in Fig.\ref{fig_env_int} (right). 
The two populations are 
clearly separated in the $n_e$-$R$ relation plotted, 
where the grey points are for regions below the envelope, and the black points 
are on the envelope. The electron density decreases with the radius of the region 
for the regions below the envelope, and increases for the regions 
on the envelope, where the H$\alpha$ luminosity depends 
super-linearly on the velocity dispersion (Equation \ref{eqenv}). 

\subsection{Virial parameter}

\begin{figure}
 \centering

\epsfig{file=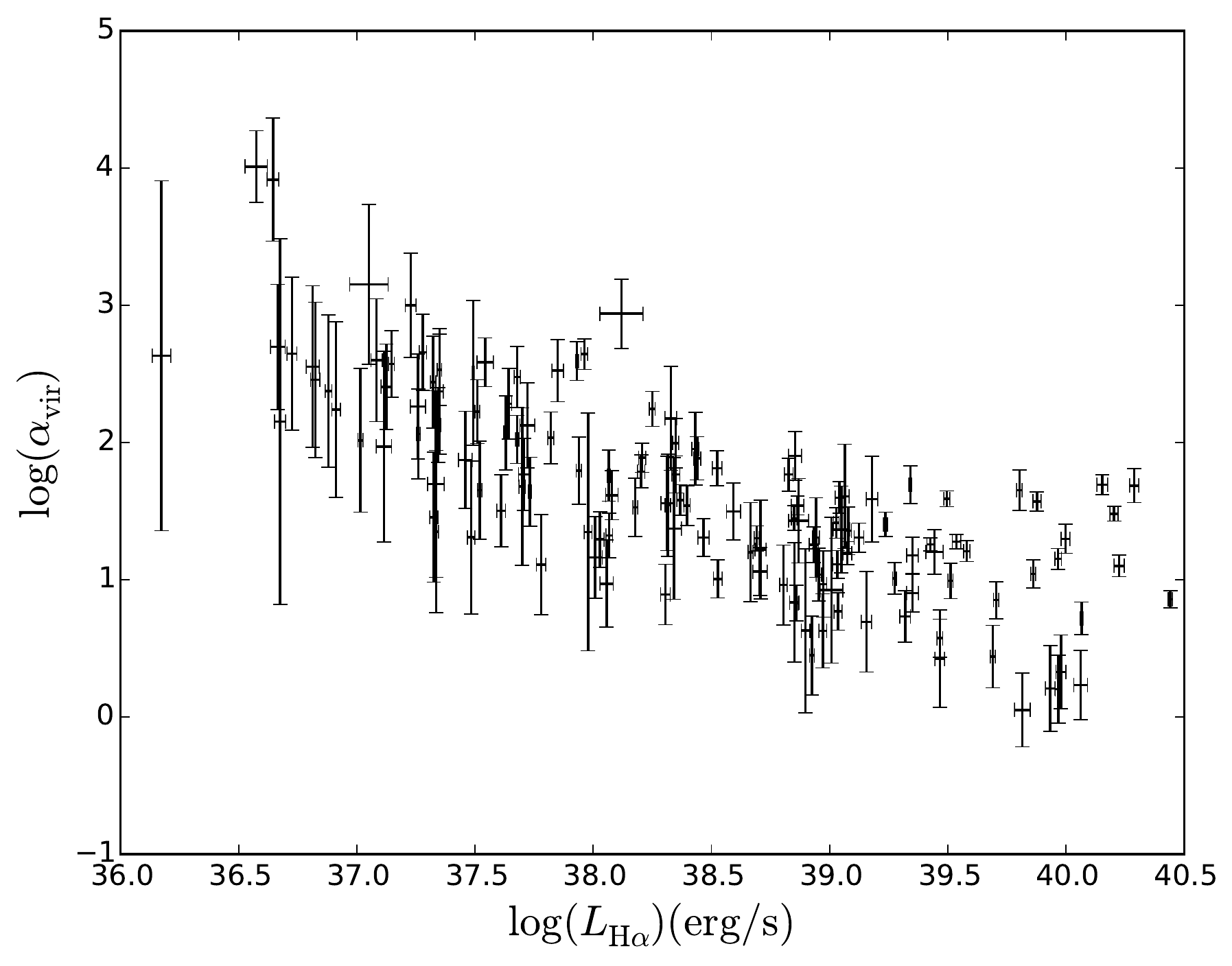,width=0.9\linewidth}
\caption{The virial parameter, $\alpha_{\rm{vir}}= 5 \frac{\sigma_{v}^2 R_{\rm{HII}}}{GM_{\rm{gas}}}$ 
versus H$\alpha$ luminosity, $L_{\rm H\alpha}$. We have used the mass of the 
ionized gas as the gas mass (in fact it should be a constant fraction of the total mass, 
but our qualitative conclusions do not depend on the value of this fraction, 
so we have not included it) $M_{\rm{gas}}=M_{\rm{HII}}$.}
\label{fig_vir_int}
\end{figure}

We have estimated the virial parameter, $\alpha_{\rm{vir}}= 5 \frac{\sigma_{v}^2 R_{\rm{HII}}}{GM_{\rm{gas}}}$, 
which is the ratio between the kinetic and gravitational energy \citep{1992ApJ...395..140B}, 
substituting the ionized gas mass, $M_{\rm{HII}}$, for the total gas mass, 
as a practical first approximation. The comparison 
between ionized and molecular gas in the Antennae galaxies \citep{zaragoza14} 
yields the result that the ionized gas mass  must be a nearly constant fraction of the total gas mass. 
Since the virial parameter is affected by a further constant factor due to the density inhomogeneities
and the non-sphericity of gas clouds, the use of the ionized gas mass yields a 
constant offset for the virial parameter. As we are making a comparative analysis 
of the two sets of H{\sc ii} regions, we will neglect the numerical values of these offsets in the first instance. 

We plot $\alpha_{\rm{vir}}$ versus $L_{\rm H\alpha}$ in Fig.\ref{fig_vir_int} for all 
the H{\sc ii} regions together. 
We have used a result from the literature which shows that the H{\sc ii} regions from the \textgravedbl electron 
density decreasing \textacutedbl regime are pressure confined 
\citep{gutierrez11}, and now we can check if the values of the virial parameter agree with this. 
$\alpha_{\rm{vir}}\gg1$ values, are for gas clumps confined by the external pressure. This explains  
 why the gas density decreases with size 
 since the external pressure 
decreases with the distance from the galactic plane, with a scale height comparable with the 
sizes of the H{\sc ii} regions, so the bigger the region, 
the lower is the effective external pressure. 
Values of $\alpha_{\rm{vir}}\sim1$ are for clouds where the self gravity is the 
dominant force. The fact that the more massive clouds take values of this order 
 implies that for clouds with masses sufficiently large the bigger the region, 
the stronger is the gravitational field so the density is an increasing function of radius.
In Fig.\ref{fig_vir_int} we can see 
that for the  H{\sc ii} regions from the lower luminosity regime, the virial parameter 
lies in a range significantly greater than unity, as predicted for 
pressure confined H{\sc ii} regions. In contrast for the H{\sc ii} regions in the 
high luminosity regime, the virial parameter lies close to a nearly constant 
value with mass, whose value is, apparently different from unity.  
 However, if we take into account the results from \cite{zaragoza14} which 
give a nearly constant ionized gas fraction of $\sim0.05$, substituting the 
total gas mass for the ionized gas mass the virial 
parameter does lie close to unity, implying that the brightest H{\sc ii} regions are 
gravitationally bound.

\section{H{\sc ii} regions in isolated galaxies}
\begin{figure*}
 \centering
\epsfig{file=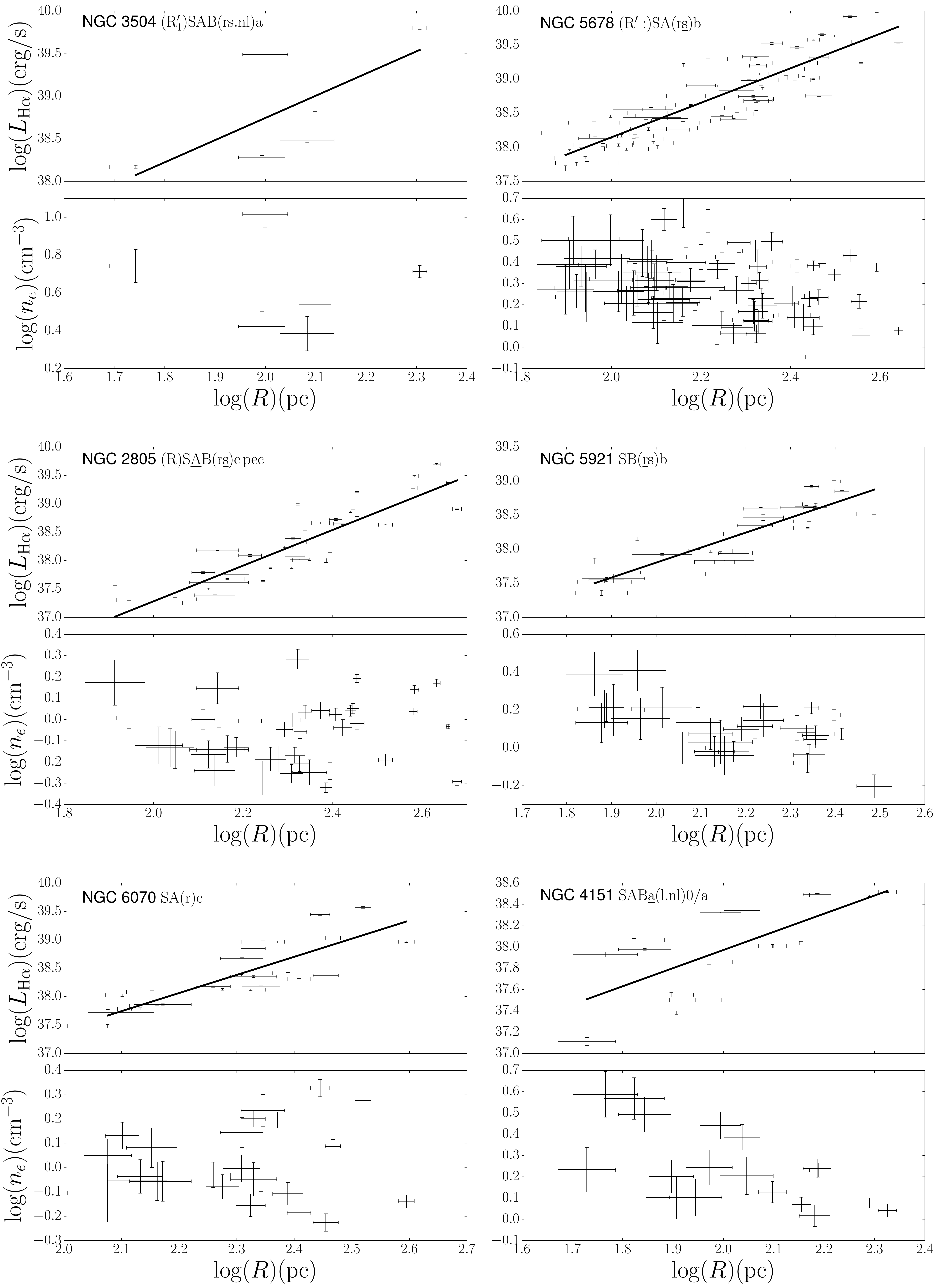,width=0.94\linewidth}
% \begin{tabular}{cc}
% 
% \epsfig{file=l-r_iso.pdf,width=0.45\linewidth} &
% \epsfig{file=dens_iso.pdf,width=0.45\linewidth}\\
% \end{tabular}
% \caption{Left: $L_{\rm H\alpha}$ versus $R$ of H{\sc ii} regions in isolated galaxies. T
\caption{H$\alpha$ luminosity, $L_{\rm H\alpha}$, versus H{\sc ii} size (radius, $R$) 
for isolated galaxies. The result of the double (or single) linear fit is drawn as a 
 solid line. The $n_e$ versus $R$ of H{\sc ii} regions in isolated galaxies is plotted below each $L_{\rm H\alpha}$-$R$ plot. }
\label{scalrel_iso1}
\end{figure*}

\begin{figure*}

 \centering
\epsfig{file=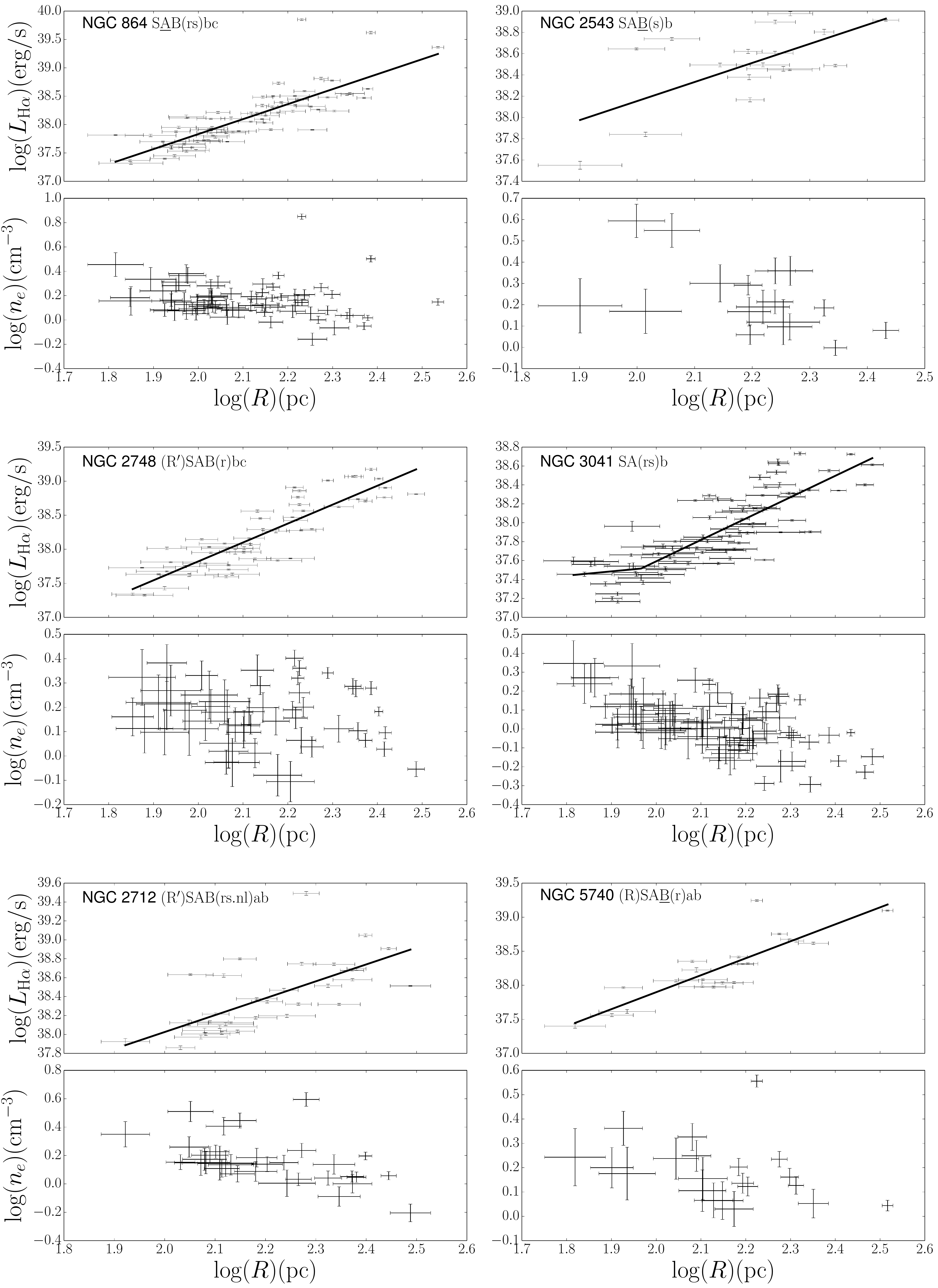,width=0.94\linewidth}
\contcaption{}
\end{figure*}

\begin{figure*}

 \centering
\epsfig{file=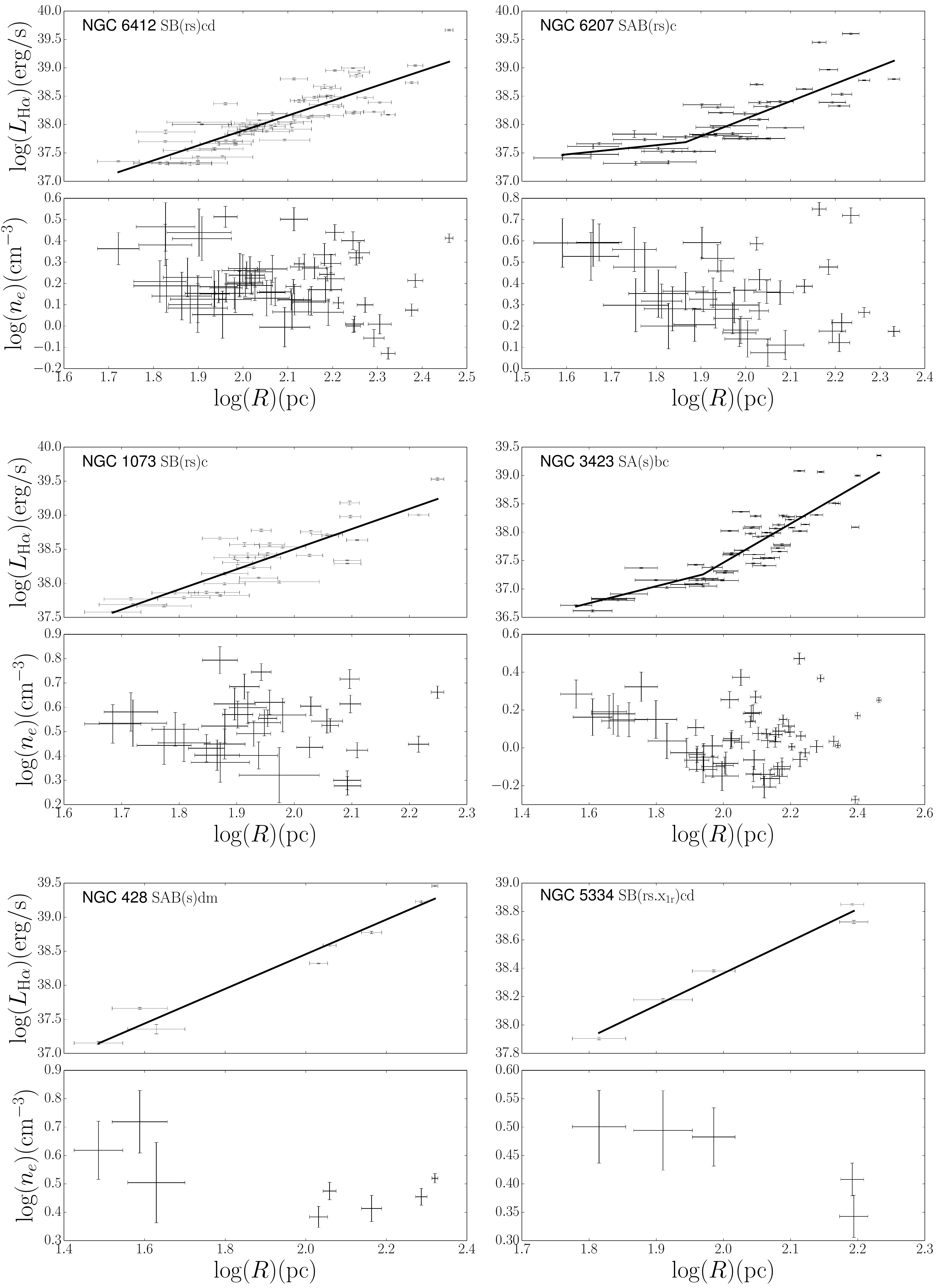,width=0.94\linewidth}
\contcaption{}
\end{figure*}

\begin{figure*}

 \centering
\epsfig{file=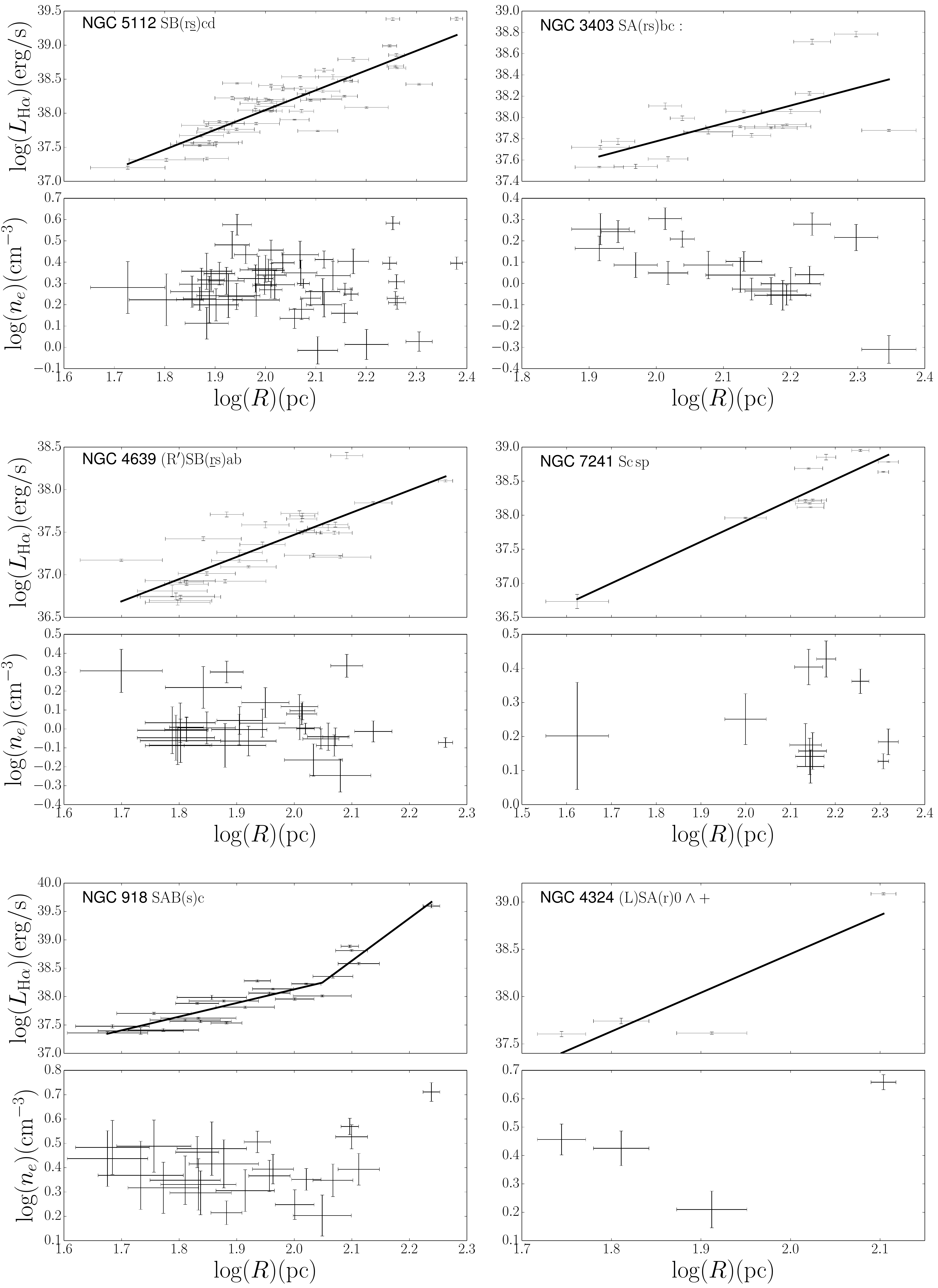,width=0.94\linewidth}
\contcaption{}
\end{figure*}

\begin{figure*}

 \centering
\epsfig{file=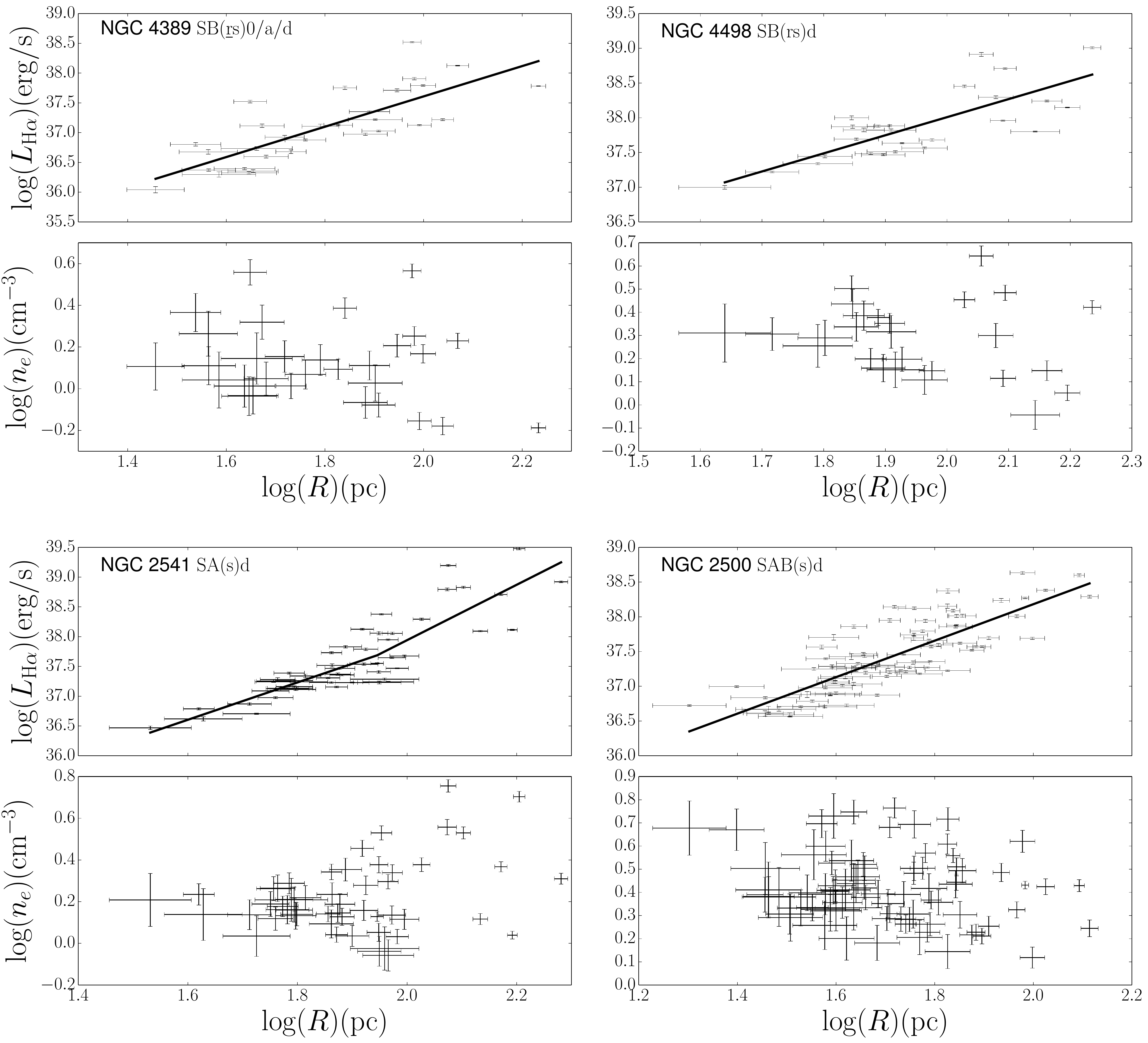,width=0.94\linewidth}
\contcaption{}
\end{figure*}
We derived the parameters of the H{\sc ii} regions in isolated galaxies using 
observational data from the sample of 28 isolated galaxies  in \citet{2015arXiv150406282E} observed with the same 
instrument, GH$\alpha$FaS, and on the same telescope, WHT, as our sample of interacting galaxies. We present 
the parameters in Table \ref{table_hii_iso}2 for 1054 H{\sc ii} regions. Again, we have used in the following analysis 
of H{\sc ii} regions in isolated galaxies, those regions with a fractional error in radius less than a $15\%$ since 
the radius of the regions is the most sensitive parameter. After applying this 
criterion we retained 1018 H{\sc ii} regions from the isolated galaxy sample. 

\subsection{Scaling relations}

\begin{table}
 \begin{tabular}{cccccc}
\hline
Name&$N_1$&$L_1$&$N_2$& $L_2$&$\log R_1$\\
\hline
NGC 3504 & 2.6 & 33.5 &  &  & \\  
NGC 5678 & 2.5 & 33.0 &  &  & \\  
NGC 2805 & 3.1 & 30.9 &  &  & \\  
NGC 5921 & 2.2 & 33.4 &  &  & \\  
NGC 6070 & 3.1 & 31.0 &  &  & \\  
NGC 4151 & 1.7 & 34.5 &  &  & \\  
NGC 864 & 2.6 & 32.5 &  &  & \\  
NGC 2543 & 1.7 & 34.5 &  &  & \\  
NGC 2748 & 2.7 & 32.2 &  &  & \\  
NGC 3041 & 0.4 & 36.6 & 2.2 & 33.0 & 1.9\\  
NGC 2712 & 1.7 & 34.4 &  &  & \\  
NGC 5740 & 2.4 & 32.9 &  &  & \\  
NGC 6412 & 2.6 & 32.6 &  &  & \\  
NGC 6207 & 0.8 & 36.1 & 3.0 & 31.9 & 1.8\\  
NGC 1073 & 2.9 & 32.5 &  &  & \\  
NGC 3423 & 1.4 & 34.3 & 3.4 & 30.6 & 1.9\\  
NGC 428 & 2.5 & 33.3 &  &  & \\  
NGC 5334 & 2.2 & 33.8 &  &  & \\  
NGC 5112 & 2.9 & 32.2 &  &  & \\  
NGC 3403 & 1.6 & 34.4 &  &  & \\  
NGC 4639 & 2.6 & 32.2 &  &  & \\  
NGC 7241 & 3.0 & 31.8 &  &  & \\  
NGC 918 & 2.4 & 33.3 & 7.4 & 22.8 & 2.0\\  
NGC 4324 & 4.0 & 30.2 &  &  & \\  
NGC 4389 & 2.5 & 32.5 &  &  & \\  
NGC 4498 & 2.6 & 32.8 &  &  & \\  
NGC 2541 & 3.1 & 31.5 & 4.6 & 28.6 & 1.9\\  
NGC 2500 & 2.6 & 32.9 &  &  & \\

\hline
\end{tabular}
\caption{Results of the single and double (when applicable) linear fits for isolated galaxies.
as defined in eq. \ref{polyfit_hii}}
\label{tab_fit_iso}
\end{table}

We plot $L_{\rm H\alpha}$ versus $R$ for the H{\sc ii} regions in isolated 
galaxies in fig. \ref{scalrel_iso1}. 

 Again,  we have chosen between a continuous double linear fit or a single one 
depending on the $\chi_{\mathrm{red}}^2$ value, taking into account the addition of 
two free parameters more in the double linear fit. 
The results of the linear fits are in Table\ref{tab_fit_iso}. We find an exponent
 $N_1$ and/or $N_2$ larger than three in the 
$L_{\rm H\alpha}-R$ relation in only 6 of 28 isolated galaxies.
 Thus, the 
regime where the ionized gas density increases with size (and luminosity) 
in the isolated galaxies 
is much less frequent than in the interacting galaxies.

In Fig. \ref{scalrel_iso1} (below each $L_{\rm H\alpha}-R$ relation) we plot the electron density 
versus the radius of the H{\sc ii} regions for each individual galaxy. For the regions in 
the isolated galaxies we find in some cases a regime where the electron density increases 
with radius but where this occurs there are relatively few H{\sc ii} regions and the phenomenon is less frequent. 
Fig. \ref{scalrel_iso1}  and Table\ref{tab_fit_iso} are sorted 
by absolute magnitude (see \citet{2015arXiv150406282E}), from brighter to fainter.
The regime where the exponent is larger than three is again independent on the 
absolute magnitude of the galaxy.

% We did not find any differences as a function of  morphological 
% type for the isolated galaxies (see the types published in \citet{2015arXiv150406282E}).

\subsection{$L_{\rm H\alpha}$-$\sigma_v$ envelope}

\begin{figure*}
 \centering
 \begin{tabular}{cc}
\epsfig{file=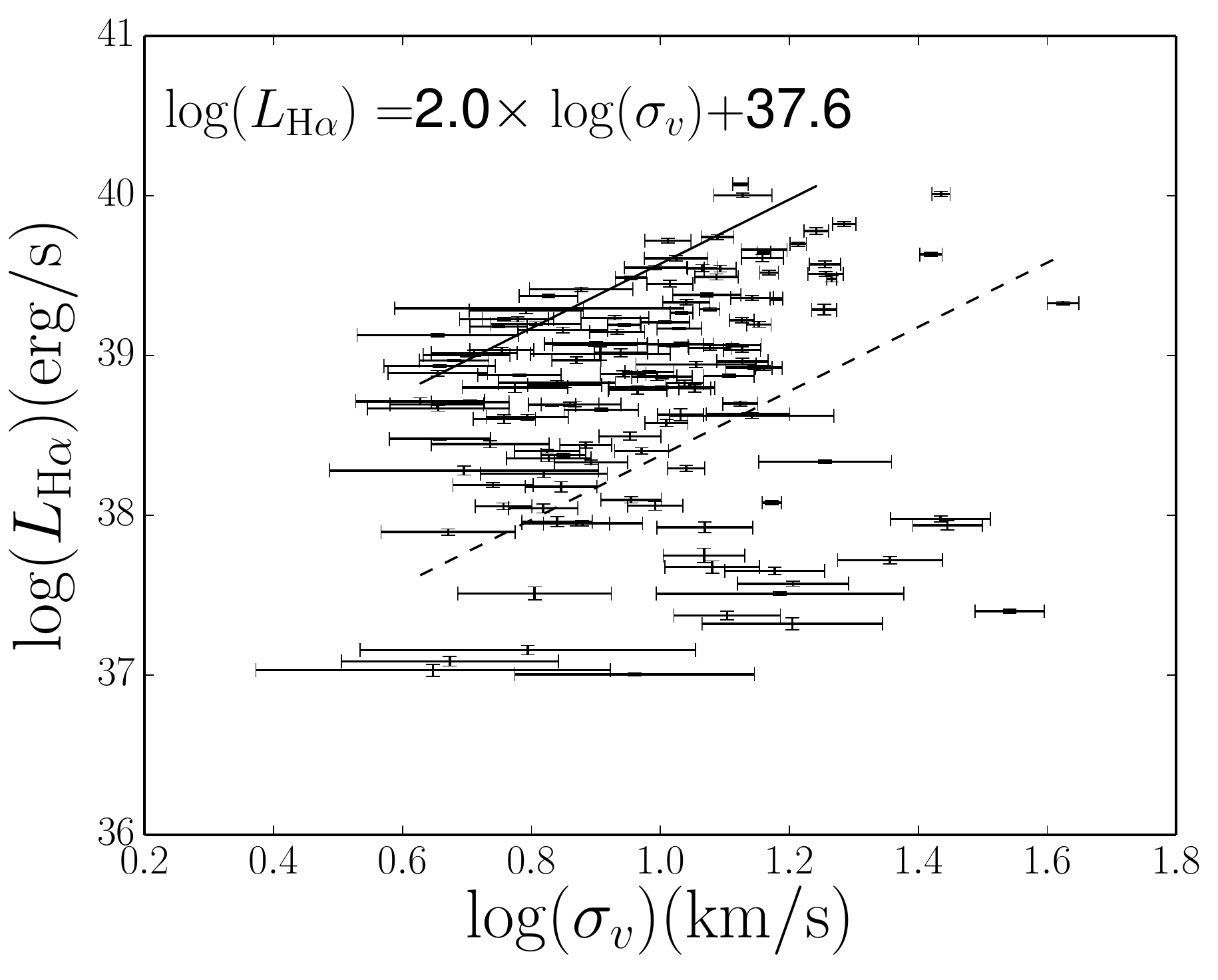,width=0.45\linewidth}&
\epsfig{file=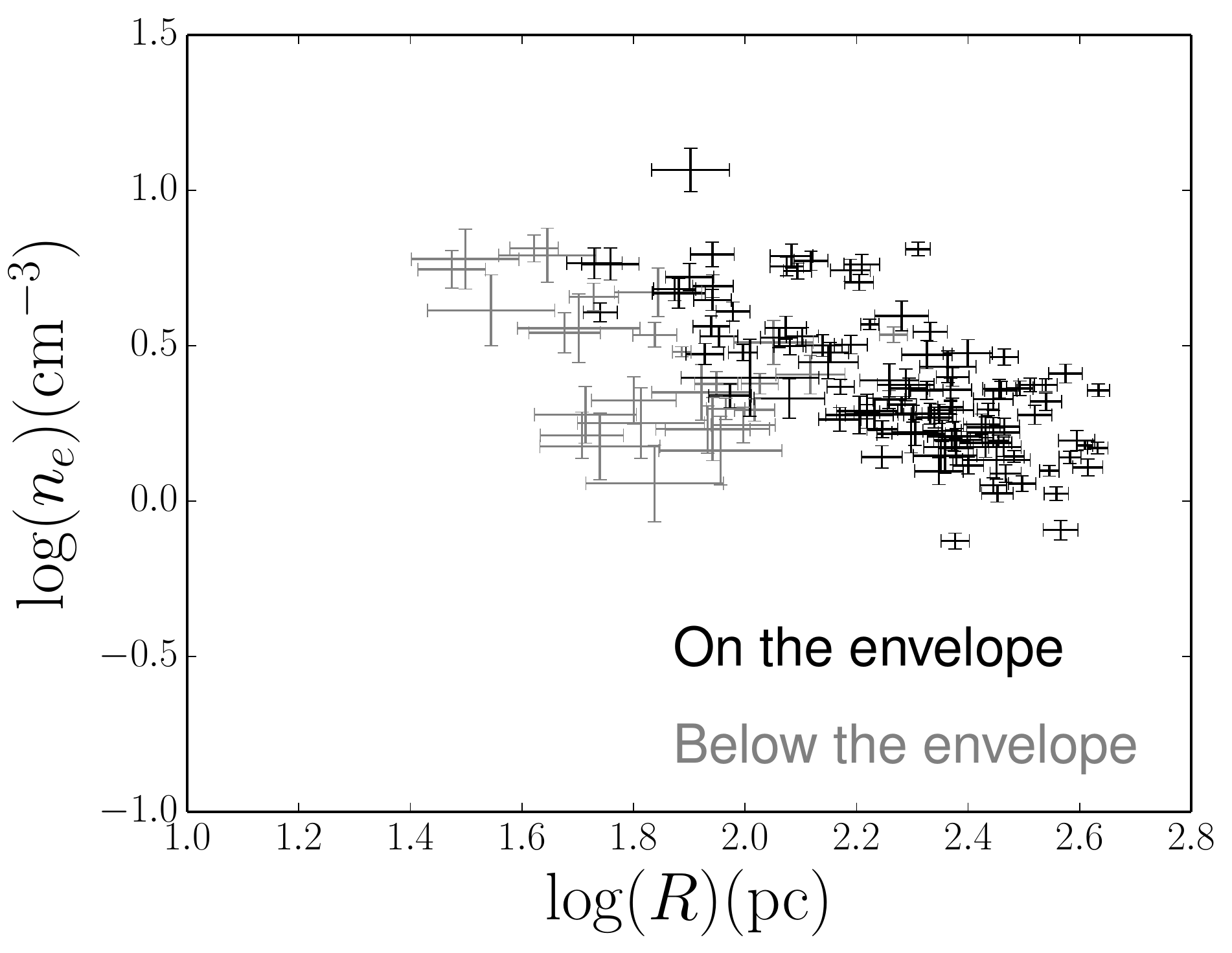,width=0.51\linewidth}\\
  \end{tabular}
\caption{Left: H$\alpha$ luminosity, $L_{\rm H\alpha}$, versus velocity dispersion, $\sigma_v$, for H{\sc ii} regions 
in isolated galaxies. We have plotted the fitted envelope as a solid line, and 
the displaced envelope condition explained in the text, as a dashed line. Right: $n_e$ versus 
$R$ for H{\sc ii} regions on the envelope (black) and those regions below the envelope (grey).}
\label{fig_env_iso}
\end{figure*}

Following the same procedure as in the previous section in interacting galaxies, we plot the 
H$\alpha$ luminosity versus velocity dispersion for H{\sc ii} regions in isolated galaxies 
in Fig. \ref{fig_env_iso} (left). Again, we plot the envelope to the data

\begin{equation}
\log(L_{\rm{H\alpha}\thinspace \mathrm{env}})=2.0\times \thinspace \log(\sigma_{v})+37.6
\label{eqenv_iso}
\end{equation}

implying that the SFR of the regions on the envelope depends super-linearly on the velocity dispersion. 

We plot the electron density versus the radius of the regions, separating those regions on 
the envelope (black) from those regions below the envelope (grey) in Fig.\ref{fig_env_iso} (right). 
For the set of isolated galaxies there is a deficiency of H{\sc ii} regions with a big enough velocity dispersion 
to be measured reliably by our observations, compared with H{\sc ii} regions in interacting galaxies. This is clear when we 
 compare Figs. \ref{fig_env_int} (left) and \ref{fig_env_iso} (left). We conclude that  
the velocity dispersions in the H{\sc ii} regions of interacting galaxies are larger presumably due to star formation feedback 
effects which enhance the dispersion. 

For the isolated galaxies we cannot detect the two regimes in the electron density-radius relation when 
we separate the regions on and below the envelope because we do not have enough 
regions below the envelope (Fig.\ref{fig_env_iso} right), implying that the effects of star-formation 
feedback are significantly less than for the interacting galaxies, as we would expect. 

\subsection{Virial parameter}

\begin{figure}
 \centering

\epsfig{file=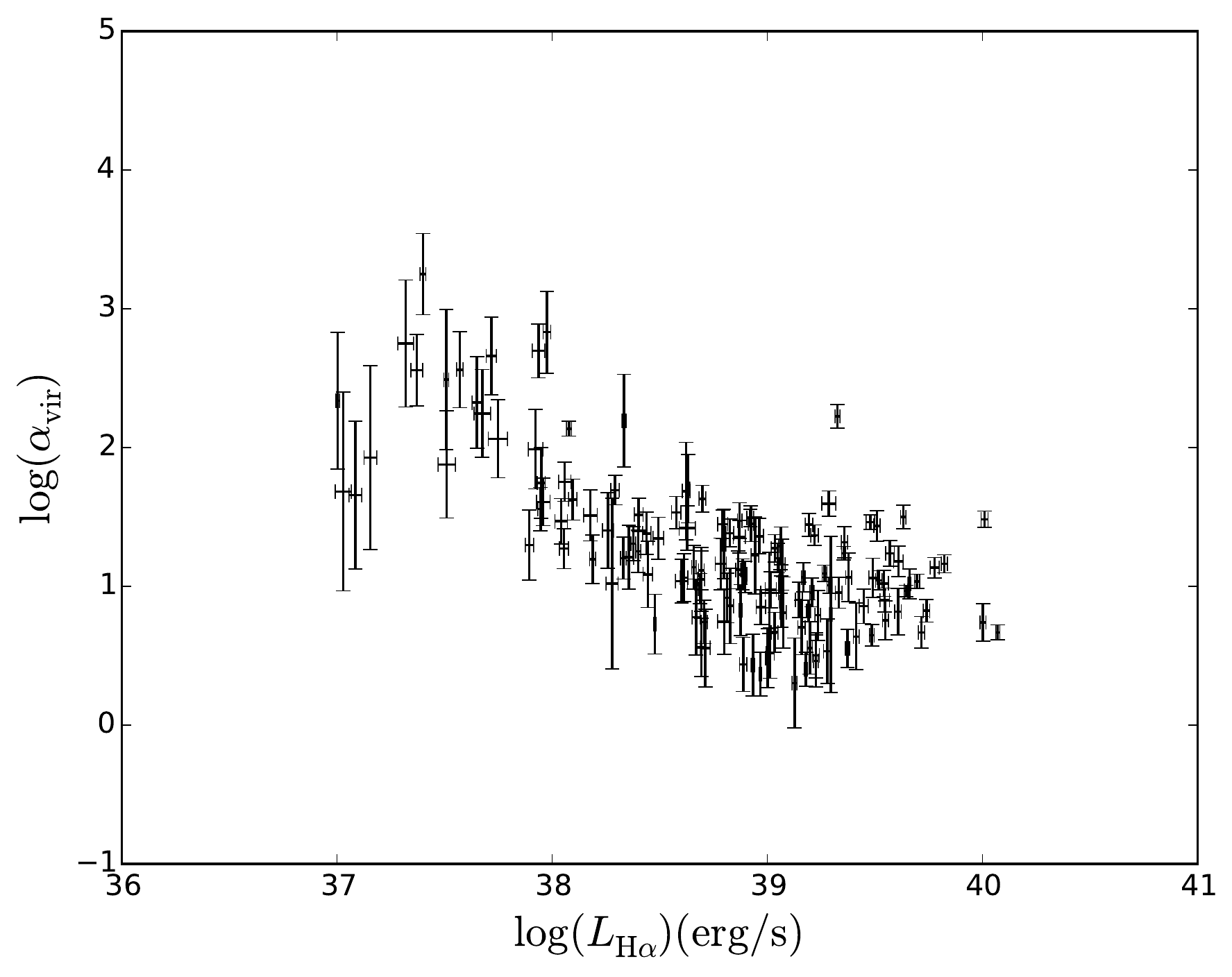,width=0.9\linewidth}
\caption{The virial parameter, $\alpha_{\rm{vir}}= 5 \frac{\sigma_{v}^2 R_{\rm{HII}}}{GM_{\rm{gas}}}$ 
versus H$\alpha$ luminosity, $L_{\rm H\alpha}$. We have used for the gas mass, the 
ionized gas, $M_{\rm{gas}}=M_{\rm{HII}}$.}
\label{fig_vir_iso}
\end{figure}

We plot the virial parameter versus H$\alpha$ luminosity in Fig. \ref{fig_vir_iso}. We see 
the same general behaviour as before, for the H{\sc ii} regions in isolated galaxies the virial parameter 
decreases with luminosity, implying that the brightest H{\sc ii} regions are 
dominated by self-gravity rather than being pressure confined. However, 
we can see that there is clearly a smaller proportion of H{\sc ii} regions in the brighter regime compared 
to the number of H{\sc ii} regions in the brighter regime in interacting galaxies.

\section{Comparison}

\begin{figure}
 \centering

\epsfig{file=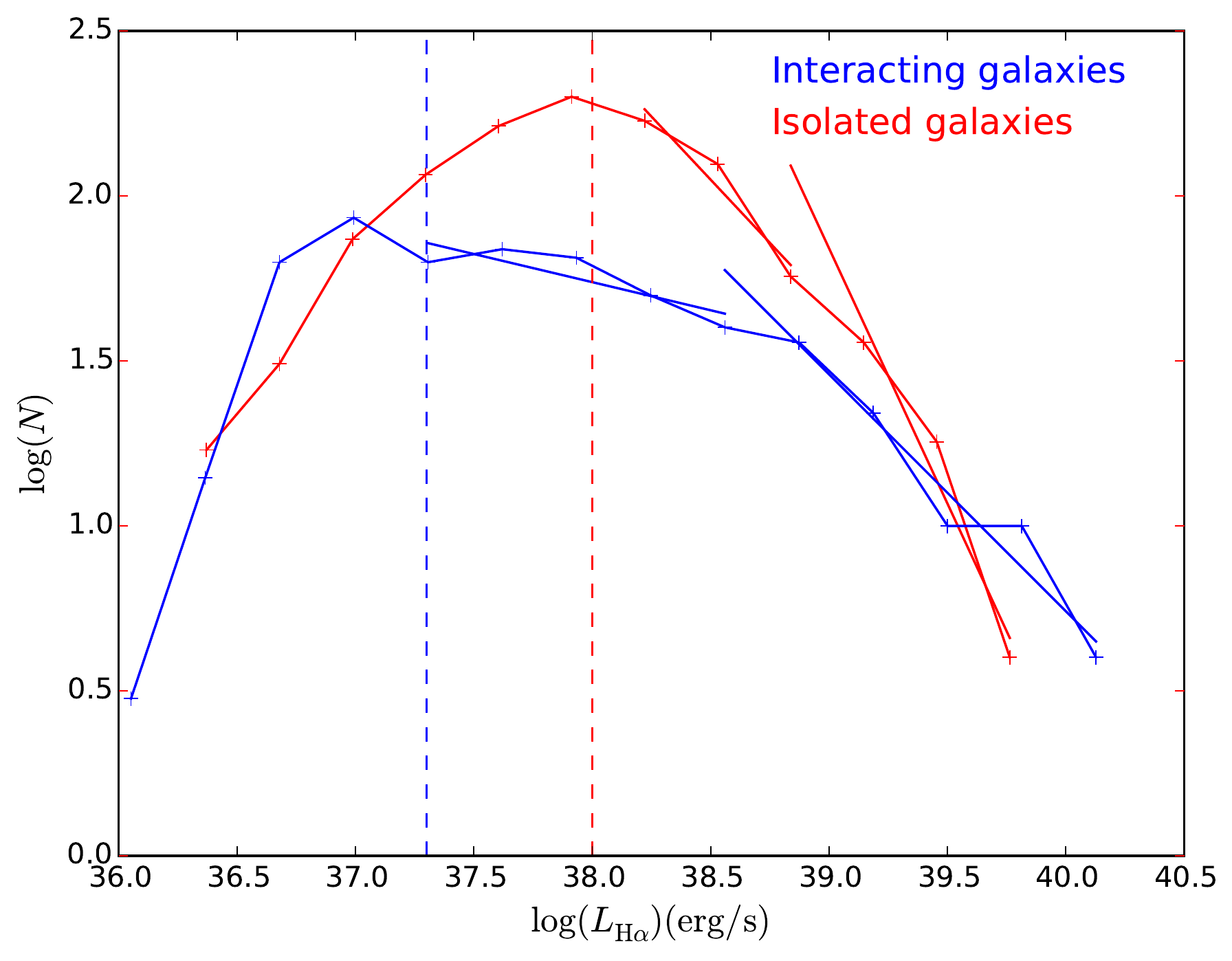,width=0.9\linewidth}
\caption{H{\sc ii} region luminosity function for interacting galaxies (blue) and 
isolated galaxies (red). The completeness limit is drawn as a 
dashed vertical line in blue for interacting galaxies, and in red for isolated galaxies. The 
double fit is plotted as a solid line in blue for interacting galaxies, and in red for isolated galaxies.}
\label{fig_lumfunc}
\end{figure}

We have studied the differences in the populations of H{\sc ii} regions in the two 
samples of galaxies. We have plotted the H{\sc ii} region luminosity 
functions for interacting galaxies and isolated galaxies, which are well represented by a power law 

\begin{equation}
 N(L)\mathrm{d}L=AL^{a}\mathrm{d}L
 \label{eq_lumfunc}
\end{equation}

where $N(L)\mathrm{d}L$ is the number of H{\sc ii} regions with luminosity in the range $[L,L+\mathrm{d}L]$. 
\cite{1989ApJ...337..761K,2000AJ....119.2728B,2006A&A...459L..13B} showed that   for a galaxy with a 
sufficiently large population of H{\sc ii} regions, 
the luminosity function is best fitted by a double power law.

We plot the H{\sc ii} region luminosity functions in Fig. \ref{fig_lumfunc}, showing 
in blue the H{\sc ii} regions of the interacting galaxies, and in red those of  the isolated galaxies, 
as well as the double power law fits.

We have estimated the completeness in both samples dividing 
each sample into two subsamples characterized by the distance to the galaxies. Thus, we have 
two subsamples for two ranges of smaller and larger distances respectively, and we estimated 
the completeness limit where the slope of the luminosity function 
converge to the same value for both subsamples. The completeness 
limit for interacting galaxies is $L_{\rm H\alpha}=37.3\thinspace\mathrm{dex}$ while for 
isolated galaxies it is $L_{\rm H\alpha}=38\thinspace\mathrm{dex}$.

For consistency with previous studies, we 
need to subtract 1 from the slopes of the fits in Fig.\ref{fig_lumfunc} to estimate 
the $a$ value appearing in Equation \ref{eq_lumfunc} since we have binned 
the data in logarithmic bins rather than linear bins. The results of the exponent 
in the luminosity function fit are 

\begin{equation}
\label{eq_lumfuncint}
 \begin{split}
  a  &= 
(-1.17\pm0.10) \thinspace \mathrm{;}%\\ 
 \thinspace\thinspace\thinspace\thinspace \mathrm{for}\thinspace \log(L_{\rm H\alpha})<38.6\thinspace\mathrm{dex}
\\
  a  &= 
(-1.71\pm0.17) \thinspace \mathrm{;}%\\ 
 \thinspace\thinspace\thinspace\thinspace \mathrm{for}\thinspace \log(L_{\rm H\alpha})>38.6\thinspace\mathrm{dex}
 \end{split}
\end{equation}

for interacting galaxies, while for isolated galaxies the result is

\begin{equation}
\label{eq_lumfunciso}
 \begin{split}
  a  &= 
(-1.8\pm0.2) \thinspace \mathrm{;}%\\ 
 \thinspace\thinspace\thinspace\thinspace \mathrm{for}\thinspace \log(L_{\rm H\alpha})<38.8\thinspace\mathrm{dex}
\\
a  &= 
(-2.5\pm0.3) \thinspace \mathrm{;}%\\
 \thinspace\thinspace\thinspace\thinspace \mathrm{for}\thinspace \log(L_{\rm H\alpha})>38.8\thinspace\mathrm{dex}
 \end{split}
\end{equation}

We draw the conclusion that the H{\sc ii} region luminosity functions 
for the two samples are different 
within the uncertainties. The H{\sc ii} region population in the interacting galaxies 
is generally brighter in H$\alpha$ 
(and therefore the SFR$(L_{\rm H\alpha})$ is higher) compared to isolated galaxies.  
 The break in the luminosity function 
is compatible with those found in previous studies 
\citep{1989ApJ...337..761K,2006A&A...459L..13B,zaragoza13,zaragoza14}.
\cite{zaragoza13,zaragoza14} claimed that this double population of H{\sc ii} regions is 
due to different scaling relations in the $L_{\rm H\alpha}$-$R$ relation.  In this study 
the break in the scaling relation for different galaxies 
is not constant due to the scatter in 
the data from which the change in scaling relations is derived.

We have applied the statistical Kolmogorov-Smirnov test to derive the 
probability that the H$\alpha$ luminosity distributions we have 
obtained for interacting and isolated galaxies are the same. The result is 
a probability of $10^{-4}\%$.  We can therefore conclude that 
the H$\alpha$ luminosity distribution in the sample of interacting galaxies 
is different from that of the isolated galaxies sample, and as Fig. \ref{fig_lumfunc} shows, 
that the H{\sc ii} regions interacting galaxies are brighter compared to those 
in isolated galaxies, because the luminosity function extends to higher values of the luminosity.

\begin{figure*}
 \centering
\begin{tabular}{cc}

\epsfig{file=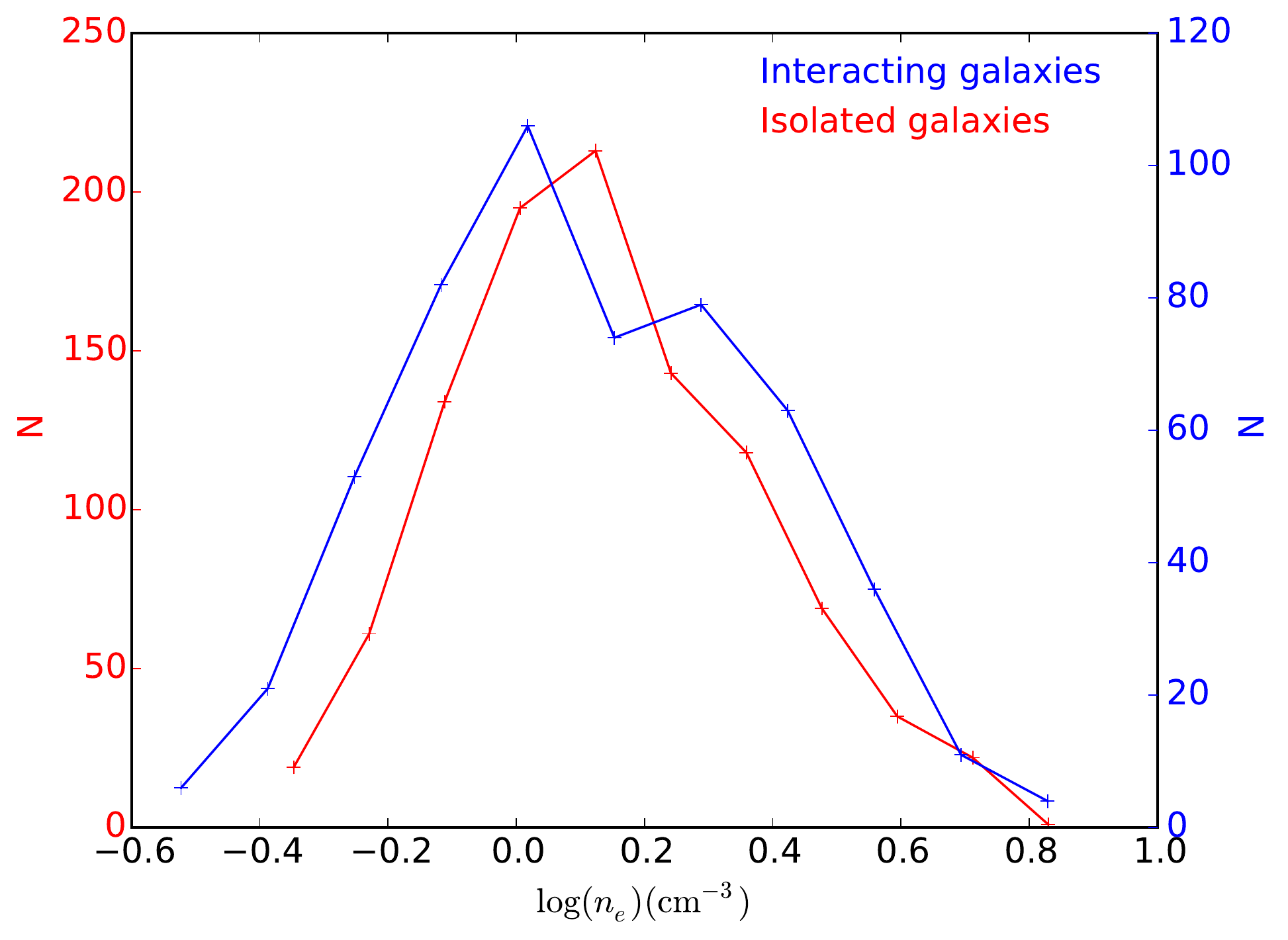,width=0.45\linewidth} &
\epsfig{file=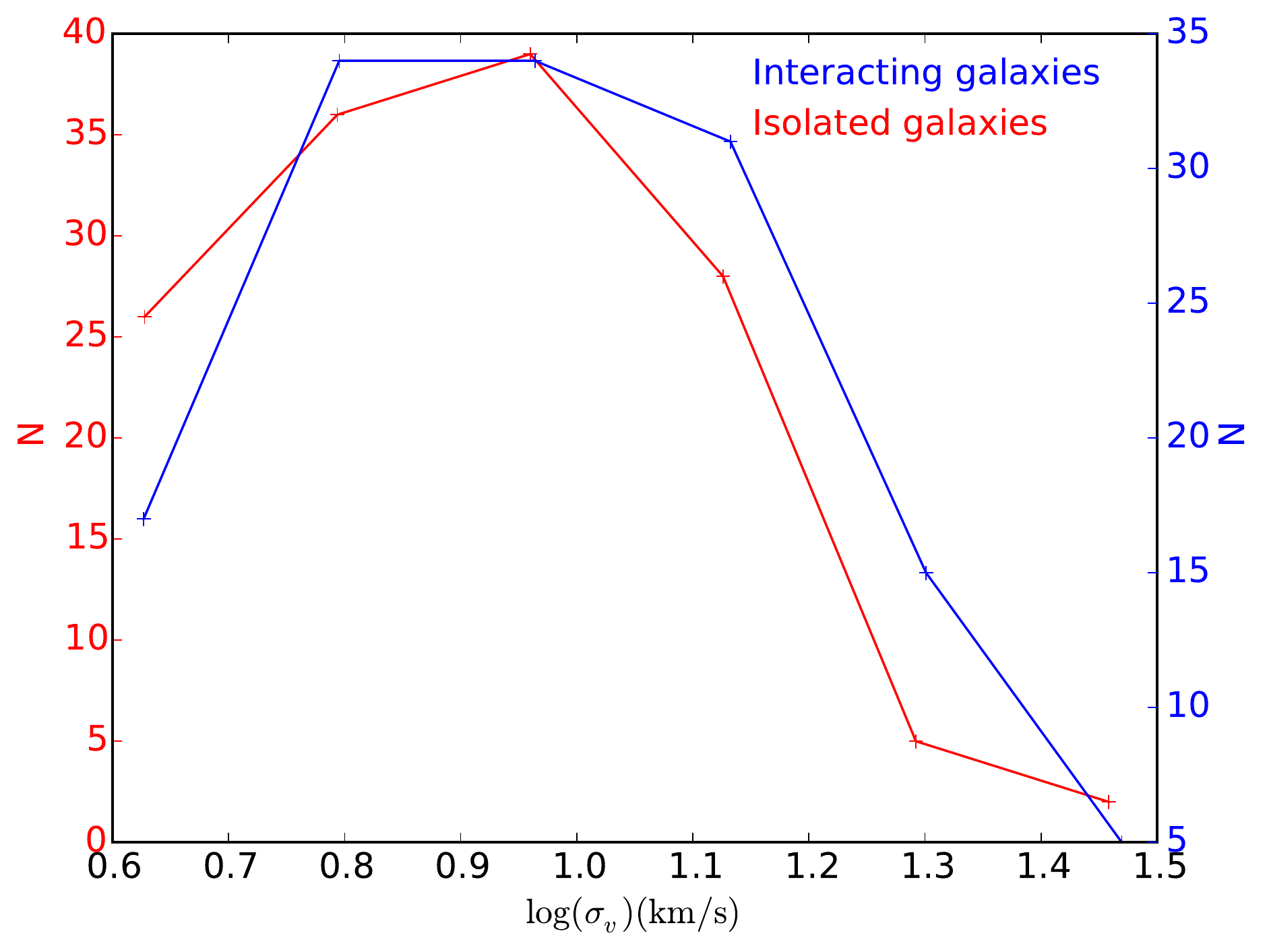,width=0.45\linewidth}\\
\end{tabular}

\caption{Left: Distribution of electron densities $n_e$ in H{\sc ii} regions of 
interacting galaxies (blue) and 
isolated galaxies (red). 
Right: Distribution of velocity dispersion $\sigma_v$ in H{\sc ii} regions of interacting 
galaxies (blue) and 
isolated galaxies (red). 
In both figures we plot two scales, one in blue at the right for interacting 
galaxies, and one at the left for isolated galaxies, 
in order to easily compare the different distributions. 
}
\label{dens_sv_func}
\end{figure*}

We plot the distribution of $n_e$ in Fig.\ref{dens_sv_func} (left). We can roughly 
compare the ionized gas density, $n_e$, with the molecular gas density \citep{zaragoza14}.
The gas density in the H{\sc ii} regions from the interacting galaxies sample is clearly enhanced compared 
to those in the isolated galaxies. Since the SFR 
depends super-linearly on the gas density, the enhancement in the gas density 
in the interacting galaxies sample implies an strong enhancement in the SFR, 
and indeed of the star formation efficiency.

We commented in the previous section that there is a deficiency of H{\sc ii} regions with enough 
velocity dispersion to be measured by our observations. 
We also plot the distribution of $\sigma_v$ in Fig.\ref{dens_sv_func} (right)
 where we can see that 
the interaction of galaxies increases the turbulence in H{\sc ii} regions. The two 
main mechanisms which can contribute to 
produce this effect are, firstly the continued accretion of gas onto the H{\sc ii} 
regions even after the first massive stars have been formed,  producing more 
massive regions, and therefore increasing the turbulence. The other mechanism 
is the star formation feedback by stellar winds and supernova explosions, 
implying that the feedback is stronger in galaxy interactions due to higher rates of 
star formation, and also the formation of more massive stars.

\subsection{Age distribution of H{\sc{ii}} regions}

\begin{figure*}
\centering
\begin{tabular}{cc}
\epsfig{file=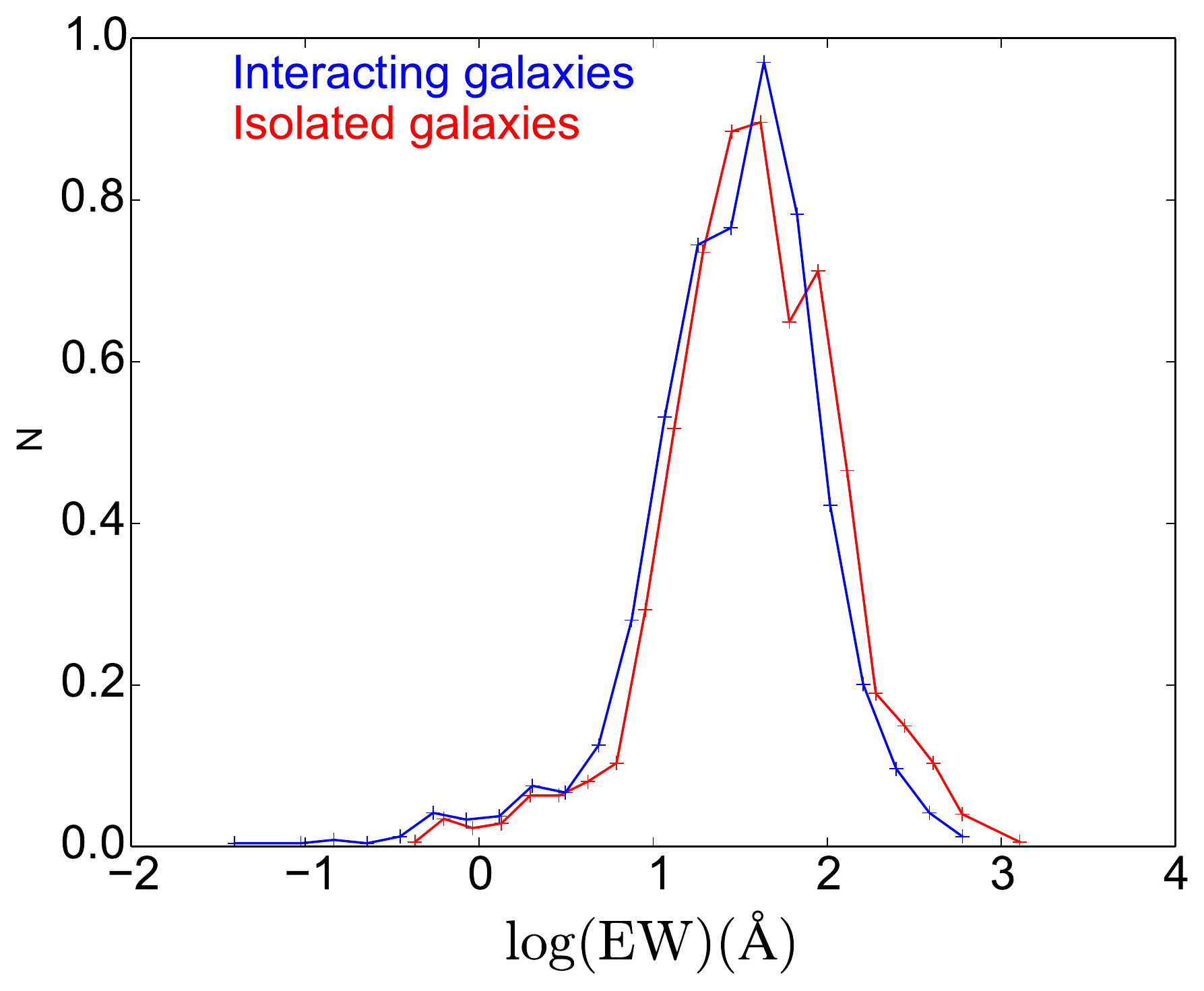,width=0.42\linewidth}&
\epsfig{file=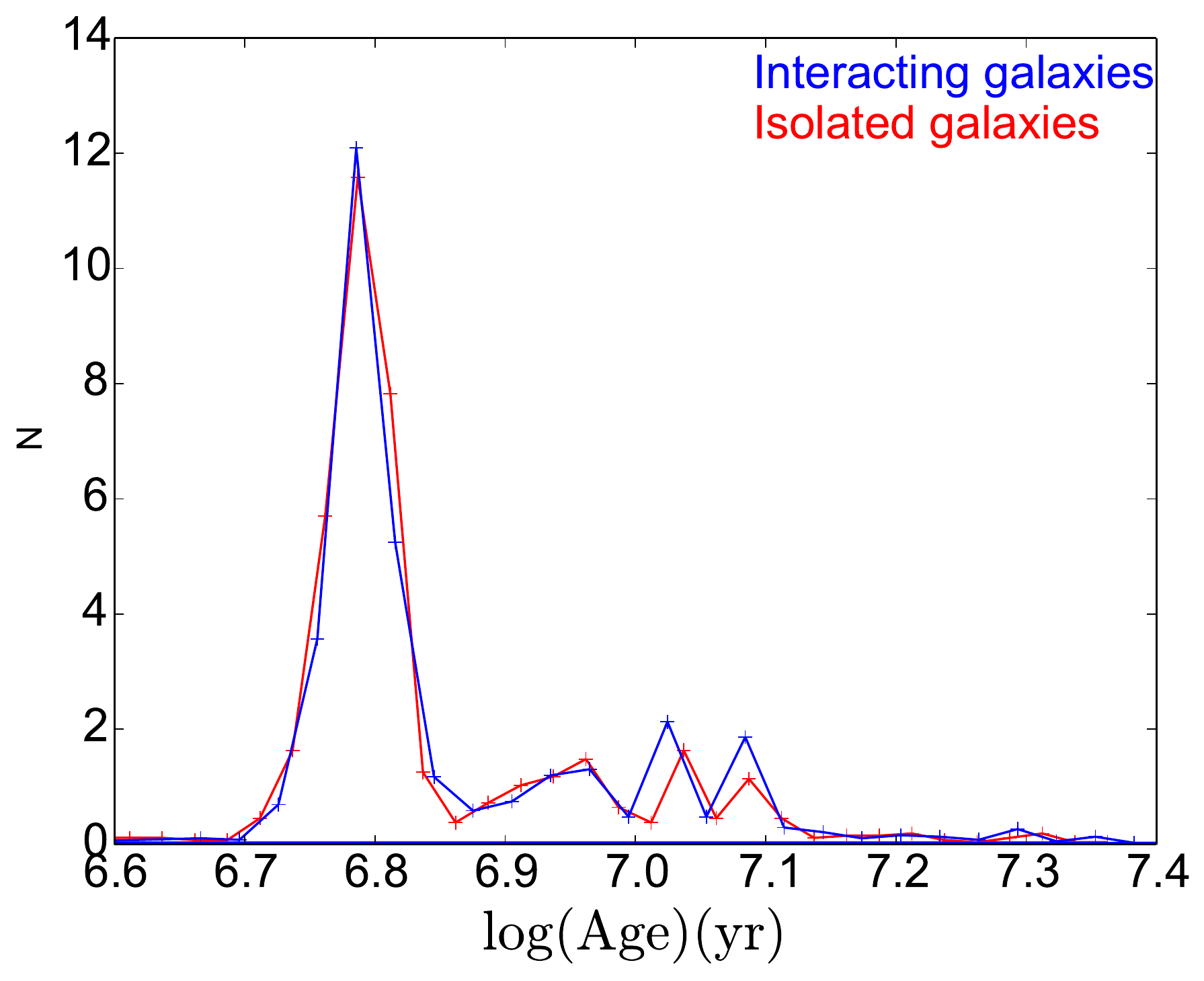,width=0.42\linewidth}\\
\end{tabular}
\caption{Left: Normalized fistribution of H{\sc{ii}} region equivalent widths, $\rm{EW}$. Right: Normalized distribution of ages 
{\bf (under the assumption of instantanous star formation)} derived using the equivalent 
	 width and Starburst99 model from \citet{1999ApJS..123....3L}.}
\label{fig_eqw}

\end{figure*}

Using the H$\alpha$ narrow band and continuum observations made with ACAM, we estimate the H$\alpha$ equivalent widths, $\rm{EW}$, of the 
H{\sc ii} regions extracted in this study, in the same way as in \cite{zaragoza13,zaragoza14}. 
Assuming a direct relation between $\rm{EW}(\mathrm H\alpha)$ and the age of the region 
\citep{1999ApJS..123....3L}, we derive also the age of the H{\sc ii} regions  in this study, assuming solar metallicity {\bf and instantanous star formation}. 
The results are included in Tables \ref{table_hii_int}1 and 
\ref{table_hii_int}2. We plot the normalized distribution of $\rm{EW}$s and ages in Fig. \ref{fig_eqw} (left and right respectively). 
The histograms for the  interacting and the isolated galaxies coincide, so there is no differenciation 
of the age distributions of H{\sc ii} regions in interacting and isolated galaxies. This could be taken as a point 
against our result of triggered star formation in interacting galaxies since the H{\sc ii} regions
are forming on the same timescales for both samples of galaxies. However, the age range in the distribution 
traced by H$\alpha$ $\rm{EW}$ is much smaller than the typical merger 
time scale, $\sim \rm{Gyr}$ \citep{2009MNRAS.399L..16C}.

\section{Off-nuclear peaks of star formation}

Galaxy interactions produce a reduction in axisymmetry, causing gas flows 
towards the central parts of galaxies 
\citep{1985AJ.....90..708K,1996ApJ...464..641M,2011EAS....51..107B}, giving a well 
understood explanation of the nuclear 
starbursts observed in interacting galaxies. However, 
there are many examples of merging galaxies where there are intense star forming regions at 
 sites away from the nuclei of the two galaxies \citep{2007AJ....133..791S,2011EAS....51..107B}, such as 
NGC 4038/9 (the Antennae galaxies), and NGC 4676 (the Mice). 

The two populations of H{\sc ii} regions seem to be related to 
two underlying populations of GMCs which are converted to ionized gas having similar 
mass distributions as the placental molecular gas \citep{zaragoza14}, at least 
in the more massive regions.

We can imagine an inflow in a galaxy induced by the interaction with 
a companion galaxy, with velocity, $v_{\rm{in}}$, towards 
the central parts, 
since galaxy interactions can produce gas inflows.
If a gas cloud is close to the inflow trajectory, 
the gas of the inflow will be accreted if the escape velocity, $v_{\rm{esc}}$, 
from the gas cloud is larger than the velocity of the inflow, $v_{\rm{in}}$.

In order to estimate the escape velocity of the gas inflow from the gas cloud, 
we use a very simplistic model where we neglect the influence 
of the galaxy which is producing the inflow towards the center. 
A more realistic model in which the competing gravitational pulls of the cloud towards its 
centre and of the galaxy towards its centre would give more accurate results, but is outside 
the scope of the present article. However, it is easy to show that for a major fraction of 
the original galactic disc the galactocentric force is significantly less than the 
perturbing force of individual clouds with radii greater than, say, 50 pc so the 
results presented here can be taken as qualitatively valid.
 The escape velocity from the surface of a gas cloud is

\begin{equation}
 \label{vesc}
 v_{\mathrm{esc}}=\sqrt{\frac{2\thinspace G\thinspace M}{b}}
\end{equation}

where $b$ is the impact parameter of the inflow with respect to the centre of the cloud. Let us assume 
that the mass of a region is related to the radius by $ M(R)=M_0\thinspace R^N$ 

\begin{equation}
 \label{vesc_exp}
 v_{\mathrm{esc}}=\sqrt{\frac{2\thinspace G\thinspace M_0 \thinspace R^{N}}{b}}=v_0\thinspace R^{\frac{N}{2}}.
\end{equation}

We have seen that for the H{\sc ii} regions we can divide the scaling 
relations between luminosity and radius into two. The regions with 
high luminosity, significantly more abundant in the interacting galaxies, 
show exponents of 3 or greater, while the regions with lower luminosity 
show exponents closer to 2. Previous work on molecular clouds has shown 
\citep{2010ApJ...723..492R,2010A&A...519L...7L,2010ApJ...716..433K,2014ApJ...784....3C}  
that molecular clouds in general have a relation between gas mass and cloud radius 
with an exponent close to 2, which is consistent with the value found by \cite{1981MNRAS.194..809L}, 
frequently quoted as one of the Larson laws. For the interacting galaxy pair the Antennae, 
\cite{zaragoza14} found an exponent of 2.6 for the mass-radius relation of 
its most massive molecular clouds. It is straightforward to show that if the 
Larson law relating the velocity dispersion $\sigma_v$ to the cloud mass $M_{\odot}$, 
($\sigma_v\propto M^{0.2}$) 
holds good, for a sequence of  molecular clouds of increasing mass where the exponent 
in the mass-radius relation is 2 or greater, there will be a mass above which the 
clouds become gravitationally bound. In \cite{zaragoza14} we showed 
that this would account for the tendency of the virial parameter to approach 
unity for molecular clouds of high enough mass (for the Antennae this implies 
cloud masses greater than some $10^6.5M_{\odot}$). We also argued that any clouds with 
masses of this order in the galaxies prior to interaction would tend to accrete 
mass from the gas flows induced by the interaction. To pursue this argument 
further we can estimate, using somewhat simplified assumptions, the escape 
velocity as a function of cloud size, taking two different scaling relations, 
one for the lower mass clouds, and the other for the high mass clouds. The 
corresponding relations can be expressed as:

\begin{figure}
 \centering

\epsfig{file=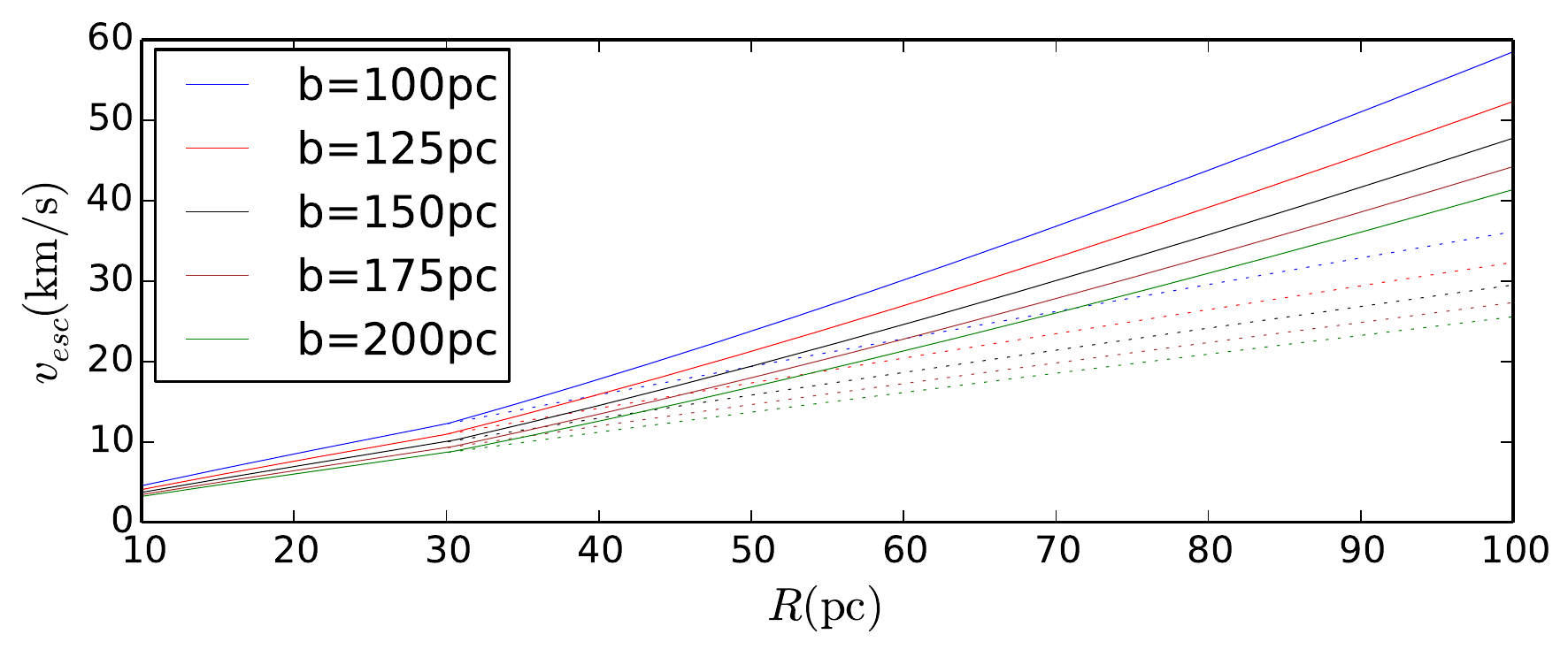,width=0.9\linewidth}
\caption{Escape velocity, $v_{\rm{esc}}$ ,of the clouds as a function of their radius, $R$, for 
different values of the impact parameter, $b$. The solid lines represent 
the case of a double regime in the mass-radius relation (Equation \ref{eq_escape}) while 
the dotted lines represent the case of the single mass-radius relation with 
a canonical exponent of $N=2$.}
\label{fig_vesc}
\end{figure}

\begin{equation}
\begin{tabular}{c}
 $M=3817R^{2};\thinspace \mathrm{for}\thinspace R<30\mathrm{pc}$ \\
$M=251R^{2.61};\thinspace \mathrm{for}\thinspace R>30\mathrm{pc}
$
\end{tabular}
\label{eq_escape}
\end{equation}

We introduce Equation \ref{eq_escape} into Equation \ref{vesc_exp} for 
different impact parameters obtaining the plots shown in Fig.\ref{fig_vesc}, where 
we can clearly see the escape velocity enhancement in the case 
of the double mass-radius regime (solid lines) compared to the 
case of single mass-radius regime (dotted lines).  

An inflow of gas of $v_{\rm{in}}=27\thinspace\mathrm{km/s}$ was observed for the 
interacting galaxy NGC 3396 \citep{zaragoza13}. As an example, with this gas inflow velocity value 
we can conclude from figure \ref{fig_vesc} that  regions 
with radius larger than $\sim60\mathrm{pc}$ would accrete gas from the inflow in the double 
mass-radius regime, while only regions with radius larger than $\sim80\mathrm{pc}$ would 
accrete gas in the single mass-radius regime. 
The latter may not be gravitationally bound prior to the merging process, 
but as they acquire more gas they become so.

We suggest that our study, showing that the largest clouds are bound by their own gravitation, 
strongly supports the scenario where the gas accreted from the flowing gas passing near the  these clouds, (which 
are already present in the isolated galaxies prior to interaction), 
 is the cause of the star formation enhancement 
in interacting galaxies and can give rise to the off-nuclear peaks of star formation.
Thus, although we still do not, at this stage, have a model for how the 
gravitationally dominated gas clouds form initially within the pre-merging galaxies, the inflow 
of gas in the interacting galaxies from the outer parts should increment the number 
of gravitationally dominated clouds.

\section{Conclusions}

We have presented new Fabry-Perot observations for 10 interacting galaxies: the 
H$\alpha$ surface brightness, velocity dispersion, and velocity dispersion maps. 
All of these values are available through CDS. 
We have amalgamated our tables of the properties of H{\sc ii} regions for the sample of interacting galaxies 
presented here together with two interacting systems, Arp 270, and the Antennae galaxies 
from our previous publications \citep{zaragoza13,zaragoza14}. We give here 
the catalogue of 1259 H{\sc ii} regions, including the 
radius, $H\alpha$ luminosity, velocity dispersion, electron density, 
ionized gas mass, the virial parameter, and the corresponding 
errors for each parameter. 

In order to compare the properties of H{\sc ii} regions in interacting galaxies with 
those in isolated galaxies, we have included the parameters 
of H{\sc ii} regions in a sample of 28 isolated galaxies observed with the same 
instrument \citep{2015arXiv150406282E}, obtaining these parameters (the same as those for the interacting galaxies) 
for 1054 H{\sc ii} regions. We present a subsample of the 
brightest H{\sc ii} regions of the catalogue 
in Table\ref{table_hii_int}1 in for the interacting galaxies, and 
in Table\ref{table_hii_iso}2 for the isolated galaxies. The full 
catalogues of the H{\sc ii} region samples are available as 
a machine readable Tablein the electronic version of the article, as 
well as through CDS.   

The scaling relations, $L_{\rm H\alpha}-R$, obtained here for interacting galaxies 
are different from previous studies for isolated galaxies \citep{1981MNRAS.195..839T,gutierrez10} in most 
of the cases, 
and in agreement with recent results for the two interacting systems whose H{\sc ii} regions we have included here, 
\cite{zaragoza13,zaragoza14}. They differ in the exponent of 
the relation $L_{\rm H\alpha}\propto R^N$. 
 We have found that for almost all the interacting galaxies, the $L_{\rm H\alpha}-R$ is best represented 
by a double line, one for the smallest H{\sc ii} regions ($\lesssim100\mathrm{pc}$) similar to 
previous studies where the exponent $N$ is smaller 
than three, and other for the largest H{\sc ii} regions ($\gtrsim100\mathrm{pc}$) where the 
exponent $N$ is larger than three,  except for two galaxies where we do not have 
enough fainter H{\sc ii} regions to fit a double linear fit and we find that the 
exponent $N$ is larger than three, independently of the radius.
If $N>3$, then the electron (or the ionized gas mass) density derived from equations 
\ref{eq:density} and \ref{eq2} increases 
with size, in contrast with previous results in isolated galaxies 
where the electron density in H{\sc ii} regions decreases with size \citep{1981MNRAS.195..839T,gutierrez10}.

In the sample of isolated galaxies, we have found that the scaling relation $L_{\rm H\alpha}-R$ is 
similar to that found in previous studies in 22 of our 28 isolated galaxy sample, with slopes smaller than three.  

The differences in the exponents of the $L_{\rm H\alpha}-R$ relation are independent of the 
absolute magnitude of the galaxies.

The $L_{\rm H\alpha}-\sigma_{v}$ plot shows an envelope which has been detected in previous published work, whose authors suggested that 
the regions on this envelope are virialized \citep{2005A&A...431..235R}. In the case 
of interacting galaxies, the regions on the envelope are the ones where 
the gas density increases with size. In the case of isolated galaxies we do 
not have sufficient regions with large enough velocity dispersion to reach a clear conclusion. 
The H$\alpha$ luminosity, and the derived star formation rate 
of the H{\sc ii} regions on the envelope depend 
super-linearly on the velocity dispersion.   
We have obtained the virial parameter $\alpha_{\rm{vir}}$ which is an estimate 
of the ratio between the kinetic and the gravitational energy \citep{1992ApJ...395..140B}, and have used this to show  
that the brighter the region, the more clearly it is held together by self gravity rather than 
external pressure. 

We have quantified the differences between the two populations of H{\sc ii} regions, 
those from interacting galaxies, and those from the isolated galaxies. 
Comparing the luminosity functions we find that 
the H{\sc ii} regions in the 
 interacting galaxies are on average brighter than the H{\sc ii} regions 
from the isolated galaxies. The K-S test performed 
on the two distributions of H$\alpha$ luminosities 
gives a probability of $10^{-4}\%$ for the hypothesis 
that the two populations of H{\sc ii} regions are really the same.

The histograms of electron density, and 
velocity dispersion 
show that on average the H{\sc ii} regions from interacting galaxies are denser and 
more turbulent than those from isolated galaxies. Since 
the star formation rate depends super-linearly 
on gas density, the enhancement in the gas density 
in interacting galaxies implies a significant enhancement in the 
star formation efficiency.  

 However the distributions of equivalent widths and ages of H{\sc ii} region coincide for the two 
samples of galaxies, showing that the formation of the young massive stellar component traced by H$\alpha$ extends over a similar range in time for 
interacting and isolated galaxies. 

However the sample of interacting galaxies is, on average, 
brighter than that of the isolated galaxies.  We could conclude 
that the presence of the population of brighter H{\sc ii} regions in the 
interacting galaxies might be simply due to their greater masses. 
Nevertheless, the number of H{\sc ii} regions in an interacting galaxy 
is bigger than the number in an isolated galaxy with the same absolute magnitude,  
suggesting that interactions do in fact increase the star formation rate.

Based on the results from \cite{2012ApJ...750..136W} and \cite{zaragoza14}, we claim that 
the two populations of H{\sc ii} regions are related to two populations 
of GMC's.  \cite{2014MNRAS.442.1230R} claim that Toomre stability criteria depends on the values 
of exponents in the Larson laws. In the case of the exponent in the mass-size 
relation, largest values of this exponent implies that larger scales are unstable, in 
agreement with the presence of larger star forming regions in the high mass regime. 
The presence of massive clumps of star formation in nearby galaxies as those studied here 
can improve our understanding about the clumpy star forming discs observed 
at higher redshifts \citep{2007ApJ...658..763E}.

We claim that the brightest and most massive star forming regions can 
accrete gas that is inflowing towards the central parts of the galaxies, or has suffered a perturbation in its original orbit, 
induced by interactions.  \cite{javiblasco13} found in their sample of 11 starburst galaxies that 8 of them show 
evidence of a recent merger while they are still dominated by rotation, in agreement with a picture 
where the interactions triggers flows and star formation without messing up the host galaxy's kinematics.
This could be a plausible scenario for the 
off-nuclear peaks of star formation produced in galaxy collisions, 
a scenario which requires further observational and theoretical exploration.

\section*{Acknowledgments}
Based on observations made with the William Herschel Telescope operated on the island of La Palma
by the Isaac Newton Group of Telescopes in the Spanish Observatorio del Roque de los
Muchachos of the
Instituto de Astrofísica de Canarias. This research has been supported by the Spanish
Ministry of Economy and Competitiveness (MINECO) under the grants
AYA2007-67625-CO2-01, AYA2009-12903 and  AYA2012-39408-C02-02.
JEB acknowledge financial support to the DAGAL network from the People Programme
(Marie Curie Actions) of the European Union's Seventh Framework Programme FP7/2007-2013/
under REA grant agreement number PITN-GA-2011-289313.

This research made use of \textsc{Astropy}, a community-developed core Python package for Astronomy \citep{2013A&A...558A..33A},
\textsc{APLpy}, an open-source plotting package for Python hosted at http://aplpy.github.com, and \textsc{astrodendro},
a Python package to compute dendrograms of Astronomical data (http://www.dendrograms.org/). 
We acknowledge the usage of the HyperLeda database (http://leda.univ-lyon1.fr).

We thank Roberto Terlevich, the referee of the paper, whose comments have led to important improvements on the original version of the paper.

\bibliographystyle{mn2e}

\begin{thebibliography}{}

\bibitem[Arp(1966)]{1966ApJS...14....1A} Arp, H.\ 1966, \apjs, 14, 1 

\bibitem[Astropy Collaboration et
al.(2013)]{2013A&A...558A..33A} Astropy Collaboration, Robitaille, T.~P., Tollerud, E.~J., et al.\ 2013, \aap, 558, A33



\bibitem[Beckman et al.(2000)]{2000AJ....119.2728B} Beckman, J.~E., Rozas, 
M., Zurita, A., Watson, R.~A., \& Knapen, J.~H.\ 2000, \aj, 119, 2728 

\bibitem[Benn et al.(2008)]{2008SPIE.7014E..6XB} Benn, C., Dee, K., 
\& Ag{\'o}cs, T.\ 2008, \procspie, 7014, 70146X 



\bibitem[Bennert et al.(2008)]{Bennert08} Bennert, N., Canalizo, G., Jungwiert, B., Stockton, A.,
Schweizer, F., Peng, C. Y.,  Lacy, M. 2008, ApJ, 677, 846

\bibitem[Bessiere et al.(2012)]{Bessiere12} Bessiere, P. S., Tadhunter, C. N., Ramos Almeida, C., Villar-Mart\' in, M. 2012,
MNRAS, 426, 276


\bibitem[Bertoldi
\& McKee(1992)]{1992ApJ...395..140B} Bertoldi, F., \& McKee, C.~F.\ 1992, \apj, 395, 140



\bibitem[Blasco-Herrera et al.(2010)]{2010MNRAS.407.2519B} Blasco-Herrera, 
J., Fathi, K., Beckman, J., et al.\ 2010, \mnras, 407, 2519 


\bibitem[Blasco-Herrera et al.(2013)]{javiblasco13} Blasco-Herrera, 
J., Fathi, K., {\"O}stlin, G., Font, J., 
\& Beckman, J.~E.\ 2013, \mnras, 435, 1958 


\bibitem[Bournaud et al.(2010)]{2010MNRAS.409.1088B} Bournaud, F., 
Elmegreen, B.~G., Teyssier, R., Block, D.~L., 
\& Puerari, I.\ 2010, \mnras, 409, 1088 


\bibitem[\protect\citeauthoryear{Bournaud}{2011}]{2011EAS....51..107B}
Bournaud F., 2011, EAS, 51, 107


\bibitem[\protect\citeauthoryear{Bradley et
al.}{2006}]{2006A&A...459L..13B} Bradley T.~R., Knapen J.~H., Beckman J.~E., Folkes S.~L., 2006, A\&A, 459, L13


\bibitem[Brandl et al.(2005)]{2005ApJ...635..280B} Brandl, B.~R., Clark,
D.~M., Eikenberry, S.~S., et al.\ 2005, \apj, 635, 280


\bibitem[Calzetti et al.(2004)]{2004AJ....127.1405C} Calzetti, D., Harris, 
J., Gallagher, J.~S., III, et al.\ 2004, \aj, 127, 1405 

\bibitem[Camps-Fari{\~n}a et al.(2015)]{2015MNRAS.447.3840C} 
Camps-Fari{\~n}a, A., Zaragoza-Cardiel, J., Beckman, J.~E., et al.\ 2015, 
\mnras, 447, 3840 

\bibitem[Canalizo et al.(2007)]{Canalizo07} Canalizo, G., Bennert, N., Jungwiert, B., Stockton, A., Schweizer, F., Lacy, M.,
Peng, C. 2007, ApJ, 669, 801

\bibitem[Cedr{\'e}s et al.(2013)]{2013ApJ...765L..24C} Cedr{\'e}s, B., 
Beckman, J.~E., Bongiovanni, {\'A}., et al.\ 2013, \apjl, 765, LL24 




\bibitem[Ch{\'a}vez et al.(2014)]{chavez} Ch{\'a}vez, R., 
Terlevich, R., Terlevich, E., et al.\ 2014, \mnras, 442, 3565 

\bibitem[Chu 
\& Kennicutt(1994)]{1994ApJ...425..720C} Chu, Y.-H., \& Kennicutt, R.~C., Jr.\ 1994, \apj, 425, 720 



\bibitem[\protect\citeauthoryear{Cisternas et 
al.}{2011}]{2011ApJ...726...57C} Cisternas M., et al., 2011, ApJ, 726, 57 


\bibitem[Colombo et al.(2014)]{2014ApJ...784....3C} Colombo, D., Hughes, 
A., Schinnerer, E., et al.\ 2014, \apj, 784, 3 


\bibitem[Conselice(2009)]{2009MNRAS.399L..16C} Conselice, C.~J.\ 2009, 
\mnras, 399, L16 

\bibitem[Daigle et al.(2006)]{2006MNRAS.368.1016D} Daigle, O., Carignan, 
C., Hernandez, O., Chemin, L., \& Amram, P.\ 2006, \mnras, 368, 1016 

\bibitem[Dyson(1979)]{1979A&A....73..132D} Dyson, J.~E.\ 1979, \aap, 73, 132 


\bibitem[Elmegreen et al.(2007)]{2007ApJ...658..763E} Elmegreen, D.~M., 
Elmegreen, B.~G., Ravindranath, S., \& Coe, D.~A.\ 2007, \apj, 658, 763 



\bibitem[Erroz-Ferrer et al.(2012)]{2012MNRAS.427.2938E} Erroz-Ferrer, S.,
Knapen, J.~H., Font, J., et al.\ 2012, \mnras, 427, 2938


\bibitem[Erroz-Ferrer et al.(2015)]{2015arXiv150406282E} Erroz-Ferrer, S., 
Knapen, J.~H., Leaman, R., et al.\ 2015, arXiv:1504.06282 


 \bibitem[Georgakakis et al.(2009)]{Georgakakis09} Georgakakis, A., et al. 2009, MNRAS, 397, 623

 
\bibitem[Giammanco et 
al.(2004)]{2004A&A...424..877G} Giammanco, C., Beckman, J.~E., Zurita, A., \& Rela{\~n}o, M.\ 2004, \aap, 424, 877 


 
\bibitem[Guti{\'e}rrez 
\& Beckman(2010)]{gutierrez10} Guti{\'e}rrez, L., \& Beckman, J.~E.\ 2010, \apjl, 710, L44 

\bibitem[Guti{\'e}rrez 
\& Beckman(2010)]{2010ihea.book...77G} Guti{\'e}rrez, L., \& Beckman, J.\ 2010, The Impact of HST on European Astronomy, 77 



\bibitem[Guti{\'e}rrez et al.(2011)]{gutierrez11} Guti{\'e}rrez,
L., Beckman, J.~E., \& Buenrostro, V.\ 2011, \aj, 141, 113

\bibitem[Hernandez et al.(2008)]{2008PASP..120..665H} Hernandez, O., Fathi, 
K., Carignan, C., et al.\ 2008, \pasp, 120, 665 

\bibitem[Hippelein 
\& Fried(1984)]{1984A&A...141...49H} Hippelein, H., \& Fried, J.~W.\ 1984, \aap, 141, 49 


\bibitem[Hong et al.(2011)]{2011ApJ...731...45H} Hong, S., Calzetti, D., 
Dopita, M.~A., et al.\ 2011, \apj, 731, 45 

\bibitem[Hopkins et al.(2013)]{2013arXiv1311.2073H} Hopkins, P.~F., Keres, 
D., Onorbe, J., et al.\ 2013, arXiv:1311.2073 


\bibitem[Karachentsev et al.(1985)]{1985BICDS..29...87K} Karachentsev, I., 
Lebedev, V., 
\& Shcherbanovski, A.\ 1985, Bulletin d'Information du Centre de Donnees Stellaires, 29, 87 

\bibitem[Kauffmann et al.(2010)]{2010ApJ...716..433K} Kauffmann, J.,
Pillai, T., Shetty, R., Myers, P.~C.,
\& Goodman, A.~A.\ 2010, \apj, 716, 433


\bibitem[Keel et al.(1985)]{1985AJ.....90..708K} Keel, W.~C., Kennicutt, 
R.~C., Jr., Hummel, E., \& van der Hulst, J.~M.\ 1985, \aj, 90, 708 



\bibitem[\protect\citeauthoryear{Kennicutt, Edgar,
\& Hodge}{1989}]{1989ApJ...337..761K} Kennicutt R.~C., Jr., Edgar B.~K., Hodge P.~W., 1989, ApJ, 337, 761

\bibitem[\protect\citeauthoryear{Larson}{1981}]{1981MNRAS.194..809L} Larson
R.~B., 1981, MNRAS, 194, 809

\bibitem[Leitherer et al.(1999)]{1999ApJS..123....3L} Leitherer, C., 
Schaerer, D., Goldader, J.~D., et al.\ 1999, \apjs, 123, 3 


\bibitem[Lombardi et 
al.(2010)]{2010A&A...519L...7L} Lombardi, M., Alves, J., \& Lada, C.~J.\ 2010, \aap, 519, L7 


\bibitem[Mihos 
\& Hernquist(1996)]{1996ApJ...464..641M} Mihos, J.~C., \& Hernquist, L.\ 1996, \apj, 464, 641 


\bibitem[Moiseev et al.(2014)]{2014arXiv1405.5731M} Moiseev, A.~V., 
Tikhonov, A.~V., \& Klypin, A.\ 2014, arXiv:1405.5731 

\bibitem[\protect\citeauthoryear{Paturel et al.}{2003}]{2003A&A...412...45P} Paturel G., Petit C., Prugniel P., Theureau G., Rousseau J., Brouty M., Dubois P., Cambr{\'e}sy L., 2003, A\&A, 412, 45



\bibitem[Ramos Almeida et al.(2011)]{Ramos11} Ramos Almeida, C., Tadhunter, C. N., Inskip, K. J., Morganti, R., Holt, J., Dicken, D. 2011a, MNRAS, 410, 1550


\bibitem[Ramos Almeida et al.(2012)]{Ramos12} Ramos Almeida, C., et al. 2012, MNRAS, 419, 687


\bibitem[Rela{\~n}o et
al.(2005)]{2005A&A...431..235R} Rela{\~n}o, M., Beckman, J.~E., Zurita, A., Rozas, M., \& Giammanco, C.\ 2005, \aap, 431, 235


\bibitem[Roman-Duval et al.(2010)]{2010ApJ...723..492R} Roman-Duval, J.,
Jackson, J.~M., Heyer, M., Rathborne, J.,
\& Simon, R.\ 2010, \apj, 723, 492

\bibitem[Romeo 
\& Agertz(2014)]{2014MNRAS.442.1230R} Romeo, A.~B., \& Agertz, O.\ 2014, \mnras, 442, 1230 

\bibitem[Rosa 
\& Solf(1984)]{1984A&A...130...29R} Rosa, M., \& Solf, J.\ 1984, \aap, 130, 29 


\bibitem[Rosolowsky
\& Leroy(2006)]{2006PASP..118..590R} Rosolowsky, E., \& Leroy, A.\ 2006, \pasp, 118, 590

\bibitem[Rosolowsky et al.(2008)]{2008ApJ...679.1338R} Rosolowsky, E.~W.,
Pineda, J.~E., Kauffmann, J., \& Goodman, A.~A.\ 2008, \apj, 679, 1338

\bibitem[Smith et al.(2007)]{2007AJ....133..791S} Smith, B.~J., Struck, C., 
Hancock, M., et al.\ 2007, \aj, 133, 791 

\bibitem[Smith et al.(2014)]{2014AJ....147...60S} Smith, B.~J., Soria, R., 
Struck, C., et al.\ 2014, \aj, 147, 60 



\bibitem[Somerville et al.(2008)]{Somerville08} Somerville, R. S., Hopkins, P. F., Cox, T. J., Robertson, B. E.,  Hernquist, L.
2008, MNRAS, 391, 481


\bibitem[\protect\citeauthoryear{Spitzer}{1978}]{1978ppim.book.....S}
Spitzer L., 1978, ppim.book,

\bibitem[Tadhunter et al.(2011)]{Tadhunter11} Tadhunter, C., et al. MNRAS, 412, 960

\bibitem[Terlevich
\& Melnick(1981)]{1981MNRAS.195..839T} Terlevich, R., \& Melnick, J.\ 1981, \mnras, 195, 839

\bibitem[Toomre(1977)]{1977egsp.conf..401T} Toomre, A.\ 1977, Evolution of 
Galaxies and Stellar Populations, 401 

% \bibitem[Torres-Flores et al.(2014)]{2014MNRAS.442.2188T} Torres-Flores, 
% S., Amram, P., Mendes de Oliveira, C., et al.\ 2014, \mnras, 442, 2188 


\bibitem[de Vaucouleurs et al.(1976)]{1976RC2...C......0D} de Vaucouleurs, 
G., de Vaucouleurs, A., 
\& Corwin, J.~R.\ 1976, Second reference catalogue of bright galaxies, 1976, Austin: University of Texas Press., 0 

\bibitem[Vorontsov-Velyaminov et al.(2003)]{2003yCat.7236....0V} 
Vorontsov-Velyaminov, B.~A., Noskova, R.~I., 
\& Arkhipova, V.~P.\ 2003, VizieR Online Data Catalog, 7236, 0 

\bibitem[Wei et al.(2012)]{2012ApJ...750..136W} Wei, L.~H., Keto, E.,
\& Ho, L.~C.\ 2012, \apj, 750, 136

\bibitem[White 
\& Rees(1978)]{1978MNRAS.183..341W} White, S.~D.~M., \& Rees, M.~J.\ 1978, \mnras, 183, 341 

\bibitem[Wisnioski et al.(2014)]{2014arXiv1409.6791W} Wisnioski, E., 
F{\"o}rster Schreiber, N.~M., Wuyts, S., et al.\ 2014, arXiv:1409.6791 

\bibitem[Yang et al.(1996)]{1996AJ....112..146Y} Yang, H., Chu, Y.-H., 
Skillman, E.~D., \& Terlevich, R.\ 1996, \aj, 112, 146 

\bibitem[Zaragoza-Cardiel et al.(2013)]{zaragoza13}
Zaragoza-Cardiel, J., Font-Serra, J., Beckman, J.~E., et al.\ 2013, \mnras,
432, 998

\bibitem[Zaragoza-Cardiel et al.(2014)]{zaragoza14} 
Zaragoza-Cardiel, J., Font, J., Beckman, J.~E., et al.\ 2014, \mnras, 445, 
1412 

\bibitem[Zurita et 
al.(2002)]{2002A&A...386..801Z} Zurita, A., Beckman, J.~E., Rozas, M., \& Ryder, S.\ 2002, \aap, 386, 801 


\end{thebibliography}

\appendix
\section{Properties of the H{\sc ii} regions}
%\LongTables

\begin{table*}
\centering
%  \begin{minipage}{140mm}

  \caption{Physical properties of the brightest H{\sc ii} regions in 
interacting galaxies derived as described in section $\S4$. 
The whole Table
is available as a machine readable Tablein the electronic version of the paper and through CDS. }
\resizebox{1.\linewidth}{!}{
  \begin{tabular}{@{}ccccccccccccc@{}}

  \hline
 N &Galaxy &RA&Dec & $\log(L_{\rm H\alpha}) $ & $R_{\rm{HII}}$ & $\log(M_{\rm{HII}})$ & $\rho_{HII}$  &$\sigma_{v}$& $\alpha_{\rm{vir}}$ & $\rm{EW_{H\alpha}(\rm{\AA})}$&$\log(\rm{Age})(\rm{yr})$\\
   &  &  hh:mm:ss         &     $^{\circ}$ $\mathrm{\prime}$ $\mathrm{\prime\prime}$ &  $\mathrm{erg/s}$   &   pc  &$\mathrm{M_{\odot}}$ & $\mathrm{M_{\odot}/pc^3}$ &         $\, \mathrm{km/s}$    &                                   \\
1 & NGC2146 & 6:18:22.7 & 78:22:25.6 &  $39.04\pm0.04$ &  $114\pm4$ &  $5.9\pm0.06$ & $0.127\pm0.008$ & $15.0\pm1.0$ & $40.0\pm9.0$ & 39.9&6.82\\ 
2 & NGC2146 & 6:18:31.1 & 78:21:39.6 &  $38.83\pm0.04$ &  $116\pm4$ &  $5.8\pm0.07$ & $0.097\pm0.007$ & $17.0\pm1.0$ & $60.0\pm10.0$ & 117.71&6.78\\ 
1 & Arp244 & 12:1:55.6 & -18:52:52.0 &  $40.2\pm0.03$ &  $280\pm5$ &  $7.06\pm0.05$ & $0.126\pm0.006$ & $32.8\pm0.7$ & $30.0\pm3.0$ & 202.62&6.76\\ 
2 & Arp244 & 12:1:53.9 & -18:52:42.7 &  $39.97\pm0.03$ &  $253\pm6$ &  $6.88\pm0.05$ & $0.111\pm0.006$ & $19.1\pm0.7$ & $14.0\pm2.0$ & 63.77&6.8\\ 
1 & NGC520 & 1:24:35.6 & 3:47:19.0 &  $39.12\pm0.04$ &  $255\pm9$ &  $6.46\pm0.07$ & $0.042\pm0.003$ & $14.1\pm0.8$ & $20.0\pm4.0$ & 2.1&7.19\\ 
2 & NGC520 & 1:24:35.5 & 3:47:30.2 &  $38.57\pm0.03$ &  $187\pm8$ &  $5.99\pm0.08$ & $0.035\pm0.003$ & $<4.0 $ & $  $ & 10.93&7.05\\ 
1 & UGC3995 & 7:44:11.3 & 29:15:13.1 &  $40.29\pm0.04$ &  $270\pm10$ &  $7.09\pm0.08$ & $0.14\pm0.01$ & $43.0\pm3.0$ & $50.0\pm10.0$ & 17.66&6.95\\ 
2 & UGC3995 & 7:44:11.1 & 29:14:34.6 &  $40.06\pm0.06$ &  $390\pm30$ &  $7.2\pm0.1$ & $0.064\pm0.009$ & $8.0\pm1.0$ & $1.7\pm0.9$ & 11.48&7.05\\ 
1 & NGC3788 & 11:39:42.5 & 31:54:34.0 &  $39.7\pm0.02$ &  $280\pm10$ &  $6.81\pm0.07$ & $0.072\pm0.005$ & $11.9\pm0.9$ & $7.0\pm2.0$ & 23.29&6.88\\ 
2 & NGC3788 & 11:39:42.3 & 31:54:32.6 &  $39.58\pm0.04$ &  $200\pm20$ &  $6.5\pm0.1$ & $0.1\pm0.01$ & $<4.0 $ & $  $ & 14.92&7.02\\ 
1 & NGC3786 & 11:39:45.0 & 31:55:30.7 &  $39.86\pm0.02$ &  $300\pm10$ &  $6.93\pm0.08$ & $0.078\pm0.006$ & $16.5\pm0.7$ & $11.0\pm2.0$ & 18.88&6.88\\ 
2 & NGC3786 & 11:39:45.3 & 31:55:24.0 &  $39.69\pm0.02$ &  $240\pm10$ &  $6.71\pm0.09$ & $0.088\pm0.008$ & $7.0\pm1.0$ & $3.0\pm1.0$ & 8.79&7.09\\ 
1 & NGC2782 & 9:14:5.1 & 40:6:43.8 &  $39.8\pm0.02$ &  $166\pm6$ &  $6.52\pm0.07$ & $0.17\pm0.01$ & $28.0\pm3.0$ & $40.0\pm10.0$ & 63.41&6.8\\ 
2 & NGC2782 & 9:14:4.1 & 40:6:47.2 &  $39.34\pm0.01$ &  $128\pm5$ &  $6.12\pm0.06$ & $0.15\pm0.01$ & $21.0\pm2.0$ & $50.0\pm10.0$ & 21.63&6.96\\ 
1 & NGC2993 & 9:45:48.2 & -14:22:7.8 &  $40.07\pm0.01$ &  $258\pm7$ &  $6.94\pm0.04$ & $0.121\pm0.005$ & $12.0\pm1.0$ & $5.0\pm1.0$ & 135.88&6.78\\ 
2 & NGC2993 & 9:45:50.8 & -14:23:48.2 &  $39.21\pm0.02$ &  $175\pm9$ &  $6.26\pm0.09$ & $0.081\pm0.007$ & $<4.0 $ & $  $ & 0.92&7.31\\ 
1 & NGC2992 & 9:45:42.5 & -14:19:30.5 &  $39.18\pm0.05$ &  $100\pm8$ &  $5.9\pm0.1$ & $0.18\pm0.03$ & $16.0\pm3.0$ & $40.0\pm20.0$ & 10.78&7.05\\ 
2 & NGC2992 & 9:45:42.2 & -14:19:22.6 &  $39.16\pm0.04$ &  $127\pm9$ &  $6.0\pm0.1$ & $0.12\pm0.02$ & $6.0\pm2.0$ & $5.0\pm4.0$ & 18.63&6.96\\ 
1 & NGC3991 & 11:57:29.6 & 32:19:49.4 &  $40.44\pm0.02$ &  $470\pm8$ &  $7.52\pm0.03$ & $0.076\pm0.003$ & $20.9\pm0.8$ & $7.2\pm0.9$ & 47.72&6.81\\ 
2 & NGC3991 & 11:57:29.6 & 32:19:51.7 &  $39.47\pm0.02$ &  $350\pm10$ &  $6.84\pm0.06$ & $0.039\pm0.002$ & $8.0\pm0.7$ & $<4.0\pm1.0$ & 68.51&6.8\\ 
1 & Arp270 & 10:49:54.1 & 32:59:44.4 &  $40.22\pm0.04$ &  $229\pm4$ &  $6.94\pm0.05$ & $0.175\pm0.009$ & $20.4\pm0.9$ & $13.0\pm2.0$ & 2.43&7.17\\ 
2 & Arp270 & 10:49:50.4 & 32:59:2.4 &  $40.15\pm0.04$ &  $146\pm2$ &  $6.62\pm0.05$ & $0.31\pm0.01$ & $35.0\pm1.0$ & $49.0\pm7.0$ & 96.73&6.79\\ 
1 & NGC3769 & 11:37:51.8 & 47:52:46.4 &  $38.31\pm0.05$ &  $66\pm3$ &  $5.17\pm0.09$ & $0.12\pm0.01$ & $<4.0 $ & $  $ & 227.02&6.76\\ 
2 & NGC3769 & 11:37:51.0 & 47:52:51.5 &  $38.14\pm0.05$ &  $70\pm4$ &  $5.1\pm0.1$ & $0.09\pm0.01$ & $<4.0 $ & $  $ & 148.42&6.77\\ 

  \hline
\\

  \end{tabular}
}
%     \end{minipage}
 \label{table_hii_int}
\end{table*}

\begin{table*}

\centering

  \caption{Physical properties of the brightest H{\sc ii} regions in 
isolated galaxies derived as described in section $\S4$. 
The whole Table
is available as a machine readable Tablein the electronic version of the paper and through CDS. }
\resizebox{1.\linewidth}{!}{
  \begin{tabular}{@{}ccccccccccccc@{}}

  \hline
 N &Galaxy &RA&Dec & $\log(L_{\rm H\alpha}) $ & $R_{\rm{HII}}$ & $\log(M_{\rm{HII}})$ & $\rho_{HII}$  &$\sigma_{v}$& $\alpha_{\rm{vir}}$ & $\rm{EW_{H\alpha}(\rm{\AA})}$&$\log(\rm{Age})(\rm{yr})$\\
   &  &  hh:mm:ss         &     $^{\circ}$ $\mathrm{\prime}$ $\mathrm{\prime\prime}$ &  $\mathrm{erg/s}$   &   pc  &$\mathrm{M_{\odot}}$ & $\mathrm{M_{\odot}/pc^3}$ &         $\, \mathrm{km/s}$    &                              \\
 1 & NGC3504 & 11:3:10.4 & 27:58:59.4 &  $39.61\pm0.04$ &  $162\pm5$ &  $6.41\pm0.06$ & $0.145\pm0.009$ & $14.4\pm0.9$ & $15.0\pm3.0$ & 17.03&6.98\\ 
2 & NGC3504 & 11:3:10.0 & 27:58:47.2 &  $39.3\pm0.01$ &  $80\pm7$ &  $5.8\pm0.1$ & $0.29\pm0.04$ & $6.0\pm3.0$ & $6.0\pm7.0$ & 17.89&6.96\\ 
1 & NGC5678 & 14:32:12.6 & 57:53:49.0 &  $40.07\pm0.02$ &  $430\pm9$ &  $7.28\pm0.04$ & $0.057\pm0.002$ & $13.3\pm0.3$ & $4.7\pm0.5$ & 29.49&6.83\\ 
2 & NGC5678 & 14:32:5.8 & 57:55:6.5 &  $40.0\pm0.03$ &  $380\pm10$ &  $7.15\pm0.06$ & $0.064\pm0.004$ & $13.0\pm1.0$ & $6.0\pm1.0$ & 5.58&7.1\\ 
1 & NGC5921 & 15:21:56.6 & 5:4:12.4 &  $39.22\pm0.02$ &  $320\pm10$ &  $6.67\pm0.06$ & $0.033\pm0.002$ & $<4.0 $ & $  $ & 1.99&7.2\\ 
2 & NGC5921 & 15:21:53.6 & 5:3:22.3 &  $39.15\pm0.03$ &  $289\pm9$ &  $6.56\pm0.06$ & $0.036\pm0.002$ & $<4.0 $ & $  $ & 16.31&7.0\\ 
1 & NGC6070 & 16:9:55.2 & 0:42:3.3 &  $39.57\pm0.04$ &  $331\pm9$ &  $6.86\pm0.06$ & $0.047\pm0.003$ & $18.0\pm0.9$ & $17.0\pm3.0$ & 79.3&6.8\\ 
2 & NGC6070 & 16:9:56.4 & 0:42:27.8 &  $39.45\pm0.04$ &  $279\pm9$ &  $6.68\pm0.07$ & $0.053\pm0.004$ & $10.4\pm0.7$ & $7.0\pm2.0$ & 120.82&6.78\\ 
1 & NGC4151 & 12:10:32.6 & 39:24:21.1 &  $38.21\pm0.01$ &  $148\pm5$ &  $5.65\pm0.06$ & $0.033\pm0.002$ & $<4.0 $ & $  $ & 126.03&6.78\\ 
2 & NGC4151 & 12:10:32.6 & 39:24:20.8 &  $38.18\pm0.02$ &  $107\pm6$ &  $5.43\pm0.09$ & $0.052\pm0.005$ & $<4.0 $ & $  $ & 138.64&6.78\\ 
1 & NGC864 & 2:15:27.6 & 6:0:8.8 &  $40.01\pm0.03$ &  $204\pm4$ &  $6.76\pm0.04$ & $0.162\pm0.007$ & $27.2\pm0.8$ & $30.0\pm4.0$ & 54.09&6.81\\ 
2 & NGC864 & 2:15:28.7 & 6:0:43.3 &  $39.78\pm0.04$ &  $291\pm6$ &  $6.88\pm0.05$ & $0.073\pm0.004$ & $17.5\pm0.7$ & $14.0\pm2.0$ & 27.61&6.83\\ 
1 & NGC2543 & 8:12:56.2 & 36:14:42.7 &  $39.13\pm0.04$ &  $220\pm20$ &  $6.4\pm0.1$ & $0.052\pm0.007$ & $<4.0 $ & $  $ & 36.42&6.82\\ 
2 & NGC2543 & 8:12:57.2 & 36:14:48.6 &  $39.07\pm0.02$ &  $320\pm10$ &  $6.59\pm0.08$ & $0.027\pm0.002$ & $<4.0 $ & $  $ & 32.49&6.82\\ 
1 & NGC2748 & 9:13:32.6 & 76:27:55.0 &  $39.33\pm0.03$ &  $292\pm7$ &  $6.66\pm0.05$ & $0.043\pm0.002$ & $11.0\pm0.8$ & $9.0\pm2.0$ & 9.14&7.09\\ 
2 & NGC2748 & 9:13:30.6 & 76:27:41.9 &  $39.23\pm0.03$ &  $270\pm9$ &  $6.55\pm0.07$ & $0.043\pm0.003$ & $6.0\pm0.7$ & $3.0\pm1.0$ & 9.95&7.07\\ 
1 & NGC3041 & 9:53:2.4 & 16:39:41.7 &  $39.02\pm0.03$ &  $293\pm7$ &  $6.5\pm0.06$ & $0.03\pm0.002$ & $4.5\pm0.7$ & $2.2\pm0.8$ & 22.31&6.93\\ 
2 & NGC3041 & 9:53:6.5 & 16:41:35.4 &  $39.02\pm0.02$ &  $381\pm7$ &  $6.67\pm0.03$ & $0.0202\pm0.0007$ & $<4.0 $ & $  $ & 1.4&7.25\\ 
1 & NGC2712 & 8:59:33.4 & 44:55:27.5 &  $39.65\pm0.04$ &  $230\pm10$ &  $6.7\pm0.1$ & $0.09\pm0.009$ & $12.2\pm0.8$ & $9.0\pm2.0$ & 51.86&6.81\\ 
2 & NGC2712 & 8:59:32.7 & 44:55:14.4 &  $39.21\pm0.03$ &  $300\pm8$ &  $6.61\pm0.05$ & $0.036\pm0.002$ & $11.9\pm0.8$ & $12.0\pm3.0$ & 38.87&6.82\\ 
1 & NGC5740 & 14:44:23.7 & 1:40:52.4 &  $39.33\pm0.02$ &  $185\pm5$ &  $6.35\pm0.05$ & $0.086\pm0.004$ & $42.0\pm2.0$ & $170.0\pm30.0$ & 39.41&6.82\\ 
2 & NGC5740 & 14:44:25.3 & 1:40:51.6 &  $39.18\pm0.02$ &  $361\pm9$ &  $6.72\pm0.04$ & $0.026\pm0.001$ & $5.6\pm0.5$ & $2.5\pm0.6$ & 15.91&6.92\\ 
1 & NGC6412 & 17:29:27.3 & 75:41:25.2 &  $39.96\pm0.03$ &  $404\pm7$ &  $7.18\pm0.04$ & $0.055\pm0.002$ & $19.3\pm0.7$ & $12.0\pm2.0$ & 2.48&7.17\\ 
2 & NGC6412 & 17:29:27.4 & 75:41:25.4 &  $39.33\pm0.02$ &  $340\pm10$ &  $6.75\pm0.06$ & $0.035\pm0.002$ & $6.4\pm0.9$ & $3.0\pm1.0$ & 2.39&7.17\\ 
1 & NGC6207 & 16:43:3.5 & 36:50:10.5 &  $39.51\pm0.03$ &  $154\pm6$ &  $6.33\pm0.07$ & $0.14\pm0.01$ & $18.0\pm1.0$ & $27.0\pm6.0$ & 120.66&6.78\\ 
2 & NGC6207 & 16:43:3.8 & 36:50:3.4 &  $39.36\pm0.03$ &  $131\pm4$ &  $6.15\pm0.06$ & $0.148\pm0.009$ & $13.9\pm0.9$ & $21.0\pm5.0$ & 0.51&7.38\\ 
1 & NGC1073 & 2:43:38.3 & 1:23:38.6 &  $39.22\pm0.03$ &  $124\pm3$ &  $6.04\pm0.05$ & $0.138\pm0.007$ & $13.4\pm0.5$ & $23.0\pm3.0$ & 1868.64&6.39\\ 
2 & NGC1073 & 2:43:39.6 & 1:23:43.3 &  $38.87\pm0.06$ &  $87\pm3$ &  $5.64\pm0.08$ & $0.16\pm0.01$ & $9.9\pm0.6$ & $23.0\pm5.0$ & 521.43&6.59\\ 
1 & NGC3423 & 10:51:19.2 & 5:51:8.5 &  $39.64\pm0.02$ &  $407\pm5$ &  $7.03\pm0.03$ & $0.038\pm0.001$ & $14.5\pm0.3$ & $9.3\pm0.8$ & 17.46&6.95\\ 
2 & NGC3423 & 10:51:18.9 & 5:51:19.4 &  $39.37\pm0.02$ &  $236\pm8$ &  $6.54\pm0.06$ & $0.063\pm0.004$ & $6.7\pm0.6$ & $<4.0\pm1.0$ & 16.14&6.9\\ 
1 & NGC428 & 1:12:53.2 & 0:56:47.5 &  $39.27\pm0.02$ &  $167\pm2$ &  $6.26\pm0.03$ & $0.093\pm0.003$ & $10.8\pm0.4$ & $12.0\pm1.0$ & 28.12&6.83\\ 
2 & NGC428 & 1:12:53.2 & 0:56:47.5 &  $39.03\pm0.03$ &  $155\pm4$ &  $6.09\pm0.06$ & $0.08\pm0.005$ & $5.7\pm0.6$ & $5.0\pm1.0$ & 28.22&6.83\\ 
1 & NGC5334 & 13:52:55.8 & -1:5:18.9 &  $38.93\pm0.01$ &  $171\pm6$ &  $6.11\pm0.06$ & $0.061\pm0.004$ & $<4.0 $ & $  $ & 57.57&6.8\\ 
2 & NGC5334 & 13:52:52.3 & -1:6:46.8 &  $38.81\pm0.02$ &  $172\pm7$ &  $6.05\pm0.07$ & $0.053\pm0.004$ & $<4.0 $ & $  $ & 29.0&6.83\\ 
1 & NGC5112 & 13:22:2.4 & 38:43:15.2 &  $39.55\pm0.04$ &  $288\pm7$ &  $6.75\pm0.06$ & $0.057\pm0.003$ & $11.6\pm0.5$ & $8.0\pm1.0$ & 155.01&6.77\\ 
2 & NGC5112 & 13:22:2.0 & 38:43:15.0 &  $39.54\pm0.04$ &  $215\pm6$ &  $6.56\pm0.06$ & $0.088\pm0.005$ & $12.4\pm0.6$ & $11.0\pm2.0$ & 137.97&6.78\\ 
1 & NGC3403 & 10:53:56.6 & 73:41:27.7 &  $38.78\pm0.05$ &  $200\pm10$ &  $6.1\pm0.1$ & $0.041\pm0.005$ & $9.2\pm0.8$ & $15.0\pm5.0$ & 13.86&7.03\\ 
2 & NGC3403 & 10:53:58.5 & 73:41:19.6 &  $38.71\pm0.05$ &  $171\pm9$ &  $6.0\pm0.1$ & $0.048\pm0.005$ & $4.2\pm0.8$ & $<4.0\pm2.0$ & 39.29&6.82\\ 
1 & NGC4639 & 12:42:55.5 & 13:14:53.2 &  $38.63\pm0.08$ &  $160\pm9$ &  $5.9\pm0.1$ & $0.047\pm0.006$ & $10.8\pm0.8$ & $26.0\pm8.0$ & 50.94&6.81\\ 
2 & NGC4639 & 12:42:49.1 & 13:16:7.4 &  $38.33\pm0.03$ &  $238\pm6$ &  $6.02\pm0.05$ & $0.0186\pm0.0009$ & $7.8\pm0.9$ & $16.0\pm5.0$ & 75.81&6.8\\ 
1 & NGC7241 & 22:15:49.3 & 19:13:49.7 &  $38.64\pm0.03$ &  $126\pm5$ &  $5.76\pm0.07$ & $0.069\pm0.005$ & $<4.0 $ & $  $ & 130.24&6.78\\ 
2 & NGC7241 & 22:15:49.7 & 19:13:48.1 &  $38.54\pm0.08$ &  $106\pm5$ &  $5.6\pm0.1$ & $0.08\pm0.008$ & $8.0\pm2.0$ & $20.0\pm10.0$ & 94.91&6.79\\ 
2 & NGC918 & 2:25:49.6 & 18:30:25.2 &  $38.58\pm0.04$ &  $87\pm3$ &  $5.49\pm0.07$ & $0.111\pm0.007$ & $10.2\pm0.7$ & $34.0\pm8.0$ & 52.28&6.81\\ 
1 & NGC4324 & 12:23:6.2 & 5:15:1.7 &  $39.31\pm0.02$ &  $165\pm5$ &  $6.28\pm0.05$ & $0.1\pm0.005$ & $<4.0 $ & $  $ & 5.74&7.1\\ 
2 & NGC4324 & 12:23:5.1 & 5:14:53.4 &  $37.97\pm0.06$ &  $84\pm5$ &  $5.2\pm0.1$ & $0.058\pm0.007$ & $<4.0 $ & $  $ & 10.56&7.06\\ 
1 & NGC4389 & 12:25:31.5 & 45:43:5.6 &  $38.98\pm0.02$ &  $161\pm6$ &  $6.09\pm0.07$ & $0.071\pm0.005$ & $13.0\pm1.0$ & $25.0\pm6.0$ & 12.86&7.04\\ 
2 & NGC4389 & 12:25:31.4 & 45:43:5.1 &  $38.59\pm0.01$ &  $199\pm9$ &  $6.03\pm0.07$ & $0.033\pm0.002$ & $4.6\pm0.7$ & $<4.0\pm2.0$ & 15.44&6.94\\ 
1 & NGC4498 & 12:31:39.5 & 16:51:22.2 &  $38.7\pm0.03$ &  $120\pm3$ &  $5.76\pm0.06$ & $0.079\pm0.005$ & $13.3\pm0.7$ & $43.0\pm8.0$ & 30.39&6.83\\ 
2 & NGC4498 & 12:31:41.4 & 16:50:51.4 &  $38.6\pm0.06$ &  $80\pm3$ &  $5.44\pm0.09$ & $0.13\pm0.01$ & $5.7\pm0.6$ & $11.0\pm3.0$ & 70.1&6.8\\ 
1 & NGC2541 & 8:14:40.4 & 49:2:30.7 &  $39.48\pm0.04$ &  $160\pm3$ &  $6.34\pm0.05$ & $0.127\pm0.006$ & $18.4\pm0.3$ & $29.0\pm3.0$ & 44.26&6.81\\ 
2 & NGC2541 & 8:14:36.3 & 49:3:7.6 &  $39.19\pm0.03$ &  $119\pm3$ &  $6.0\pm0.06$ & $0.143\pm0.009$ & $14.2\pm0.5$ & $28.0\pm5.0$ & 90.61&6.79\\ 
1 & NGC2500 & 8:1:46.9 & 50:44:3.3 &  $38.44\pm0.04$ &  $76\pm4$ &  $5.3\pm0.1$ & $0.12\pm0.01$ & $7.7\pm0.6$ & $24.0\pm7.0$ & 214.74&6.76\\ 
2 & NGC2500 & 8:1:59.7 & 50:45:57.8 &  $38.4\pm0.04$ &  $99\pm2$ &  $5.49\pm0.05$ & $0.075\pm0.004$ & $9.4\pm0.8$ & $33.0\pm8.0$ & 20.71&6.96\\

  \hline
\\

  \end{tabular}
  }

 \label{table_hii_iso}
\end{table*}

\label{lastpage}

\end{document}